%% file: main.tex
\documentclass[11pt]{article}
\usepackage[a4paper,margin=24mm]{geometry}
\usepackage{amsmath,amssymb,graphicx,booktabs,natbib}
\usepackage{url}
\usepackage{placeins}
\usepackage{float}
\usepackage{adjustbox}
\let\SECApreprintTabular\tabular
\let\SECApreprintEndTabular\endtabular
\renewenvironment{tabular}[2][]{%
  \adjustbox{max width=\textwidth}\bgroup\SECApreprintTabular[#1]{#2}%
}{\SECApreprintEndTabular\egroup}
\usepackage{alphalph}
\newif\ifTOG\TOGfalse
\newcommand{\Description}[1]{}
\input{common/dep_numbers}
\title{SECA: Strict-Elastic Compact Activation for Differentiable Elastoplastic Deformation}
\author{Yu Ren \and Shuangjiu Xiao \and Deli Dong\\
Shanghai Jiao Tong University, China}
\date{October 2026}
\begin{document}
\maketitle
\begin{abstract}
\input{common/abstract}
\end{abstract}
\begin{figure}[t]
  \centering
  \includegraphics[width=0.853\linewidth]{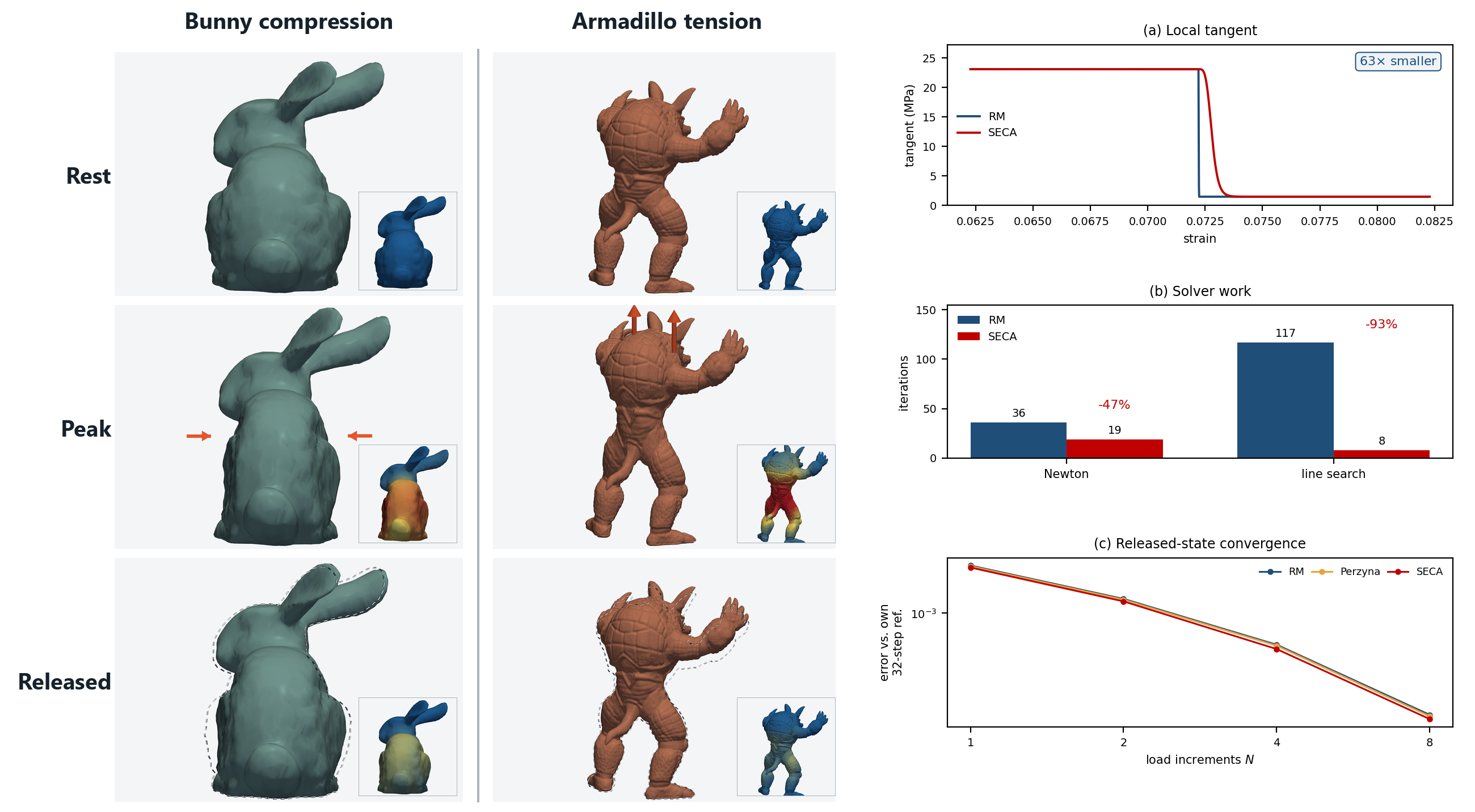}
  \caption{\textbf{SECA for elastoplastic deformation: macroscopic load--release
  response and numerical regularity across the elastic--plastic transition.}
  \textbf{Left:} Bunny compression and Armadillo tension, each stepping from rest
  through peak loading to release; every state carries a small Kirchhoff von Mises
  inset and the insets of one model share a colour scale. \textbf{Right:} SECA
  against classical return mapping (RM) and linear-overstress Perzyna in the
  local, solver-work and released-state comparisons recorded in the supplement's
  protocol table. Panel (a) shows the local tangent near yield.
  At the measurement window of Section~\ref{sec:material-point-comparison} the
  measured secant-slope change $\Delta C_{\mathrm{tr}}$ of SECA is about
  $1/\depTangJumpRatio$ of RM's; panel (b) gives the nonlinear solver work of the
  matched transition-crossing run, where in the tested coarse cube-indentation run
  the counts cover the whole history while its load phase is a single increment, so
  SECA requires \depNewtonSaved\% fewer Newton
  iterations and \depLineSearchSaved\% fewer line-search backtracking halvings;
  and panel (c) reports released-state convergence on the controlled cube
  indentation, each update measured against its own 32-increment reference.}
  \Description{Left: rest, peak, and released states of the Stanford Bunny and
  Armadillo, each with a Kirchhoff von Mises inset. Right: local response
  tangent, solver work, and released-state convergence.}
  \label{fig:authoring}
\end{figure}
% Body: the section list of the journal manuscript.
\input{common/introduction}
\input{common/related_work}
\input{common/method}
\input{common/regularity_bias}
\input{common/finite_method}
\input{common/contact_sensitivity}
\section{Results}\label{sec:results}
\input{common/results_local}
\input{common/results_cost}

\input{common/results_derivative}
\input{common/macroscopic}
\input{common/discussion}
\input{common/conclusion}
% Appendices: the material the journal manuscript carries in its supplement, kept
% here so the preprint is self-contained; the last two files are arXiv-only.
\input{common/appendix}
\input{common/appendix_moves_a}
\input{common/appendix_moves_b}

\input{common/appendix_moves_main}
\input{common/appendix_n1}
\input{common/appendix_n2}

% Past the 26th appendix the class's own \Alph runs out; alphalph prints AA, AB,
% ... while the letters already used above are unchanged.
\renewcommand{\thesection}{\AlphAlph{\value{section}}}
\input{common/appendix_selection}
\input{common/appendix_contact_diagnostics}
\bibliographystyle{plainnat}
\bibliography{references}
\end{document}

%% file: common/dep_numbers.tex
\ExplSyntaxOn
\NewDocumentCommand{\depsci}{m}{
  \tl_set:Nx \l_tmpa_tl {#1}
  \regex_replace_once:nnN {e\+?0*(\d+)} {\c{times}10\cB\{\1\cE\}} \l_tmpa_tl
  \ensuremath{\l_tmpa_tl}
}
\ExplSyntaxOff

\newcommand{\depTangJumpRM}{1.527e+07}
\newcommand{\depTangJumpPerzyna}{1.585e+07}
\newcommand{\depTangJumpDEP}{2.442e+05}
\newcommand{\depTangJumpRatio}{63}

\newcommand{\depNewtonRM}{36}
\newcommand{\depLineSearchRM}{117}
\newcommand{\depCoarseErrRM}{1.782e-03}
\newcommand{\depFineErrRM}{0.000e+00}
\newcommand{\depNewtonPerzyna}{27}
\newcommand{\depLineSearchPerzyna}{37}
\newcommand{\depCoarseErrPerzyna}{1.766e-03}
\newcommand{\depFineErrPerzyna}{0.000e+00}
\newcommand{\depNewtonDEP}{19}
\newcommand{\depLineSearchDEP}{8}
\newcommand{\depCoarseErrDEP}{1.739e-03}
\newcommand{\depFineErrDEP}{0.000e+00}

\newcommand{\depNewtonSaved}{47}
\newcommand{\depLineSearchSaved}{93}
\newcommand{\depBunnyPeak}{21.294}
\newcommand{\depBunnyRetained}{12.716}
\newcommand{\depBunnyRecovery}{8.578}
\newcommand{\depArmadilloPeak}{29.062}
\newcommand{\depArmadilloRetained}{9.104}
\newcommand{\depArmadilloRecovery}{19.958}
\newcommand{\depLoadCount}{5}
\newcommand{\depLoadDepthMin}{0.05}
\newcommand{\depLoadDepthMax}{0.45}
\newcommand{\depLoadDentMin}{0.00}
\newcommand{\depLoadDentMax}{19.72}

\newcommand{\depYieldLo}{0.300}
\newcommand{\depYieldHi}{2.000}
\newcommand{\depYieldFirst}{8.893}
\newcommand{\depYieldLast}{0.000}

\newcommand{\depWidthCoupledLo}{0.010}
\newcommand{\depWidthCoupledHi}{0.240}
\newcommand{\depWidthCoupledFirst}{8.661}
\newcommand{\depWidthCoupledLast}{8.481}

\newcommand{\depWidthFixedLo}{0.010}
\newcommand{\depWidthFixedHi}{0.240}
\newcommand{\depWidthFixedFirst}{8.642}
\newcommand{\depWidthFixedLast}{8.559}

\newcommand{\depHardLo}{0.25}
\newcommand{\depHardHi}{4.00}
\newcommand{\depHardStressFirst}{0.653}
\newcommand{\depHardStressLast}{0.865}
\newcommand{\depHardStressRel}{32.6}
\newcommand{\depHardFirst}{11.717}
\newcommand{\depHardLast}{9.693}

\newcommand{\depFamilyCount}{4}

\newcommand{\depFamilyStressMin}{2.091}
\newcommand{\depFamilyStressMax}{8.826}
\newcommand{\depFamilyYoungMin}{0.50}
\newcommand{\depFamilyYoungMax}{4.00}
\newcommand{\depFamilyYieldMin}{0.60}
\newcommand{\depFamilyYieldMax}{2.50}

\newcommand{\depFamilyTrendMin}{56.45}
\newcommand{\depFamilyTrendMax}{78.37}

\newcommand{\depCubeTetrahedra}{384}
\newcommand{\depCubeNodes}{125}

\newcommand{\depCubeMinimumDetF}{0.5285}

\newcommand{\depSphereMinimumDetF}{0.7229}

\newcommand{\depTorusMinimumDetF}{0.6850}

\newcommand{\depBeamMinimumDetF}{0.9408}

\newcommand{\depCubePeakPct}{25.05}
\newcommand{\depCubeReleasedPct}{18.30}
\newcommand{\depCubeRetainedPct}{73.05}

\newcommand{\depSpherePeakPct}{34.57}
\newcommand{\depSphereReleasedPct}{24.24}
\newcommand{\depSphereRetainedPct}{70.11}

\newcommand{\depTorusPeakPct}{25.00}
\newcommand{\depTorusReleasedPct}{15.51}
\newcommand{\depTorusRetainedPct}{62.05}

\newcommand{\depBeamPeakPct}{25.00}
\newcommand{\depBeamReleasedPct}{4.48}
\newcommand{\depBeamRetainedPct}{17.93}

\newcommand{\depNewtonCrossingRM}{32}
\newcommand{\depNewtonCrossingDEP}{10}
\newcommand{\depLineSearchCrossingRM}{117}
\newcommand{\depLineSearchCrossingDEP}{7}
\newcommand{\depDampingRM}{3}
\newcommand{\depDampingPerzyna}{4}

\newcommand{\depMpAlphaRM}{0.15595}
\newcommand{\depMpAlphaPerzyna}{0.15523}
\newcommand{\depMpAlphaDEP}{0.15140}
\newcommand{\depMpResidualRM}{-1.9162}
\newcommand{\depMpResidualPerzyna}{-1.9213}
\newcommand{\depMpResidualDEP}{-1.9500}
\newcommand{\depMpZeroStressRM}{0.119493}
\newcommand{\depMpZeroStressPerzyna}{0.119240}
\newcommand{\depMpZeroStressDEP}{0.117950}

\newcommand{\depNewtonSavedCrossing}{69}
\newcommand{\depLineSearchSavedCrossing}{94}
\newcommand{\depNewtonRatioRefinedMin}{1.02}
\newcommand{\depNewtonRatioRefinedMax}{1.10}

\newcommand{\depTangJumpPctRM}{1.6}
\newcommand{\depTangJumpPctPerzyna}{1.5}

\newcommand{\depMpResidualGainRM}{1.8}
\newcommand{\depMpResidualGainPerzyna}{1.5}
\newcommand{\depMpZeroStressBelowRM}{1.3}

%% file: common/abstract.tex
Return mapping keeps an exact elastic domain but switches its response
tangent at yield, while smoothing laws with a sub-yield tail admit plastic
flow below the threshold. We present
SECA, a Strict-Elastic Compact Activation method for differentiable elastoplastic
simulation. SECA prescribes a bounded $C^2$ activation over a finite
positive-overstress interval and integrates it into a Perzyna-type flow law, so the
implicit update returns exactly zero plastic increment for sub-yield trials while
smoothing the onset. For scalar and proportional isotropic $J_2$ updates we
derive an exact decomposition of the plastic-increment discrepancy into
finite-width and viscosity contributions, and an explicit algebraic budget for the
transition width. The update is embedded in finite-deformation
loading histories with contact and complete unloading, and differentiated to
released equilibria. Controlled indentation results stay closely matched. In
the tested coarse cube-indentation run, where the
transition is crossed without intermediate increments, SECA reduces whole-run
Newton iterations by \depNewtonSaved\% and backtracking halvings by
\depLineSearchSaved\% against return mapping. A four-method comparison evaluates
released-shape error and computational cost across loading partitions, and
on a contact scene the path derivative drives a bounded control task
to its target in three iterations. Further examples show elastic recovery and
residual deformation under compression, tension and bending.
% Neighbour-comparison note: the four-method sentence refers to the comparison
% recorded in SECA_CL2.0/derived/n1_scenes.json; run counts, reference level and
% per-increment numbers are in Section~\ref{sec:n1-comparison}.
% Control-task note: the clause above reports SECA_CL2.0/derived/n2_control.json;
% the measured band, the derivative cross-check and the recorded curves are in
% Section~\ref{sec:n2-control}.

%% file: common/introduction.tex
\section{Introduction}
Elastoplastic load--release simulation has to linearise a history-dependent local
update at every iteration of an accepted time partition, carry committed plastic
history through the loading history, and return the released equilibrium rather
than the loaded pose \citep{Li2020,Li2022ECI}. The local update enters that chain
throughout: the solver
linearises it during each Newton iteration, the committed plastic multiplier fixes
the stress-free configuration that unloading relaxes towards, and the released
deformation a user finally sees follows from the whole evolution under the given
simulation settings. Gradient-based editing and inverse design add one more
requirement, because the released quantity must be differentiated with respect to
the prescribed tool controls through the loading history and the complete
unloading that follows it.
Figure~\ref{fig:authoring} shows both sides of the paper: the macroscopic
load--release evolution with its stress distributions, and the numerical
properties of the transition, through which irreversible history is accumulated.

Three properties are in tension once the update is discretised, and the nearest
constitutive methods each give one of them up. Exact return mapping keeps an exact
elastic domain, but its algorithmic tangent switches branch at yield, so the
linearisation the solver uses is locally nonsmooth and sensitivities to loads and
material parameters jump there. Linear-overstress Perzyna returns exactly zero
plastic multiplier below yield, because its positive part vanishes identically,
but that positive part leaves a kink at the onset: the transition has no finite
width and the local derivative stays discontinuous. Smoothing laws whose tail does
not vanish at negative overstress, such as the softplus form, remove the hard onset
but admit a positive plastic multiplier below the threshold, so states that should
be recoverable keep accumulating plastic strain. Other nonlinear rate functions
that are both strictly elastic and locally smooth also exist, and the controlled
comparison in this work is designed to include one such neighbour rather than
presenting the choice as lying between return mapping and smoothing.
% N1-PLACEHOLDER: when the neighbour comparison of plan section 8 is complete,
% name the fourth rate function here and state its role; report no outcome before
% the records exist.

Our construction prescribes the slope of the overstress law before defining the
flow itself. The activation vanishes below yield, varies over a finite
positive-overstress interval, and reaches unity above it. Integrating this
activation produces a flow law that retains an exact elastic domain and has an
explicit post-transition offset; for the chosen quintic activation this offset is
$\delta/2$, which enables an exact local separation of finite-width and viscosity
contributions.

Embedding the local computation in a finite-deformation solver connects it to the
multiplicative plastic history, the element residual and contact, and
differentiating an accepted time partition gives the derivatives of the released
equilibrium with respect to the tool controls,
\[
D_\delta\longrightarrow\gamma\longrightarrow\boldsymbol F_p
\longrightarrow\boldsymbol P\longrightarrow\boldsymbol R\longrightarrow
\frac{\partial Q_{\mathrm{rel}}}{\partial\boldsymbol\theta}.
\]
The controlled comparison runs exact return mapping, linear-overstress Perzyna and
SECA on one mesh, one tool path, one time step, one tolerance and one line search,
so the constitutive method is the only difference. The tested scope covers a
controlled indentation benchmark, simple bodies under compression, tension,
bending and grip loading, two character-scale models, and the local and
released-state derivatives that are measured; the solver-work saving belongs to
the coarse schedule that crosses the transition without intermediate increments.

Our contributions are three.
{\emergencystretch=2em
\begin{itemize}
\item \emph{C1: Strict-elastic transition regularisation.} We replace the
activation through which an overstress drives plastic flow by a bounded $C^2$ gate
$D_\delta$ that is identically zero for $f\leq0$ and identically one for
$f\geq\delta$, integrate that gate into the primitive $r_\delta$, and take the
implicit multiplier of the resulting flow law. Strict elasticity is a property of
the discrete update: a sub-yield trial returns a plastic increment of exactly
zero, not merely a small one.
\item \emph{C2: Exact local regularity and bias characterisation.} At a common
trial state in the scalar and proportional isotropic $J_2$ setting, the
plastic-increment discrepancy against linear-overstress Perzyna obeys the
two-sided bracket
\[
0\leq\gamma_L-\gamma
\leq\frac{\delta}{2\,(B+\eta/\Delta t)},
\]
and its algebraic corollary for a prescribed local plastic-increment budget
$\epsilon_\gamma$ at fixed $B$, $\eta$ and $\Delta t$ is
$\delta\leq2\,(B+\eta/\Delta t)\,\epsilon_\gamma$. Both statements are local: they
bound the multiplier of one trial state through the elastic and hardening
constants, they are not a global geometric error guarantee, and they do not by
themselves select an optimal width.
\item \emph{C3: Finite-deformation integration and targeted evaluation.} We carry
the same local computation through multiplicative plastic history, committed
history and released-equilibrium differentiation, and evaluate it against exact
return mapping and linear-overstress Perzyna on common loading histories, with a
fourth, strictly elastic reference update added only for the error-and-cost
comparison.
\end{itemize}
At fixed viscosity and time discretisation, the transition width controls local
regularity and finite-width bias while preserving exact sub-yield inactivity. It
does not remove every trade-off.\par}

%% file: common/related_work.tex
\section{Related Work}
\subsection{Constitutive Integration and Regularisation}
{\emergencystretch=2em%
Return mapping is a classical and widely used numerical procedure for
elastoplastic simulation, together with a consistent algorithmic tangent
\citep{SimoHughes1998,SimoTaylor1985}. It enforces
the yield condition to machine precision; its radial form for isotropic $J_2$ is
cheap, and its tangent is exact wherever the active set is constant. It is the
reference implementation and our baseline throughout, not a straw man. Its cost
is structural: the active set is a discrete object, so the tangent is one value
inside the elastic domain and another on the plastic branch, and it is
discontinuous at the yield surface. A Newton solve crossing that surface in one
increment sees this less smooth linearisation, the behaviour
Section~\ref{sec:material-point-comparison} measures.%
\par}

Perzyna's formulation makes irreversible flow depend on overstress
\citep{Perzyna1963,Perzyna1966}, and implicit integration with
algorithm-consistent tangents for viscous plasticity is established
\citep{HeeresDeBorst2002}; finite-strain viscoplastic model and discretisation
errors have been studied in a goal-oriented hierarchy
\citep{LarssonRunesson2004}. SECA uses a compact rate function in that
Perzyna-type lineage, and a smooth Perzyna implementation with the identical
$r_\delta$, viscosity, history update and time discretisation is locally the
same model, which is why we evaluate linear-overstress Perzyna under the same
protocol instead of claiming a separate family. Smoothing also joins multiple
yield surfaces: Wilkins et al.\ derive a convex $C^2$ smoothing and study its
approximation and solver behaviour \citep{Wilkins2020}. Their object is a smooth
yield envelope; SECA is a one-sided compact activation of positive overstress,
whose exact saturation gives its primitive a finite post-transition offset,
enabling the decomposition of Section~\ref{sec:regularity-bias}.

\subsection{Inelastic Graphics}
Graphics elastoplasticity includes multiplicative MPM for snow
\citep{Stomakhin2013} and Drucker--Prager sand \citep{Klar2016}, and more recent
work develops energetically consistent inelasticity for graphics time
integration \citep{Li2022ECI}. A public preprint in this research lineage used
noncompact softplus activation and a maximum-history navigation formulation
\citep{Ren2026SmoothExplicit}, whereas SECA centres compact strict elasticity,
exact finite-width analysis, and a finite-$\boldsymbol F_p$
released-equilibrium derivative.

\subsection{Differentiable Simulation and Contact}
Differentiable simulators propagate sensitivities through constitutive systems
that evolve in time \citep{Hu2020,HuangDiffSolver2024}, and generalised
differentiable constitutive models support inverse learning across elastic,
plastic, and viscous behaviour \citep{Su2023}. SECA uses established implicit
differentiation but applies it to the actual residual and every committed state
along complete loading and withdrawal, the sensitivity an inverse-design loop
consumes at the released equilibrium.
Intersection-aware contact matters when tools leave residual deformation
\citep{Li2020}. We use IPC geometric primitives for distances, barrier
derivatives, and continuous collision detection, retaining a multiplicative
Hencky elastoplastic residual. Contact remains enabling: the analysed object is
the compact activation and its propagation to released equilibrium.

%% file: common/method.tex
\section{SECA Local Constitutive Method}
\label{sec:strict-elastic}

% A dedicated counter keeps the local-update box referable without pulling in an
% algorithm package; the guard lets the same source be shared by the profiles.
\makeatletter
\@ifundefined{c@localalgorithm}{\newcounter{localalgorithm}}{}
\makeatother

One evaluation of the local method maps the trial deformation of a single material
point and the history committed at the previous accepted increment to the plastic
multiplier of this increment, the candidate history, the stress and the local
derivative of that map. SECA changes one ingredient, the \emph{activation} through
which an overstress drives plastic flow; the $J_2$ energy and yield function
\citep{SimoHughes1998}, the associative radial flow
\citep{Perzyna1963,Perzyna1966} and the implicit time integration
\citep{SimoTaylor1985,deSouzaNeto2008} are standard. The quintic below is classical;
the contribution is its use as a strictly elastic compact gate, the exact bias
analysis of its finite window (\S\ref{sec:regularity-bias}) and the embedding of
both in a released-equilibrium solver (\S\ref{sec:finite-update}).
Algorithm~\ref{alg:local} describes SECA alone; the interface shared by the three
constitutive methods of Section~\ref{sec:results} follows the box.

\subsection{State, Yield Function and Driving Stress}
For an isotropic small-strain material, let $\boldsymbol\epsilon$ be the total
strain, $\boldsymbol p$ the deviatoric plastic strain and $a\geq0$ the accumulated
equivalent plastic strain. Following the standard small-strain elastoplastic
framework of de Souza Neto et al.~\cite{deSouzaNeto2008}, we adopt a quadratic
stored free-energy density with linear isotropic hardening. With the internal
state held fixed, the stored free-energy density, the stress it defines and the
driving overstress are
\begin{equation}
\begin{aligned}
  \Psi(\boldsymbol\epsilon,\boldsymbol p,a)
  &=\tfrac12(\boldsymbol\epsilon-\boldsymbol p):\mathbb C:
    (\boldsymbol\epsilon-\boldsymbol p)+\tfrac12 H a^2 ,\\
  \boldsymbol\sigma&=\partial_{\boldsymbol\epsilon}\Psi,\qquad
  \boldsymbol s=\operatorname{dev}\boldsymbol\sigma,\qquad
  \sigma_{\mathrm{eq}}=\sqrt{3/2}\,\|\boldsymbol s\|,\\
  f&=\sigma_{\mathrm{eq}}-\sigma_y-Ha ,
\end{aligned}
\label{eq:yield}
\end{equation}
The first term is the elastic stored energy and the second is the energy stored
by isotropic hardening. Here $\mathbb C$ is the isotropic elasticity tensor built
from the Young modulus
$E$ and the Poisson ratio $\nu$ through the shear and bulk moduli
$\mu=E/[2(1+\nu)]$ and $K=E/[3(1-2\nu)]$, $H$ is the isotropic hardening modulus and
$\sigma_y$ the initial yield stress. Stress-valued quantities, including $f$ and
$\sigma_{\mathrm{eq}}$, are in pascals; strains and $a$ are dimensionless; the
viscosity $\eta$ is in pascal-seconds and the increment $\Delta t$ in seconds.
Unloading keeps $\boldsymbol p$ and $a$ separate: neither is replaced with a
reversible function of the current strain.

\subsection{Compact Activation and Primitive}
Let $\delta>0$ be the transition width, in the same unit as $f$, and let
$\xi=f/\delta$ be the normalised overstress. The implemented activation, or gate, is
\begin{equation}
  D_\delta(f)=
  \begin{cases}
    0, &f\leq0,\\
    6\xi^5-15\xi^4+10\xi^3, &0<f<\delta,\\
    1, &f\geq\delta.
  \end{cases}
  \label{eq:activation}
\end{equation}
It is bounded, monotone in overstress, globally $C^2$ and exactly zero for
$f\leq0$, so a sub-yield trial returns an exactly zero multiplier rather than a
small one. The quintic is the lowest-degree polynomial matching $P(0)=0$, $P(1)=1$
with vanishing first and second derivatives at both ends, and the supplement solves
that endpoint system. Its non-trivial variation is confined to $0<f<\delta$, the
window analysed in \S\ref{sec:regularity-bias}.

The primitive of the activation, normalised by $r_\delta(0)=0$, is
\begin{equation}
r_\delta(f)=
\begin{cases}
0,&f\leq0,\\
\delta(\tfrac52\xi^4-3\xi^5+\xi^6),&0<f<\delta,\\
f-\tfrac{\delta}{2},&f\geq\delta,
\end{cases}
\label{eq:primitive}
\end{equation}
so $r_\delta'(f)=D_\delta(f)$ wherever the derivative exists, and the coefficient
$5/2$ in the middle branch matches $r_\delta$ and $r_\delta'$ at $\xi=0$ and
$\xi=1$. It is $r_\delta$, not $D_\delta$, that enters the flow rule:
\begin{equation}
  \dot a=\frac{r_\delta(f)}{\eta},\qquad
  \dot{\boldsymbol p}=\dot a\,\frac{3\boldsymbol s}{2\sigma_{\mathrm{eq}}},
  \label{eq:flow}
\end{equation}
where $D_\delta$ is the normalised slope $r_\delta'$, the zero-stress direction is
continuous on the inactive branch, and neither function is a plastic strain or a
stiffness-degradation variable. For finite $\eta$ this flow rule is rate dependent,
so physical loading duration and transition width are different controls and are
not interchangeable; the supplement derives the dissipation identity
$\boldsymbol\sigma:\dot{\boldsymbol p}-Ha\dot a=(\sigma_y+f)\dot a\geq0$ on the
active branch.

\subsection{Implicit Multiplier and Local Derivative}
\label{sec:local-root}
For an isotropic $J_2$ trial state the multiplier of the increment, $\gamma$, solves
one scalar backward-Euler equation and the candidate history follows the
associative flow direction,
\begin{equation}
\begin{aligned}
  \gamma&=\frac{\Delta t}{\eta}
  r_\delta\!\left(f_{\mathrm{tr}}-(3\mu+H)\gamma\right),
  &\qquad
  \boldsymbol n_{\mathrm{tr}}&=\frac{3\boldsymbol s_{\mathrm{tr}}}
  {2\sigma_{\mathrm{eq,tr}}},\\
  a^{n+1}&=a^n+\gamma,
  &\qquad
  \boldsymbol p^{n+1}&=\boldsymbol p^n+\gamma\boldsymbol n_{\mathrm{tr}},
\end{aligned}
\label{eq:implicit}
\end{equation}
where $\boldsymbol n_{\mathrm{tr}}$ is not a unit-Frobenius tensor. Under the
radial-correction assumption the trial state is corrected along that fixed
direction, so $\sigma_{\mathrm{eq}}^{n+1}=\sigma_{\mathrm{eq,tr}}-3\mu\gamma$; with
$a^{n+1}=a^n+\gamma$ in Equation~\ref{eq:yield}, the candidate overstress
$f^{n+1}=f_{\mathrm{tr}}-(3\mu+H)\gamma$ is exactly the argument of the primitive,
so the scalar equation is the consistency condition of the candidate state and an
increment solves one scalar equation rather than a system. With $B=3\mu+H$,
$\alpha=\Delta t/\eta$ and $\bar f=f_{\mathrm{tr}}-B\gamma$, the function
$G(\gamma)=\gamma-\alpha r_\delta(\bar f)$ has
$G_\gamma=1+\alpha B D_\delta(\bar f)\geq1$ and therefore a unique root; the
one-dimensional reduction used in \S\ref{sec:regularity-bias} is the same
construction with $B=E+H$. For a perturbation $v$ of the trial state,
\begin{equation}
 D\gamma[v]=
 \frac{\alpha D_\delta(\bar f)}
 {1+\alpha B D_\delta(\bar f)}
 \,Df_{\mathrm{tr}}[v].
\label{eq:local-jvp}
\end{equation}
Equation~\ref{eq:local-jvp} is the directional derivative of the \emph{scalar}
multiplier at fixed committed history, material parameters, $\delta$, $\eta$ and
$\Delta t$; it is not the three-dimensional stress tangent, which also carries the
derivative of the flow direction and of the elastic stress update, nor a substitute
for the assembled three-dimensional Jacobian. Every derivative check of
Section~\ref{sec:path-derivative} perturbs the tool controls of
Equation~\ref{eq:path-sensitivity} at fixed material parameters, gate width and
viscosity, and material derivatives are not audited here. At fixed $\eta$ and
$\Delta t$, $\delta\to0$ approaches the linear-overstress root, not rate-independent
$J_2$, which requires the viscosity to vanish as well.

\subsection{Local Update}
\label{sec:local-update}
Algorithm~\ref{alg:local} states SECA as a procedure, so that the inputs, the
committed state and the outputs of one evaluation are explicit.

\noindent\fbox{\parbox{\dimexpr\columnwidth-2\fboxsep-2\fboxrule}{%
\refstepcounter{localalgorithm}%
\textbf{Algorithm \thelocalalgorithm\ (local SECA update).}\label{alg:local}\\
\emph{Input:} trial strain $\boldsymbol\epsilon$ (or $\boldsymbol F$,
\S\ref{sec:finite-update}); committed history $(\boldsymbol p^n,a^n)$; material
parameters $(E,\nu,\sigma_y,H)$; gate width $\delta$, viscosity $\eta$, increment
$\Delta t$.\\
1.\quad form the trial state and its overstress $f_{\mathrm{tr}}$
(Equation~\ref{eq:yield}).\\
2.\quad if $f_{\mathrm{tr}}\leq0$: return $\gamma=0$, unchanged history, the elastic
stress and tangent.\\
3.\quad otherwise solve Equation~\ref{eq:implicit} for $\gamma>0$ (unique root).\\
4.\quad evaluate $D_\delta(\bar f)$ and the step slope $1+\alpha B D_\delta(\bar f)$
at $\bar f=f_{\mathrm{tr}}-B\gamma$.\\
5.\quad form the candidate history $\boldsymbol p^{n+1}$ (or
$\boldsymbol F_{p,n+1}$) and $a^{n+1}=a^n+\gamma$.\\
6.\quad \emph{output} $\gamma$, the candidate history, the fixed-state stress, the
multiplier derivative (Equation~\ref{eq:local-jvp}) and, on the plastic branch, the
remaining local derivative terms.\\
\emph{Commit rule:} within one increment the candidate history is committed only
after that increment's iteration is accepted, so a rejected trial leaves the
committed history untouched. A candidate history is committed only after feasible
convergence of the global increment. A failed increment does not produce an
accepted state; any diagnostic continuation is excluded from valid released-state
error and time-to-solution comparisons. For diagnosis the comparison driver of
Section~\ref{sec:n1-comparison} does continue past a missed tolerance from the
last iterate, which is why such a run still completes and still yields a released
configuration; those records are reported as diagnostics rather than as released
solutions of converged history, and Section~\ref{sec:n1-comparison} says which
they are.}}
\vspace{2pt}

\paragraph*{Shared interface of the compared updates.}
Every update compared in this work maps the same trial state and committed history to a
multiplier, a candidate history and a stress under the same solver and commit rule; only the
local evolution equation of line~3 differs. Exact return mapping and linear-overstress
Perzyna are the two explicit roots of Equation~\eqref{eq:three-roots}, SECA is the
implicit compact-activation equation of Equation~\eqref{eq:implicit}, and these three carry
the analysis and every finite-deformation example of this paper. The error and cost
comparison of Section~\ref{sec:n1-comparison} adds one further, purely comparative update ---
a strict-elastic Perzyna flow whose rate is a quartic power of the positive overstress ---
which is defined and used for that comparison only and is not part of the method. The
rate-independent $J_2$ reference is exact return mapping itself, not a rate-dependent
model given an artificial viscosity. Sections~\ref{sec:strict-elastic}
and~\ref{sec:regularity-bias} analyse this local, small-strain reduction;
Section~\ref{sec:finite-update} embeds it in the finite-deformation formulation used
for the graphics experiments.

\begin{figure*}[!t]
  \centering
  \includegraphics[width=0.86\textwidth]{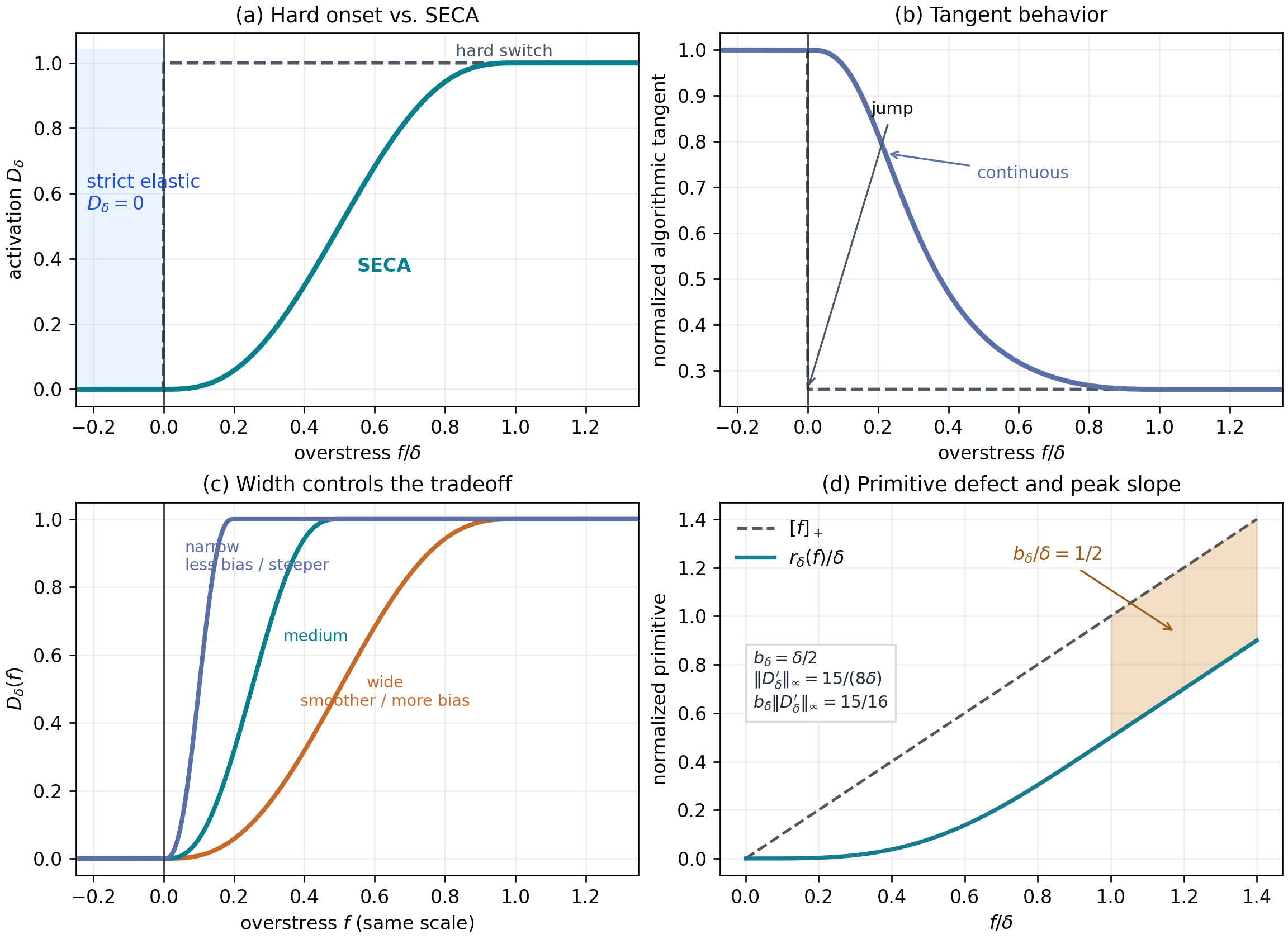}
  \caption{The SECA gate and its exact consequences. (a) Unlike the hard
  switch, the compact activation is exactly zero below yield and transitions
  continuously. (b) Its implicit one-dimensional tangent for $E=100$, $H=10$ and
  $\Delta t/\eta=0.04$, normalised by $E$, joins the elastic branch without the
  hard-onset jump. (c) Three transition widths on one overstress axis: narrowing
  the transition lowers the bias and steepens the activation, and the curves are
  not rescaled. (d) The primitive stays $\delta/2$ below the positive-part
  function after full activation, with the bias, the peak activation slope and
  their gate-specific product printed in the panel. The supplement reproduces the
  six-panel version, including the width-versus-slope and exact-product panels.}
  \Description{Hard and SECA activation, their algorithmic tangents, three
  transition widths, and the exact primitive offset with the bias, peak slope,
  and constant product printed as formulas illustrate strict elasticity and the
  regularity--bias relation.}
  \label{fig:core-activation}
\end{figure*}

%% file: common/regularity_bias.tex
\section{Regularity and Bias Analysis of the SECA Constitutive Method}
\label{sec:regularity-bias}

Every statement below concerns the local method of \S\ref{sec:strict-elastic} at one
trial state with the previous committed history fixed: the activation of
Equation~\ref{eq:activation}, its primitive of Equation~\ref{eq:primitive} and the
multiplier equation of Equation~\ref{eq:implicit}. Three objects are separated. The
\emph{primitive defect} $b_\delta$ of \S\ref{sec:regularity-bias}.1 is a property of
the activation alone and carries stress units. The \emph{multiplier difference} of
\S\ref{sec:regularity-bias}.3 also depends on $B$, on $\eta/\Delta t$ and on the trial
state. The \emph{released-shape difference} of \S\ref{sec:regularity-bias}.4 depends
further on how the history propagates and how the release solves onto a stable branch,
so a statement about the first is not a statement about the third.

\subsection{Strict Elasticity and the Root}
Strict elasticity is a statement about the discrete increment, not about the size of
the activation: for $f_{\mathrm{tr}}\leq0$ the gate is exactly zero, so $\gamma=0$, the
stress is the elastic stress of the trial state, and the increment uses the elastic
tangent. For $f_{\mathrm{tr}}>0$ the root of Equation~\ref{eq:implicit} is unique and
positive, and the supplement gives the two inequalities that establish it; the same
construction with $B=E+H$ is the one-dimensional reduction analysed here.

The finite window leaves an exactly known offset: because the gate saturates exactly
above $\delta$, the primitive lags the positive-part function for every $f\geq\delta$ by
a defect $b_\delta=\int_0^\delta[1-D_\delta(z)]\,dz$, measured in the same units as $f$.
The supplement derives that identity, which requires exact saturation and therefore does
not hold in this form for logistic or softplus tails. For the quintic of
Equation~\ref{eq:activation}, symmetry
$D_\delta(f)+D_\delta(\delta-f)=1$ inside the transition fixes the defect and the peak
slope:
\begin{equation}
\begin{aligned}
  \boxed{b_\delta=\frac{\delta}{2}},&\qquad
  D_\delta'(f)=\frac{30}{\delta}\xi^2(1-\xi)^2,\\
  \boxed{\|D_\delta'\|_\infty=\frac{15}{8\delta}},&\qquad
  \boxed{b_\delta\|D_\delta'\|_\infty=\frac{15}{16}}.
\end{aligned}
\label{eq:regularity-bias-product}
\end{equation}
The constant product is a property of the chosen construction rather than a fitted
relation: narrowing the window reduces the defect linearly while increasing the maximum
slope inversely. It is not an error bound, it does not bind any released geometry, and
it does not by itself select an optimal width; panel~(d) of
Figure~\ref{fig:core-activation} plots both relations.

\subsection{Exact Implicit Derivatives}
Write $A=1+\alpha B D_\delta(\bar f)$ with $\bar f=f_{\mathrm{tr}}-B\gamma$. Implicit
differentiation of $G(\gamma)=0$ at the root gives the partial derivatives of the
multiplier with respect to the trial overstress, the transition width, the viscosity and
the time increment; the supplement writes all four out, together with the fixed-argument
width partial of the primitive, the coupled case $\eta=c\delta$ and the multiplier
curvature. The width partial is the derivative of the same primitive with respect to the
same width at fixed overstress argument, not a new material parameter. The gate
therefore controls how sharply the multiplier tangent changes, while $\alpha=\Delta
t/\eta$ also controls its magnitude; that curvature scale is what a finite-difference
check on this update can converge against, and it is not a claim about the assembled
solver.

\subsection{Finite-Width and Viscosity Bias}
Compare the root of Equation~\ref{eq:implicit} with the two reference roots an
application would otherwise use, all three evaluated at the same trial overstress. With
$\beta=\eta/\Delta t$, they are
\begin{equation}
\gamma_L=\frac{[f_{\mathrm{tr}}]_+}{B+\beta},\qquad
\gamma_J=\frac{[f_{\mathrm{tr}}]_+}{B}.
\label{eq:three-roots}
\end{equation}
Because $b_\delta(\bar f)=\bar f_+-r_\delta(\bar f)$ satisfies
$0\leq b_\delta(\bar f)\leq\delta/2$, substituting
$f_{\mathrm{tr}}=\bar f+B\gamma$ into those roots gives the exact decomposition
\begin{align}
\gamma_L-\gamma
&=\frac{b_\delta(\bar f)}{B+\beta},\label{eq:width-bias}\\
\gamma_J-\gamma
&=\underbrace{\frac{\beta[f_{\mathrm{tr}}]_+}{B(B+\beta)}}_{\text{viscosity bias}}
+\underbrace{\frac{b_\delta(\bar f)}{B+\beta}}_{\text{finite-width bias}},
\label{eq:width-viscosity-decomposition}\\
0\leq\gamma_L-\gamma&\leq\frac{\delta}{2(B+\beta)},
\label{eq:width-bracket}\\
0\leq\gamma_J-\gamma&\leq
\frac{\beta[f_{\mathrm{tr}}]_+}{B(B+\beta)}
+\frac{\delta}{2(B+\beta)}.
\label{eq:multiplier-bound}
\end{align}
Above the transition the width term equals $\delta/[2(B+\beta)]$, the constant used in
the bracket. Both relations bound the multiplier of one trial state: they are not a
global geometric error guarantee, and they do not by themselves select an optimal width.
At fixed $\beta>0$, letting $\delta\to0$ recovers the linear-overstress root. Recovering
rate-independent $J_2$ additionally requires $\beta\to0$; the supplement works out that
limit, the coupled case
$\eta=c\delta$ and the non-uniform convergence of derivatives at yield. The formulas
assume the scalar affine correction $\bar f=f_{\mathrm{tr}}-B\gamma$, exact for the
implemented one-dimensional law and for proportional isotropic $J_2$ reduction; rotating
flow directions, anisotropy and general nonproportional finite-strain paths fall outside
them.

\subsection{Local Parameter Budget}
The bracket of Equation~\ref{eq:width-bracket} has a direct algebraic corollary. Given a
local plastic-increment error budget $\epsilon_\gamma$ at fixed $B$, $\eta$, $\Delta t$
and a common trial state, a width satisfying
\begin{equation}
\delta\leq2(B+\beta)\epsilon_\gamma
\label{eq:width-budget}
\end{equation}
keeps the finite-width part of the multiplier difference from linear-overstress Perzyna
below $\epsilon_\gamma$, and a budget taken against exact return mapping must also
include the viscosity term of Equation~\ref{eq:multiplier-bound}. This is a local budget
and not a selection rule: it does not choose a width that is optimal for all meshes and
time increments, and it is not a global geometric error guarantee.

The local differences reach a released state only under stated assumptions: a shared
initial committed history, a shared accepted time partition and contact-feature
sequence, and one continuously connected stable equilibrium branch after complete tool
removal with a nonsingular residual Jacobian. The supplement states both propagation
steps, the constants they use, and the iteration that carries a multiplier difference to
a released observable. That bridge is conditional, and a difference measured at a single
material point cannot stand for the history of a whole body; the released states of
Section~\ref{sec:results} are measured directly rather than certified by it.

%% file: common/finite_method.tex
\section{Embedding SECA in a Finite-Deformation Equilibrium Solver}
\label{sec:finite-update}

This section embeds the local method of Section~\ref{sec:strict-elastic}---the activation
$D_\delta$ of Equation~\ref{eq:activation}, the primitive $r_\delta$ of
Equation~\ref{eq:primitive} and the multiplier equation of Equation~\ref{eq:implicit}---in
the finite-deformation solver used for the graphics experiments, evaluated at a
finite-deformation trial state. Only the radial reduction that evaluates the trial
overstress is coaxial; the update remains rate dependent through $\eta$ and is not an exact
projection onto the rate-independent yield surface.

\subsection{Multiplicative State and Fixed-State Stress}
The character implementation uses reference tetrahedral volumes $V_e$,
$\boldsymbol F=\boldsymbol I+\nabla_X\boldsymbol u$, and the multiplicative state
$\boldsymbol F_e=\boldsymbol F\boldsymbol F_p^{-1}$. With the right Hencky strain
$\boldsymbol e=\log\boldsymbol U_e$ from $\boldsymbol F_e=\boldsymbol R_e\boldsymbol U_e$,
its fixed-state energy is
\begin{equation}
W=\mu\|\operatorname{dev}\boldsymbol e\|^2
  +\frac K2(\operatorname{tr}\boldsymbol e)^2+\frac H2a^2 .
\end{equation}
The implementation differentiates $W$ with the internal variables fixed, so the
Kirchhoff and first-Piola stresses are
\begin{equation}
\boldsymbol\tau=\boldsymbol R_e
 [2\mu\operatorname{dev}\boldsymbol e+K\operatorname{tr}(\boldsymbol e)\boldsymbol I]
 \boldsymbol R_e^T,\qquad
 \boldsymbol P=\boldsymbol\tau\boldsymbol F^{-T}.
\label{eq:finite_stress}
\end{equation}
Taking the \emph{total} derivative of
$W(\boldsymbol F,\boldsymbol F_p(\boldsymbol F))$ would include evolution terms and is
not the stress used by the solver; the material Jacobian instead differentiates the full
discrete stress update, including the dependence of the candidate internal state on the
current deformation.

Let $e_{\mathrm{eq}}=\sqrt{(2/3)\operatorname{dev}\boldsymbol e_{\mathrm{tr}}:
\operatorname{dev}\boldsymbol e_{\mathrm{tr}}}$, a different symbol from the discrete
overstress $\bar f$ of \S\ref{sec:local-root}, and let
$\boldsymbol N=\operatorname{dev}\boldsymbol e_{\mathrm{tr}}/e_{\mathrm{eq}}$ on an
active trial. For the isotropic Hencky model this direction is coaxial with the trial
elastic stretch, so the updated equivalent strain obeys
$e_{\mathrm{eq},n+1}=e_{\mathrm{eq,tr}}-\gamma$ and the finite update reuses the scalar
root of Equation~\ref{eq:implicit} in the form
\begin{equation}
\begin{aligned}
f_{\mathrm{tr}}&=3\mu e_{\mathrm{eq,tr}}-\sigma_y-Ha_n ,\\
\gamma&=\frac{\Delta t}{\eta}r_\delta(f_{\mathrm{tr}}-(3\mu+H)\gamma),\\
\boldsymbol F_{p,n+1}&=\exp(\gamma\boldsymbol N)\boldsymbol F_{p,n},
\qquad a_{n+1}=a_n+\gamma .
\end{aligned}
\label{eq:finite_update}
\end{equation}
Equation~\ref{eq:finite_update} is that same root under the stated radial-correction
condition, not a general projection of an arbitrary finite-strain model onto the
small-strain update. The flow direction $\boldsymbol N$ is evaluated in the frame of the
trial elastic strain and multiplies $\boldsymbol F_{p,n}$ from the left, so the
exponential update and the flow direction are implemented together and audited as a
pair. The stress of Equation~\ref{eq:finite_stress} is the energy partial at the trial
internal state of the current iterate, held fixed; committing that state is part of the
order of operations below.

\subsection{Finite-Element Residual and Order of Operations}
For node $i$, the material residual is assembled as
\begin{equation}
\boldsymbol R_i^{\mathrm{int}}
 =\sum_{e\ni i}V_e\boldsymbol P_e\nabla_X\mathcal{N}_{ei} .
\label{eq:material-residual}
\end{equation}
Here $\mathcal{N}_{ei}$ is the shape function of node $i$ in element $e$, a different
symbol from the flow direction $\boldsymbol N$ of Equation~\ref{eq:finite_update}; support
and contact terms are added to this common material residual. In the controlled cube and
simple-body experiments the iteration matrix is the consistent material tangent of
Equation~\ref{eq:finite_update} plus the analytically differentiated tool term of
\S\ref{sec:contact-sensitivity}. The character solver instead uses a modified-Newton
matrix built from a
constant rest-elasticity stiffness with a small stabilisation term, but its residual and
the derivative of that residual remain the true ones, so assembling a different Newton
matrix changes the iteration and not the residual that defines the solution.

One accepted increment then runs as follows. The trial state of every element is formed
from the current iterate and the \emph{committed} history $(\boldsymbol F_p^n,a^n)$,
through the selected local evolution equation of \S\ref{sec:local-update}, whose SECA
branch is Algorithm~\ref{alg:local}. The residual of
Equation~\ref{eq:material-residual} is assembled together with the support and contact
terms of \S\ref{sec:contact-sensitivity}, the Newton loop exits when the relative
free-dof residual falls below the tolerance of the case, and the tangent system is
solved and globalised with a line search whose rejected trials discard their candidate
history. Only a feasible converged increment commits
$(\boldsymbol F_p^{n+1},a^{n+1})$. An increment whose iteration stops without
reaching that tolerance produces no accepted state: for diagnosis the comparison
driver of Section~\ref{sec:n1-comparison} records the missed increment and
continues from its last iterate, and such a diagnostic continuation enters no
error and no time-to-solution comparison of this paper. Along that chain the local multiplier of
Equation~\ref{eq:implicit} becomes the history update of Equation~\ref{eq:finite_update},
its fixed-state stress is Equation~\ref{eq:finite_stress}, and the assembled residual of
Equation~\ref{eq:material-residual} and its history derivative of
Equation~\ref{eq:path-sensitivity} are the two objects the solver iterates.

\subsection{Supports, Tools and Loading Control}
A support is a prescribed-displacement condition on a fixed node set: those displacements
are part of the data in every state, the free degrees of freedom are their complement, and
no stage reduces a support. A tool or grip is the opposite kind of object: it enters the
residual as a penalty term with a recorded stiffness, and removing it is part of the
schedule rather than a change of the supports. In the Armadillo character the grip
stiffness is ramped linearly to exactly zero at fixed control, so the last state is a
grip-free equilibrium on the unchanged supports. Complete tool withdrawal likewise removes the true tool or grip terms
from the residual; it does not reset the committed $\boldsymbol F_p$ or $a$ and it does
not zero the control $\theta$ while keeping the old history, so a released state in this
paper is a tool-free or grip-free equilibrium under the stated supports rather than a
free-floating one. The instantaneous $D_\delta$ is the flow activation, not a
stiffness-degradation variable.

The control a user sets, $\theta$, and the trajectory the tool follows,
$\boldsymbol c(t;\theta)$, are different objects: $\theta$ is a scalar or
low-dimensional input of the schedule, while $\boldsymbol c(t;\theta)$ is the resulting
physical motion of the tool or grip, including hold and withdrawal durations. Reported
derivatives are taken along the complete accepted equilibrium-state sequence: for each
control value the full load--hold--withdraw--release rerun is executed and
differentiated, rather than differentiating the last release solve alone. Discrete events
along that sequence, such as the activation of a contact pair, a closest-feature change,
or a subdivision of the time interval, are detected and reported as event crossings and
are not differentiated as if the response were smooth there.

%% file: common/contact_sensitivity.tex
\subsection{Contact, Globalisation, and Commitment}
\label{sec:contact-sensitivity}
For analytic sphere $j$ and boundary triangle $t$, the tool gap is measured from the
point--triangle distance and the sphere radius, and positive gaps below a barrier distance
$\hat d$ contribute a logarithmic barrier energy with the standard
$-(g-\hat d)^2\log(g/\hat d)$ profile, which vanishes for $g\geq\hat d$ and is strictly
positive below it, so a separated pair carries no energy and the force grows as the gap
closes and pushes the two bodies apart. IPC Toolkit supplies point--triangle distances
and derivatives. The self-contact-capable Bunny path additionally uses IPC boundary
collision candidates, barrier energy, and force, while relative CCD limits each accepted
update for tool and deformable-surface contact.

The forward Newton system uses positive-semidefinite projection of contact blocks for
globalisation. This projected matrix is an optimisation device, not the derivative of the
residual: the residual itself is never projected, and the path derivative of
\S\ref{sec:path-derivative} below uses the exact unprojected tool and self-contact Hessian
actions. Only a feasible converged stage commits $(\boldsymbol F_p,a)$, so rejected
line-search steps do not commit history; loading, complete withdrawal, and hold durations
are preserved when an interval is subdivided.

\subsection{Actual-Residual Path Derivative}
\label{sec:path-derivative}
Let $\boldsymbol z_k=(\boldsymbol F_{p,k},a_k)$ collect every committed internal state:
\begin{equation}
\boldsymbol R_k(\boldsymbol u_k,\boldsymbol z_{k-1},\boldsymbol\theta)=0,\qquad
\boldsymbol z_k=\boldsymbol\Phi_k(
\boldsymbol u_k,\boldsymbol z_{k-1},\boldsymbol\theta).
\end{equation}
On a fixed accepted schedule and for direction $\boldsymbol v$, the path
Jacobian-vector product (JVP) satisfies
\begin{equation}
\begin{split}
R_uD\boldsymbol u_k[\boldsymbol v]
&=-R_zD\boldsymbol z_{k-1}[\boldsymbol v]-R_\theta\boldsymbol v,\\
D\boldsymbol z_k[\boldsymbol v]
&=\Phi_uD\boldsymbol u_k[\boldsymbol v]
+\Phi_zD\boldsymbol z_{k-1}[\boldsymbol v]+\Phi_\theta\boldsymbol v .
\end{split}
\label{eq:path-sensitivity}
\end{equation}
The spaces are fixed by the discretisation: $\boldsymbol u_k$ collects the free nodal
displacements of increment $k$, so the supports are eliminated rather than differentiated;
$\boldsymbol z_k$ collects $(\boldsymbol F_{p,k},a_k)$ over all elements of the mesh; and
$\boldsymbol\theta$ collects the audited controls, the tool parameters that the trajectory
is generated from. The operator combines the automatic-differentiation JVP of the actual
finite-$\boldsymbol F_p$ internal residual, analytic tool-control linearisation, and exact
unprojected tool and self-contact Hessian actions. Neither the positive-semidefinite
forward matrix nor a perturbed forward solve is substituted for $R_u$.

The controls that were audited are symmetric sphere depth and one-sphere tangential
translation. Coarse and medium algorithmic products propagate the complete $30$-stage
committed history and are compared with independently executed centred full-path replays.
Closest-feature changes, adaptive subdivision, and line-search decisions are discrete and
are not differentiated, so the check is a fixed-partition one, on one accepted schedule
rather than on two adaptively chosen ones. The measured discrepancies, the replicated
levels and the per-step table are reported with the released-state results in
Section~\ref{sec:released-sensitivity}; the adaptive forward scan described there is a
different experiment, solving the whole path at each control value instead of perturbing
one fixed discrete map, and it is not used as evidence here.

%% file: common/results_local.tex
\subsection{Experimental Setup and Local Properties}
\label{sec:setup}
\label{sec:local-regularity}
\label{sec:material-point-comparison}

Every comparison in this section runs the three constitutive methods on one
elemental interface: exact radial return-mapped $J_2$, linear-overstress
Perzyna, and SECA compact activation. They differ only in the flow rule, so the
same simulation settings isolate the constitutive branch rather than the
implementation. Those settings fix the mesh, the boundary conditions, the tool or
gripper schedule, the physical duration, the material parameters, the Newton
tolerance (a relative free-dof residual of $10^{-9}$ for the controlled cube and
the simple bodies), the Armijo line search, and the common initial committed
state; the assembled tangent and the diagonal-damping and least-squares fallbacks
are those of Section~\ref{sec:finite-update}. All reported runs are deterministic
CPU solves from a fixed initial state, so the iteration totals below are not
subject to sampling noise. The visualisations plot the Kirchhoff von Mises field
$\tau_{\mathrm{vM}}=\sqrt{\tfrac{3}{2}\boldsymbol s:\boldsymbol s}$, whereas the
residual-strength numbers quoted below are the signed material-point component
$\tau_{11}$ of the same Kirchhoff stress; the two are not interchangeable.

The evaluation metrics are the yield-transition tangent change
$\Delta C_{\mathrm{tr}}$ defined below, the released state after complete tool
removal, its discretisation error against each method's own finest reference, the
solver work in Newton and line-search iterations, and the wall time. A shared
setting is not a shared plastic history, and each method's own finest reference is
not a cross-method true error. Table~\ref{tab:method-properties} states the design
properties that distinguish the three constitutive methods before any
measurement.

\begin{table}[!t]
\centering
\small
\caption{Design properties of the three constitutive methods compared throughout
this section. The error and cost comparison of Section~\ref{sec:n1-comparison} adds a
fourth, purely comparative update whose definition, parameters and selection are given
there; it shares the strict-elastic row of this table and has no finite transition width.
\emph{Strict elastic} means the discrete update returns an identically zero
plastic multiplier below yield, not merely a small one. It separates SECA from
smooth noncompact smoothing, such as a softplus overstress law whose primitive
stays positive below the nominal threshold; the linear-overstress law compared
here also satisfies it, but neither it nor return mapping has a transition of
finite width.}
\label{tab:method-properties}
\setlength{\tabcolsep}{4pt}
\begin{tabular}{lccc}
\toprule
Property & exact RM & Perzyna & SECA \\
\midrule
plastic multiplier zero below yield & yes & yes & yes \\
finite transition window & no & no & yes ($\delta$) \\
activation smooth in overstress & no & no & $C^2$ \\
tangent continuous at onset & no & no & yes \\
rate dependent & no & yes ($\eta$) & yes ($\eta$) \\
$J_2$ limit recovered by & reference & $\eta\to0$ & $\eta,\delta\to0$ \\
\bottomrule
\end{tabular}
\end{table}

The local evaluation point is the scaled prototype configuration recorded with
these checks: yield stress $2.0$, hardening $10.0$, width $0.2$,
$\eta/\delta=0.2$, and unit loading duration. Because $G(0)=0$,
$G_\gamma=1+\alpha B D_\delta(\bar f)\geq1$ and $r_\delta=0$ for $\bar f\leq0$,
the backward-Euler root is $\gamma=0$ whenever the trial overstress is
nonpositive, so a sub-yield trial returns an exactly zero multiplier rather than a
small one. Executable scalar checks compare the closed-form primitive and implicit
derivatives with independent quadrature, bisection, and centred differences over
elastic, onset, transition, plastic, unloading, and reverse states, and the exact
width/rate decomposition in Equation~\ref{eq:width-viscosity-decomposition} is
checked as a floating-point identity, which measures round-off rather than
agreement between two derivations; the supplement's protocol section reports those
per-state discrepancies, the residual of the identity, and the difference step at
which an individual state leaves the elastic branch.

Strict elasticity also holds discretely: all three constitutive methods enter the
plastic branch at the same load step of the controlled cube benchmark, with the
same plastic volume fraction ($0.015625$ of the mesh), so the compact onset does
not shift the apparent yield point. The same comparison measures one explicitly
named noncompact law instead of asserting its behaviour: over a 200-step sub-yield
hold at $f_{\mathrm{tr}}=-5\times10^{5}$\,Pa the softplus law accumulates
$2.168\times10^{-4}$ of equivalent plastic strain that survives the release,
whereas the compact gate returns exactly zero at every step of the same hold. Both
are rate-function substitutions in the scalar kernel and neither is a reproduction
of a published solver.

The yield-transition tangent change at the tested yield point is
$\depsci{\depTangJumpRM}\,$Pa for exact return mapping,
$\depsci{\depTangJumpPerzyna}\,$Pa for Perzyna, and
$\depsci{\depTangJumpDEP}\,$Pa for SECA: $\depTangJumpPctRM\%$ of the hard
reference and $\depTangJumpPctPerzyna\%$ of the overstress reference. The ratio
$\depTangJumpRatio$ is the rounded $62.545$ of this protocol and is quoted with
it: a window of $2\times10^{-4}$ logarithmic strain, sampled as twenty steps of
$10^{-5}$ on a $2001$-point grid around the yield point, so the factor is
protocol-specific rather than a universal speed-up, and the supplement audits it
one condition at a time. $\Delta C_{\mathrm{tr}}$ is not the mathematical jump of
a discontinuous derivative: it is the absolute difference between the secant slope
of Kirchhoff stress over the $2\times10^{-4}$ window below yield and the secant
slope over the same window above yield, on the shared material-point path of
Figure~\ref{fig:numeric-comparison}(b) and the supplement's protocol table. The
overstress comparison is the informative one: that law already carries a finite
viscosity and is already rate dependent, yet its change is slightly \emph{larger}
than the hard reference's rather than smaller, because its linear positive part has
derivative one on the active side and zero on the inactive side, so viscosity alone
does not remove the onset change; SECA varies the tangent continuously over a
finite transition interval instead.

On the nonproportional material path in the same scaled prototype configuration,
reducing the width from $0.2$ to $0.05$ and $0.005$ reduces the maximum
yield-normalised stress discrepancy from 3.480\% to 0.890\% and 0.089\%; that path
lies outside the exact proportional decomposition, and the supplement records its
load-step and duration checks.

%% file: common/results_cost.tex
\subsection{Controlled Computational Benefit}
\label{sec:micro-numerics}

\begin{figure*}[!t]
\centering
\includegraphics[width=\textwidth]{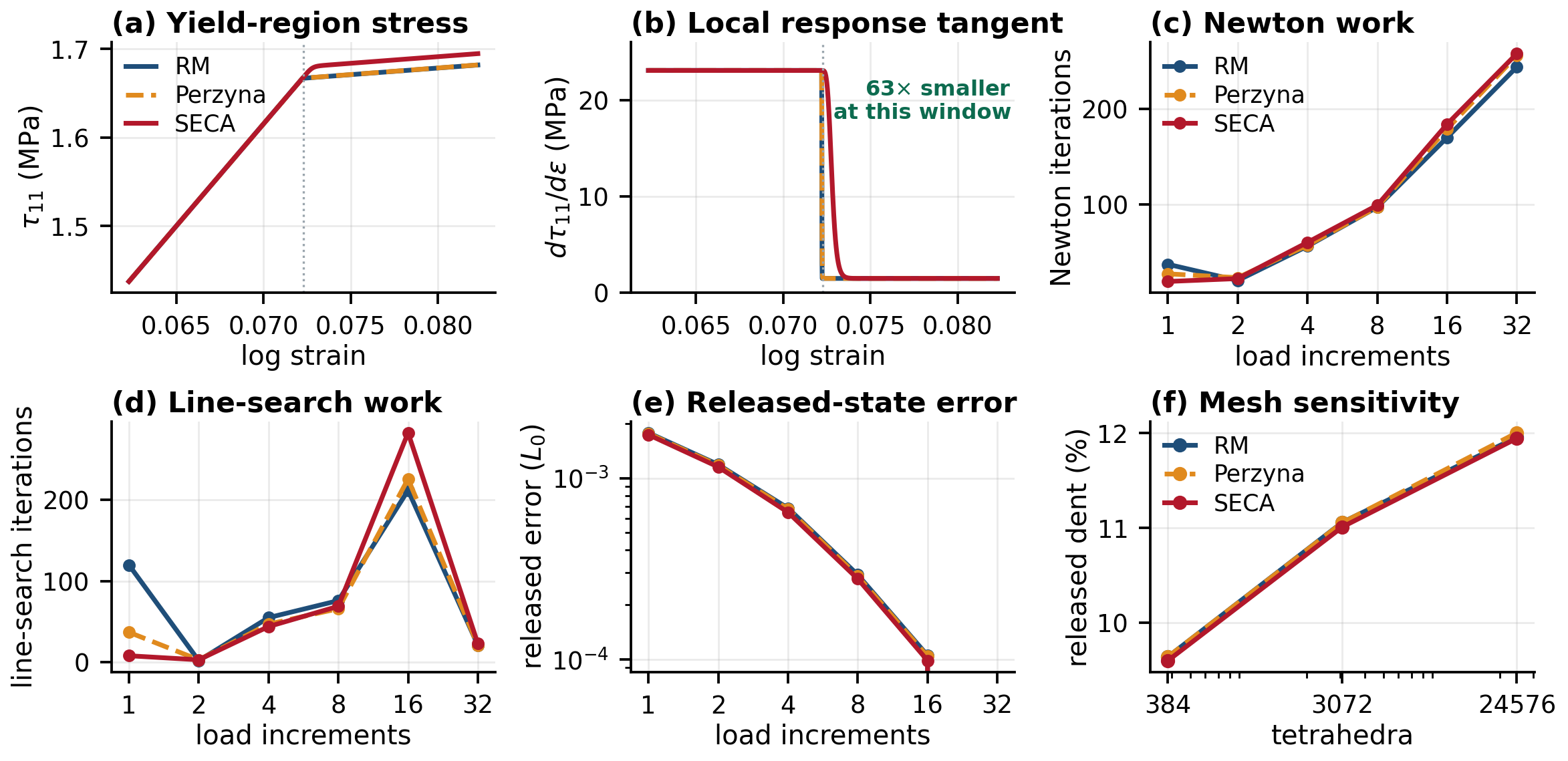}
\caption{From the local transition to the released answer, under the same
simulation conditions. (a) Kirchhoff stress and (b) the local response tangent
$\mathrm{d}\tau_{11}/\mathrm{d}\varepsilon$ through the yield window
$\varepsilon_y=0.0723$ on the shared material-point path; the yield-transition
tangent change $\Delta C_{\mathrm{tr}}$ is measured from this slope as defined
in the text. (c) Newton and
(d) line-search iterations against the number of load increments on the
controlled cube indentation, where one load increment means a single Newton solve
must cross the entire elastic--plastic transition; both counts are whole-run
totals. (e) Released-state error against the number of load
increments on one fixed mesh, each method measured against its own
$32$-increment solution, so this panel is the refinement record of the loading
partition and not a mesh refinement; the panel covers the direct run, whose
non-converged increments were rerun with the increment subdivided. (f) Released
dent against spatial mesh resolution in tetrahedra, where every mesh level shares
one loading/control schedule.}
\Description{Six panels: yield-region stress, algorithmic tangent, Newton
iterations, line-search iterations, released-state error against increments, and
released dent against mesh resolution, for three elastoplastic methods.}
\label{fig:numeric-comparison}
\end{figure*}

The compact activation is measured against the two updates it has to be judged
against on one shared loading--unloading path: exact return-mapped $J_2$ and
linear-overstress Perzyna, marched through the same implicit solve and the same
logarithmic strain history, so the differences are the constitutive difference and
nothing else. Figure~\ref{fig:numeric-comparison} collects the transition, the
nonlinear solver work, the released-state consequence and the mesh sensitivity of
that comparison.

The controlled comparison is the cube-indentation run of
Section~\ref{sec:cube-benchmark} at one load increment, and every count is a
whole-run sum over every solved stage of that run, load through final release, so
\emph{Newton} means whole-run Newton iterations and \emph{line search} means
whole-run Armijo backtracking halvings rather than line-search evaluations. Across
the whole run SECA needs \depNewtonSaved\% fewer Newton iterations and
\depLineSearchSaved\% fewer halvings than return mapping, and at one increment the
errors against the respective $32$-increment references differ by less than the
discretisation level they measure. The supplement's protocol table states those
counts together with the material-point quantities of
Section~\ref{sec:material-point-comparison}, which are not solver work.

The split across the loading history is not uniform: the load stage that crosses
the transition carries every return-mapping halving but only part of the SECA ones,
so the stage and whole-run readings are different splits of the same run, and the
supplement reports the stage-level counts. The counts are diagonal-damping aware:
the return-mapping load stage used \depDampingRM\ damped solves and the Perzyna
stage \depDampingPerzyna, while SECA used none, so the coarse comparison is not
between two identically globalised iterations. Every run classifies each step
against a relative free-dof residual of $10^{-9}$, and the counts are carried in
the retained summaries rather than reconstructed afterwards.

The comparison is not favourable to SECA at every schedule: when the load phase is
refined to two or more increments, SECA needs from
$\depNewtonRatioRefinedMin$ up to $\depNewtonRatioRefinedMax$ times the
return-mapping Newton count at every refined level, with Perzyna the cheapest of
the three at some of those levels and the most expensive at others; on the
grip-tension scene the coarse run needs more Newton work under SECA than under
return mapping, with the numbers in the supplement's grip-tension solver-work
table. The lower counts on the coarse schedule are consistent with the smoother
local linearisation and do not identify a convergence-basin mechanism.

The compact transition does not buy that regularity by suppressing residual
deformation: The same supplement table shows SECA retaining a \emph{larger}
residual stress magnitude than either reference, \depMpResidualGainRM\% and
\depMpResidualGainPerzyna\%, and reaching zero stress at strains
\depMpZeroStressRM, \depMpZeroStressPerzyna\ and \depMpZeroStressDEP,
\depMpZeroStressBelowRM\% below the hard reference. Strict elasticity also means
that some complete-unload targets genuinely have a zero gradient, so the retained
residual is a result rather than a defect, and the failure and non-convergence
records of this comparison are reported with it.

\subsubsection{Cross-method error and cost at a fixed observation level}
\label{sec:n1-comparison}

\begin{figure*}[!t]
\centering
\includegraphics[width=\textwidth]{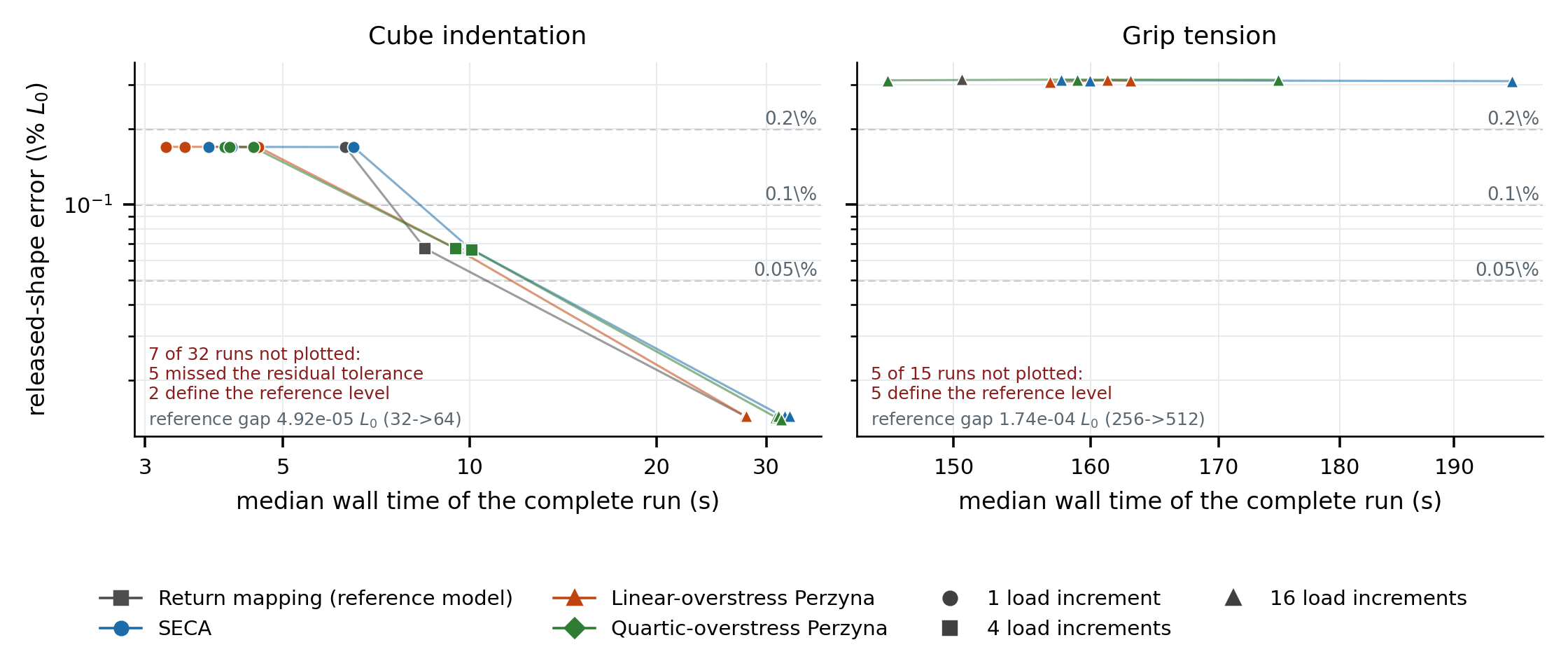}
\caption{Error against cost for the four constitutive updates on the two comparison
scenes. Marker shape is the number of load increments and colour is the update. The five cube
runs whose committed history contains an increment that missed the residual tolerance have no
released solution of the acceptance rule: they are not plotted, and the supplement locates the
increment in each of them together with its own residual and iteration counts. Dashed lines are
the budgets fixed before any cost was measured. Both
axes are logarithmic, and the cube panel carries its own reference-refinement gap. A schedule
the solver refuses outright (positive-determinant guard of the finite-strain update) has no
point in its panel and is listed with its refusal text in the supplement.}
\label{fig:n1-error-cost}
\end{figure*}

A controlled comparison asks whether strict elasticity costs accuracy or time. Under a
protocol frozen before the first three-dimensional solve, four constitutive updates ran on
two small scenes: exact return mapping (rate independent, taken as the target model),
linear-overstress Perzyna, SECA, and a strict-elastic Perzyna neighbour whose rate function is
a quartic power of the positive overstress, $r_{\mathrm{pow}}(f)=\sigma_\star([f]_+/\sigma_\star)^4$
with $\sigma_\star=\sigma_y$ and its own viscosity. All four enter through the same
constitutive interface and the same global solver, so the comparison is controlled, but
per-increment work is measured rather than assumed equal: exact return mapping and
linear-overstress Perzyna evaluate their multiplier in closed form, while the compact
activation and the quartic neighbour each solve their implicit scalar equation by a
$64$-step bisection on a bracketed interval. Each regularised update has a single accuracy knob --- the
activation width for SECA, the viscosity for the two Perzyna updates --- selected on a
material-point history at three grades $\{5,10,20\}\times10^{-4}$ before any scene run; the
supplement records the selection ladder, the realised offsets and the full parameter set.
The counts quoted earlier in this section come from the original controlled run of
Figure~\ref{fig:numeric-comparison} at that run's own parameter set, whereas the comparison
below uses the calibrated grades of Table~\ref{tab:n1-cost} and
Figure~\ref{fig:n1-error-cost}; the two series are different settings and not the same run.

Error is the weighted relative deviation
$e_{\mathrm{rel}}=\sqrt{\sum_i w_i\lVert x_i-x_i^{\mathrm{ref}}\rVert^2/\sum_i w_i}/L_0$
of the released configuration on the pre-fixed observation set, with uniform weights, $L_0$
the largest rest bounding-box extent, and no post-hoc alignment. The reference is the same
scene solved with exact return mapping on a refined partition; for cube indentation the
adjacent-level difference is $4.9\times10^{-5}L_0$ between the $32$- and $64$-increment
levels, far below the smallest budget, so the $64$-increment solution is accepted. The
budgets $0.2$, $0.1$ and $0.05\,\%\,L_0$ were fixed before any cost was measured. Cost is
the wall time of the complete load--hold--unload--release solve on one thread, the median
of three repeats, reported together with Newton iterations, line-search halvings and the
number of increments whose iteration stopped without reaching the residual tolerance. Only
records that reach the tolerance at every increment are ranked; a record whose history
advanced from a missed increment keeps its row as a diagnostic, with no error value and no
time to solution quoted for it.

\input{common/n1_cost_table}

Table~\ref{tab:n1-cost} and Figure~\ref{fig:n1-error-cost} give the outcome. On cube
indentation the error follows the loading partition rather than the constitutive update:
the twenty-five ranked runs reach $0.1692$--$0.1700\,\%\,L_0$ at one increment,
$0.0661$--$0.0669\,\%\,L_0$ at four and $0.0140$--$0.0144\,\%\,L_0$ at sixteen, and each
of the four updates reaches all three budgets at the same increment counts ($0.2\,\%$ at
one increment, $0.1\,\%$ at four and $0.05\,\%$ at sixteen). The cost ordering depends on
the partition. At one increment the regularised updates are the cheaper ones, the two
Perzyna neighbours taking $3.2$--$4.6$\,s and the compact activation $3.8$--$6.5$\,s
against $6.3$\,s for strict return mapping at the same error. At sixteen increments the
reference is the cheapest at $27.6$\,s, linear-overstress Perzyna matches it at
$27.7$--$27.8$\,s, the quartic neighbour costs $31.1$--$31.7$\,s, $12$--$15\,\%$ more,
and the compact activation $32.0$--$32.7$\,s, $16$--$19\,\%$ more, both at marginally
smaller error. Five of the thirty cube runs miss the tolerance instead of reaching it. In
every one of them the missed increment is the second of the ten recorded increments, the
second loading step, where the iteration stops at a final relative residual of
$8.0\times10^{-3}$ to $1.2\times10^{-2}$ after $44$--$60$ Newton iterations and
$544$--$884$ line-search halvings. The solver does not roll that increment back: the
remaining eight increments, including the release, start from the state it leaves behind.
Those five records are therefore diagnostics rather than released solutions of the
acceptance rule. They keep their rows in Table~\ref{tab:n1-cost}, marked $^{\dagger}$ and
without an error or a time to solution, and they are excluded from the figure, from the
ranges above and from every budget count; the supplement locates the missed increment of
each one with its own residual and iteration counts, and records the length of those runs
($19.3$--$30.5$\,s) without calling it a time to solution.

On grip tension the comparison is limited by the scene rather than by the update. The
pre-registered one- and four-increment schedules cannot be run at all: the finite-strain
update refuses them through its positive-determinant guard
(\texttt{ValueError: F and Fp must have positive determinant}), for the rate-independent
reference and for every regularised update alike, and each refusal is kept as a record with
its own text. At the only accepted level, sixteen increments, the four updates reach
$0.3074$--$0.3153\,\%\,L_0$, that is above all three budgets, and they agree to within about
$2.5\,\%$ relative, a spread of $8\times10^{-5}\,L_0$ --- smaller than the
$1.7\times10^{-4}\,L_0$ by which the finest reference level of that scene still differs from
its neighbour at the pre-registered level cap. The level, not the constitutive update, is
what these errors measure: differences below the reference's own remaining change are not
ranked, and the budgets are not reached on this scene at the increments the solver accepts;
the refused levels are reported as refusals rather than as numbers. Cost at
that level is $150.6$\,s for the reference, $157.0$--$163.1$\,s for linear-overstress
Perzyna, $145.4$--$174.9$\,s for the quartic neighbour and $157.8$--$195.3$\,s for the
compact activation, whose Newton count rises from $167$ to $169$--$199$; none of the ten
accepted tensile runs missed the residual tolerance in any increment. The reference ladder itself ($384$, $469$, $913$,
$1773$ and $3224$\,s at $32$, $64$, $128$, $256$ and $512$ increments) and its adjacent
differences are tabulated in the supplement.

The comparison is deliberately narrow: two coarse scenes, one representative
strict-elastic neighbour and three accuracy grades. It shows no accuracy advantage for any
of the four updates on either scene, and cost differences that stay moderate: at matched
increments they agree within a few percent relative in released-shape error, while wall time
on the tensile scene spans from about $3\,\%$ below to $30\,\%$ above the reference. It is
not a survey:
the neighbour reproduces a smooth power-law rate form as a fourth method rather than a
published state-of-the-art viscoplastic model, and the derivative and globalisation savings
reported in this section are properties of the solver rather than of this comparison.

%% file: common/n1_cost_table.tex
% generated by SECA_CL2.0/audit/cgf_n1_report.py
\begin{table*}[!t]
\centering
\footnotesize
\setlength{\tabcolsep}{2.5pt}
\begin{tabular}{@{}llrrrrr@{}}
\toprule
Method & $n$ & Released error (\%) $\downarrow$ & Wall time (s) $\downarrow$ & Newton iterations $\downarrow$ & Backtracking halvings $\downarrow$ & missed \\
\midrule
\multicolumn{7}{l}{\emph{Cube indentation}} \\
Return mapping (ref.) & 1 & 0.1700 & 6.30 & 36 & 117 & 0 \\
Return mapping (ref.) & 4 & 0.0669 & 8.46 & 55 & 52 & 0 \\
Return mapping (ref.) & 16 & 0.0144 & 27.64 & 170 & 212 & 0 \\
SECA & 1 & 0.1698 & 4.13 & 24 & 24 & 0 \\
SECA & 4 & 0.0668 & 10.01 & 56 & 58 & 0 \\
SECA & 16 & 0.0144 & 32.74 & 176 & 212 & 0 \\
Perzyna, linear & 1 & 0.1700 & 3.24 & 21 & 26 & 0 \\
Perzyna, linear & 4 & --- & --- & 94 & 643 & 1$^{\dagger}$ \\
Perzyna, linear & 16 & 0.0144 & 27.82 & 174 & 213 & 0 \\
Perzyna, quartic & 1 & 0.1696 & 4.10 & 24 & 24 & 0 \\
Perzyna, quartic & 4 & --- & --- & 97 & 763 & 1$^{\dagger}$ \\
Perzyna, quartic & 16 & 0.0142 & 31.05 & 177 & 212 & 0 \\
\midrule
\multicolumn{7}{l}{\emph{Grip tension}} \\
Return mapping (ref.) & 16 & 0.3153 & 150.56 & 167 & 34 & 0 \\
SECA & 16 & 0.3129 & 159.99 & 172 & 40 & 0 \\
Perzyna, linear & 16 & 0.3113 & 163.06 & 177 & 79 & 0 \\
Perzyna, quartic & 16 & 0.3143 & 174.87 & 183 & 116 & 0 \\
\bottomrule
\end{tabular}
\caption{Released-shape error ($e_{\mathrm{rel}}$, percent of $L_0$) and cost of the four constitutive updates on the two comparison scenes at the middle calibrated accuracy grade; the supplement tabulates the other two grades and the reference levels. Return mapping is the rate-independent reference model. $n$ counts loading increments, wall time is the median of three repeats of the complete load--hold--unload--release solve on one thread, and \emph{missed} counts increments whose Newton iteration stopped without reaching the residual tolerance. A run with a missed increment advanced from an iterate the acceptance rule does not admit, so it is a diagnostic record: it keeps its row, marked $^{\dagger}$, but is excluded from the error and cost columns and from every ranking; the supplement locates each missed increment. Arrows indicate the preferred direction of each metric.}
\label{tab:n1-cost}
\end{table*}

%% file: common/results_derivative.tex
\subsection{Released-State Derivative and Minimal Control Task}
\label{sec:released-sensitivity}

Structural derivatives enter three settings of this section: fixed-partition
field-valued JVPs on the Bunny hierarchy, finite-difference sensitivities of a
scalar released-displacement observable in cube indentation, and a bounded control
task on one contact scene driven by the path derivative itself
(Section~\ref{sec:n2-control}). The local checks above evaluate a constitutive
mapping at a fixed history, whereas a design loop
needs the sensitivity of a structural response that includes unloading and tool
removal: the first two settings measure that sensitivity, and the third consumes
it.

\subsubsection{Fixed-partition path-derivative audit.}
Every audited product and its two replays carry the same recorded schedule hash
and the same $30$ compared stages, with no change of the active contact stencil,
so the two sides were compared on one accepted schedule. The audit is the
algorithmic tangent--vector product of Equation~\eqref{eq:path-sensitivity}
against independently executed centred full-path replays, on the two audited
tool-control directions: the symmetric control moves both tools equally along the
surface normal, and the tangential control moves one tool alone. It is per unit
and per accepted partition, not tensor-valued, and the assembled
three-dimensional tangent is not checked in arbitrary directions anywhere in this
section, nor is any convergence order across the mesh levels claimed.

Nine of the twelve planned two-sided replays completed, yielding eighteen
measured discrepancy values, five of which exceed the prescribed $10^{-3}$ target
and are kept rather than rescaled. Each value is an area-weighted
root-mean-square difference between the released observable field and its centred
replay field, divided by the root-mean-square of that replay field. Among the
completed comparisons the largest discrepancy against replay is $0.177\%$
response-relative, on the coarse tangential control at the widest step, and the
largest torso-observable discrepancy is $0.159\%$, on the medium symmetric-depth
control at the middle step. The rows in which one side did not finish are marked
and left empty, and the fine level carries no stored summary at all, so no
missing value is inferred from a neighbouring step size.

Table~\ref{tab:path-jvp-audit} summarises the completed fixed-partition
JVP--replay comparisons by mesh and tool-control direction; the complete
step-size scans, including the incomplete replay sides, are provided in the
supplement.

\input{common/path_jvp_table}

The second setting is the released-state response sweep of the controlled cube
indentation, whose measured object is the mean $z$ displacement over a fixed
rest-space node set of the indented face in the released state,
$Q_{\mathrm{rel}}(\theta)$, with $\theta$ scaling the tool trajectory at fixed
mesh, tool radius, barrier stiffness, loading duration and material parameters,
and every evaluation integrating a complete load--hold--unload--release history
from the initial state rather than perturbing a frozen state. That sweep runs
over a pre-registered grid of $21$ tool depths, with increment subdivision when
needed to satisfy the residual tolerance; its first plastically active grid point
is $\theta=0.292857$ for all three methods and the three released curves nearly
coincide away from the onset. An independent estimate from the two bracketing
grid points agrees with the complete-rerun difference to about $1\%$ at the
plastic control and differs by $13.7\%$, $13.7\%$ and $17.3\%$ at the near-onset
control, where the widest scanned step crosses the onset; the elastic control is
at round-off level, and the response table, the per-step scans, the subdivision
and rerun records and the protocol detail are in the supplement.

The claim is limited accordingly. The derivative is that of the committed path:
contact activation, adaptive subdivision and line-search decisions are carried
through as they occurred and are not differentiated, so consecutive runs may
solve different time partitions and each entry is a difference of full solves
rather than a perturbation of one fixed discrete map. The supplement also records
a depth-only proxy calibration, an independent exploration that is not evidence
about the derivative audited here.

\subsubsection{A minimal control task on the real chain.}
\label{sec:n2-control}

The third setting asks whether the same derivative can drive a control problem rather than
audit one. The scene is the coarse symmetric Bunny indentation, the control is the common
$z$ amplitude $\theta$ of both tools, so the control vector is $[0,0,\theta,0,0,\theta]$ and
the direction $[0,0,1,0,0,1]$ is the exact chain rule from the amplitude to the tool motion,
and the measured quantity is the released gauge compression after the full history. The
derivative is taken along the stored states and calls no perturbed forward solve.

A derivative on a fixed partition is only usable where the solver accepts that partition, so
one common partition of 26 intervals is used for every run of this setting; no control is
given a partition of its own. Every control in the measured band $[0.8415,\,0.8585]$
completes on it. The band is a measured set, not an assumed bracket: refining the partition
uniformly or only around contact onset did not widen it, the controls that a fixed partition
refuses outside it stop with a relative-CCD substep request, and the band is not
symmetric about the control used to build the partition, which limits the study.

The derivative was cross-checked against centre differences of the complete measured
quantity, with full re-solves at $\theta \pm h$ for four step sizes. At $\theta = 0.85$ the
relative difference against the derivative falls as the step shrinks,
$4.13\times10^{-4}$, $3.27\times10^{-4}$, $2.11\times10^{-4}$, $1.22\times10^{-4}$. At
$\theta = 0.85255$ it is at most $1.22\times10^{-4}$ at every step size, the same level as the
finest step at the first control, so the derivative agrees with full re-solves at both sampled
controls. An earlier evaluation of this same check left a difference of about
$4.5\times10^{-3}$ at the second control that did not fall with the step size: the contact
term of the derivative was assembled at the control that built the partition instead of at
the control of the run, and it was corrected. The previous and the corrected derivative are
both kept per control in the record, and the supplement reports the diagnosis.

Two controls below the contact-active range bound where this partition is usable there. At
$\theta = 0.40$ the forward history completes (released gauge compression $0.004186$) and the
derivative is available, $0.064968$; before the correction above, that derivative stopped with
a contact-linearisation error, which is kept next to the run instead of being discarded. At
$\theta = 0.55$ the forward history itself does not complete on the common partition: it stops
in the loading phase with a relative-CCD substep request that the fixed partition does not
perform. Both records are kept in the supplement.

\begin{figure*}[!t]
\centering
\includegraphics[width=\textwidth]{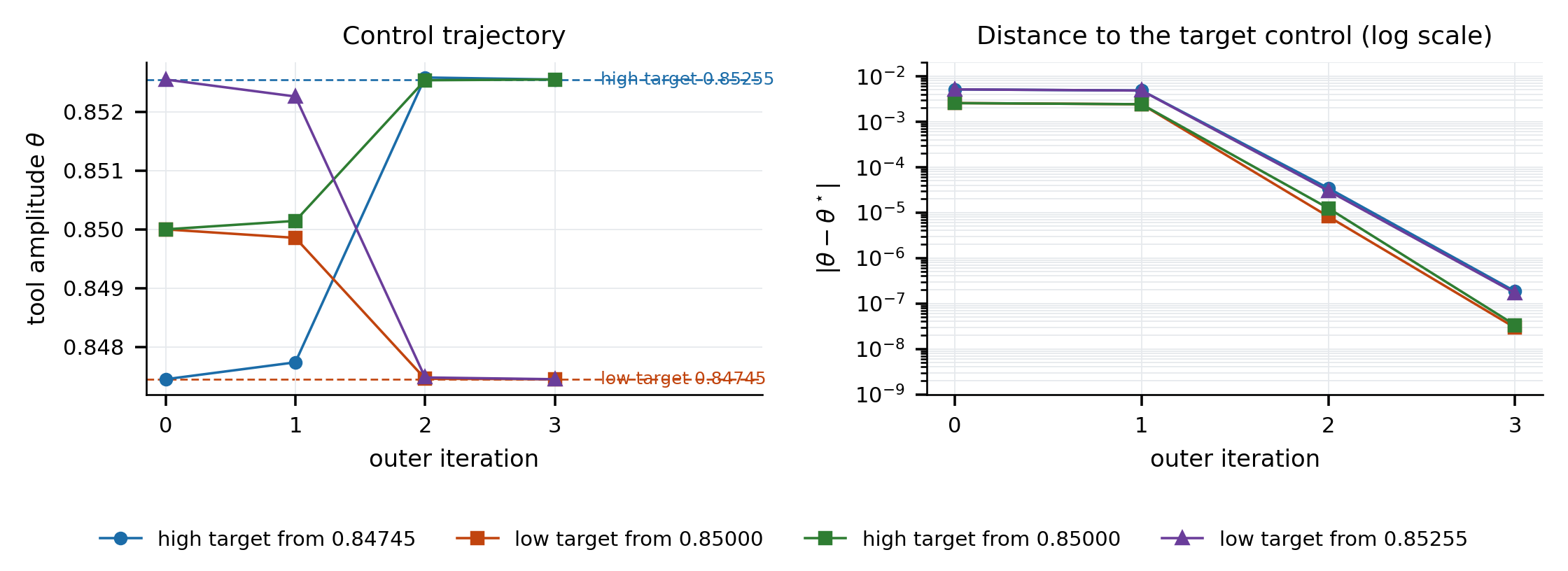}
\caption{The four non-trivial control trajectories of the bounded task of
Section~\ref{sec:n2-control}, drawn from the stored curve records without re-running any
solve. Left: control amplitude $\theta$ versus outer iteration, with target controls
shown as dashed lines. Right: absolute control-parameter error
$|\theta-\theta^{\star}|$ on a logarithmic scale; released-response discrepancies are
reported in Section~\ref{sec:n2-control}. The two remaining recorded curves start exactly
on a target, stop after one evaluation and are consistency checks; they are listed with
all six paths in the supplement's curve table.}
\label{fig:n2-trajectory}
\end{figure*}

The same derivative was then used to search on the fixed partition. Six bounded curves
(three starts, two targets in the band, twelve iterations and fourteen complete histories
allowed per curve) all stop on the projected-gradient criterion without reaching either limit.
Two start at their own target: they are consistency checks and return loss zero after a single
evaluation. The other four are the non-trivial calibrations. The four non-trivial trials use the
initial controls indicated in Figure~\ref{fig:n2-trajectory}, and each needs three iterations,
that is four complete histories and $265$--$537$~s of wall time, to drive $|\Delta Q|/L_0$ from
$6.1\times10^{-4}$--$1.2\times10^{-3}$ down to at most $8.2\times10^{-6}$ before the final
evaluation lands on the target. Every evaluation on all six
paths completed and returned a derivative, so no infeasible objective value and no derivative
failure arises inside the band. The supplement lists the curves and their evaluations.

The claim is limited accordingly. No other constitutive update of the comparison exposes a
path derivative through the same interface, so this setting measures whether the
released-state derivative of SECA is usable on a contact scene and how it behaves; it is not
a comparison of optimisation cost or convergence speed against another method, and no
optimisation advantage is claimed. The two targets are measured values of the same response
at two controls of the band, so a curve that reaches one re-measures a point already
measured, and what the curves report is the number of complete histories the derivative needs
to get there.

%% file: common/path_jvp_table.tex
% Generated by SECA_CL2.0/audit/v12_jvp_table.py -- do not edit by hand.
% Source: SECA_CL2.0/derived/g12_derivative_evidence.json (read-only)
%         SECA_CL2.0/derived/g05_retained_jvp.json (read-only)
%         SECA_CL2.0/derived/g13_table1_definitions.json
\begin{table*}[t]
\centering
\small
\caption{The actual-residual path Jacobian--vector product of
Equation~\eqref{eq:path-sensitivity} against independently executed centred
full-path replays on the self-contact-capable Bunny hierarchy, grouped by mesh
and audited control. The algorithmic side differentiates the finite-$\boldsymbol F_p$
residual and applies the exact unprojected tool and self-contact Hessian
actions; the replay side restarts from zero history on both sides of the step
$h$, a dimensionless increment of the control direction: the symmetric control
moves both tools equally along the surface normal and the tangential control
moves one tool alone. A completed replay yields two discrepancies of the
released observable field, each an area-weighted root-mean-square difference
divided by the root-mean-square of the centred replay field itself:
$q_{\mathrm{scene}}$ over all $6{,}002$ stored observation vertices and
$q_{\mathrm{obs}}$ over the $1{,}444$-vertex torso subset of those vertices.
The columns report the number of completed replays out of those planned, the
largest completed value of each observable, and how many of the measured
values exceed the target $10^{-3}$ that was fixed from the field scale before
the comparison. Three of the twelve planned replays did not complete on one
side and contribute no value, so the largest error is quoted among the
completed comparisons: $0.177\%$ response-relative on the coarse tangential
control at the widest step. The fine level carries no stored summary in the
retained record. Every step, every incomplete side and the stored aliases of
the smallest step are listed in the supplement. Linearised equilibrium
residuals stay below $2.35\times10^{-7}$ relative on every unit.}
\label{tab:path-jvp-audit}
\begin{tabular}{llrrrr}
\toprule
Mesh (tetrahedra) & Audited control & Completed / planned & max $q_{\mathrm{scene}}$ & max $q_{\mathrm{obs}}$ & Above target / measured \\
\midrule
coarse (2{,}502) & symmetric sphere depth & 1 / 3 & $4.76\times10^{-4}$ & $3.83\times10^{-4}$ & 0 / 2 \\
coarse (2{,}502) & one-sphere tangential translation & 3 / 3 & $1.77\times10^{-3}$ & $1.50\times10^{-3}$ & 2 / 6 \\
medium (4{,}992) & symmetric sphere depth & 2 / 3 & $1.48\times10^{-3}$ & $1.59\times10^{-3}$ & 2 / 4 \\
medium (4{,}992) & one-sphere tangential translation & 3 / 3 & $9.82\times10^{-4}$ & $1.56\times10^{-3}$ & 1 / 6 \\
\bottomrule
\end{tabular}
\end{table*}

%% file: common/macroscopic.tex
\subsection{Macroscopic, Parameter and Complex-Model Response}
\label{sec:macro-response}
\label{sec:cube-benchmark}

The properties above are stated for one material-point computation. This
subsection asks whether they survive assembly, contact, complete unloading and
parameter changes, and whether the same transition stays usable when the loading
mechanism, the connectivity and the geometry change. Every value is measured
under the stated simulation settings, so it describes the tested cases and does
not establish a global guarantee.

\subsubsection{Controlled spherical indentation}

The controlled benchmark is a unit cube compressed by an analytic spherical
tool, loaded, held and then completely withdrawn, with all three constitutive
methods on one mesh, one tool path, one physical duration, one Newton tolerance,
one line search and one committed-history rule, so the constitutive branch is the
only difference. The three agree on the measurable response: recovery is $48.35\%$
for SECA against $48.13\%$ and $48.14\%$, the largest element Kirchhoff von Mises
stress is $3.695\,$MPa against $3.658\,$MPa, the yielded volume is $23.44\%$ for
all three, and the smallest deformation determinant is $0.6733$ against $0.6737$,
so the compact transition neither suppresses yielding nor degrades the
deformation range. The agreement is the intended result: SECA regularises the
transition rather than producing a different resolved plastic response. Refining
the mesh from \depCubeTetrahedra\ to $24{,}576$ tetrahedra raises the released
dent from $9.602\%$ to $11.946\%$ for SECA, from $9.642\%$ to $11.958\%$ for
exact return mapping and from $9.642\%$ to $12.005\%$ for Perzyna, with the
spread between methods within $0.06$ percentage points at every level, so an
absolute indentation figure is quoted with its mesh level. Peak dent is
$18.589\%$ to $18.591\%$ of the body height and the released dent is $9.602\%$ to
$9.642\%$.

\subsubsection{Parameter control}
\label{sec:ablations}

\begin{figure*}[!t]
\centering
\includegraphics[width=\textwidth]{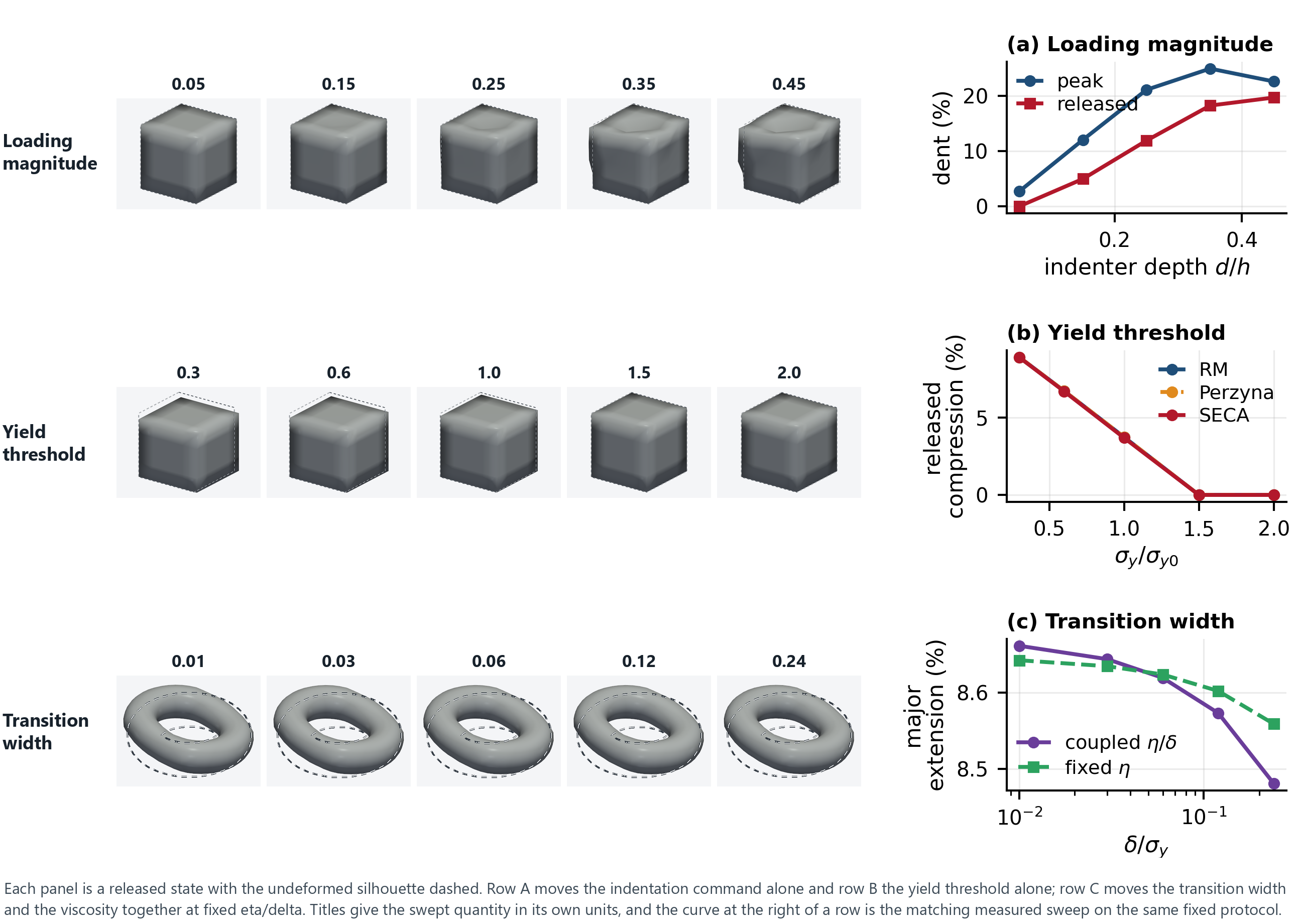}
\caption{Parameter control. Each row shows
the released state with the undeformed silhouette dashed, at the values printed
above the panels, beside the matching measured sweep on the same protocol.
Row~A varies the prescribed indentation depth of the block and plots the peak and
released response of the SECA runs; row~B varies the yield threshold of a matched
compressed cube and carries all three constitutive methods; row~C varies the
transition width of the gripped torus in two layers of the same SECA run set, one
with the viscosity coupled to the width ($\eta/\delta=0.2$) and one with the
viscosity held fixed. Within a row every quantity other than the named sweep is
held fixed; the coupled layer of row~C is the single place where two quantities
move together.}
\Description{Three rows of released shapes---block indentation at five depths,
compressed cubes at five yield thresholds, and stretched tori at five transition
widths---each beside the curve of its measured released response.}
\label{fig:parameter-control}
\end{figure*}

The finite-width bias is only useful if it is a controllable parameter rather
than an incidental artefact. Here the constitutive method is held fixed and one
named quantity moves on a fixed mesh, boundary condition, load path, time step,
tolerance and line search, so each trend belongs to the swept quantity on this
protocol; the supplement lists the swept values and the observable
read from the released state, and Figure~\ref{fig:parameter-control} places the
released shapes beside the measured sweep that produces them.
\noindent\textbf{Loading magnitude.} Indenting the block in \depLoadCount\ steps
moves the released dent from \depLoadDentMin\% to \depLoadDentMax\%, from an
entirely recoverable response to an almost entirely retained one; at the
shallowest level the released dent is zero to the reported precision, and the
stored states of that run carry a maximum accumulated plastic strain of exactly
zero and a multiplier of exactly zero at every increment rather than a small one.

\noindent\textbf{Strength.} On the matched compressed cube the released
compression of the SECA runs falls from \depYieldFirst\% to \depYieldLast\%; at
$\sigma_y/\sigma_{y,0}=2$ the released deformation is zero for all three
constitutive methods, again with a maximum accumulated plastic strain of exactly
zero rather than a small number, because the prescribed compression no longer
reaches the threshold. On the beam, hardening spans $H/H_0$ over
\depHardLo--\depHardHi\ and moves the peak Kirchhoff von Mises stress from
\depHardStressFirst\ to \depHardStressLast\ MPa while the residual tip offset
moves from \depHardFirst\% to \depHardLast\%. The material family moves modulus,
yield stress and hardening together across five presets under one normalised gate
setting; \depFamilyCount\ of the five reach a converged release and are the ones
reported, with the retained fraction rising monotonically from
\depFamilyTrendMin\% to \depFamilyTrendMax\% over those four and their peak stress
spanning \depFamilyStressMin\ to \depFamilyStressMax\ MPa. The fifth, softest
preset does not converge its final tool-free stage, so it is excluded from the
quoted range and kept in the supplement record rather than dropped; because the
presets move four parameters at once this is a joint-preset trend, not a
single-parameter ablation.

\noindent\textbf{Transition width.} Width and viscosity move together in the
coupled layer and only the width moves in the fixed-$\eta$ layer, which separates
the width bias from the rate effect that any overstress law carries. The released
major-axis extension of the gripped torus falls monotonically from
\depWidthCoupledFirst\% to \depWidthCoupledLast\% in the coupled layer and from
\depWidthFixedFirst\% to \depWidthFixedLast\% in the fixed-$\eta$ layer, so the
width is a dial rather than a source of scatter on this protocol and moving it
alone acts in the same direction as moving width and viscosity together, which is
not an additive decomposition of the released shape on other geometries. A
narrower transition stores more of the imposed extension as residual deformation,
the direction the local bias of Section~\ref{sec:regularity-bias} predicts, and
over the swept ranges of the supplement's sweep table the bias behaves as a control
rather than as scatter, the two layers separate, and no wider range is claimed.

\subsubsection{Diverse loading conditions}
\label{sec:simple-bodies}

\begin{figure*}[!t]
\centering
\includegraphics[width=0.94\textwidth]{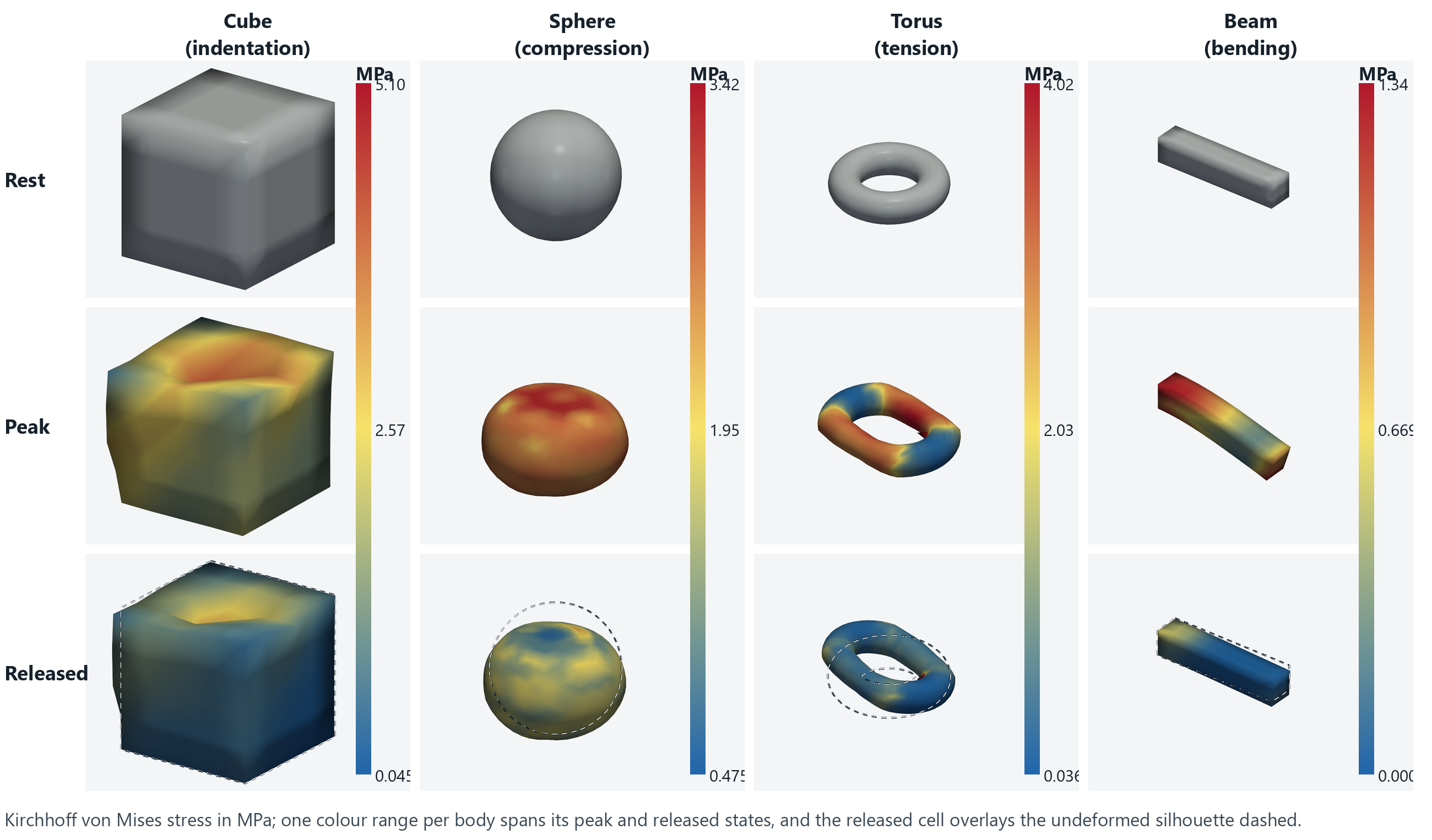}
\caption{Elastoplastic response under diverse loading conditions. The four columns
are indentation-, compression-, tension-, and bending-dominated cases. Rows are the
rest, peak, and released states; both the peak and released rows are coloured by
Kirchhoff von Mises stress, and the released row also shows the undeformed
silhouette dashed. Each column uses one camera, one scale, and one colour range
across its peak and released states.}
\Description{Three rows of undeformed, peak-stress, and released states for an
indented block, a compressed sphere, a stretched torus, and a bent cantilever.}
\label{fig:simple-bodies}
\end{figure*}

Evidence from one loading condition would not establish broader applicability, so
the same three constitutive methods run on four controlled geometries: local
indentation of a block, bilateral platen compression of a sphere, finite-area
grip tension of a genus-one torus and cantilever bending of a slender beam, with
one material, one mesh per body, one Newton tolerance, one Armijo line search and
complete removal of every tool or grip before the released state is measured
(Figure~\ref{fig:simple-bodies}). Under one SECA setting the retained fractions
are \depCubeRetainedPct\%, \depSphereRetainedPct\%, \depTorusRetainedPct\% and
\depBeamRetainedPct\%, a spread that follows from the loading geometry and not
from the constitutive update, which is identical in all four, and every column is
strictly elastic below its threshold. The supplement's released-state table collects the
measured peak, released and retained fractions with the smallest deformation
determinant on each path. Taken together the four cases measure generality rather
than accuracy: one width and one viscosity keep loading stable, recovery elastic
and the residual state physical when the loading mechanism, the connectivity and
the stress sign all change.

\subsubsection{Character-scale models}
\label{sec:characters}

These examples test whether the local transition remains robust once it is
embedded in finite deformation, contact, committed plastic history, complete
unloading and complex geometry. Two characters are loaded to a peak, held and
then released by removing the tools completely; the reported quantity is the
released equilibrium, not the pose under load, and Figure~\ref{fig:authoring}
shows each character at rest, at peak deformation and at released equilibrium.

\noindent\textbf{Bunny compression.} At dimensionless depth 0.85 the torso gauge
of the Bunny path reaches \depBunnyPeak\% peak compression under two opposed
analytic spheres and retains \depBunnyRetained\% after complete withdrawal, for
\depBunnyRecovery\ percentage points of recovery, with positive sphere--triangle
gap at every accepted state, so the loaded configuration cannot substitute for the
released equilibrium.

\noindent\textbf{Armadillo tension.} The second character uses the same
finite-$\boldsymbol F_p$ update with three pelvic supports and two
rest-area-weighted finite grip patches whose stiffness is reduced to zero after
the targets are held at peak. All 17 equilibrium stages converge at symmetric
axial control 1.00, and the gauge extends \depArmadilloPeak\% at peak and retains
\depArmadilloRetained\% after complete grip removal, for \depArmadilloRecovery\
percentage points of recovery. The same peak-versus-release distinction therefore
holds under tension and finite-area grip loading, not only under analytic-sphere
compression: a robustness result for the tested paths, not a claim that it changes
the resolved shapes.

\begin{figure*}[!t]
\centering
\includegraphics[width=\textwidth]{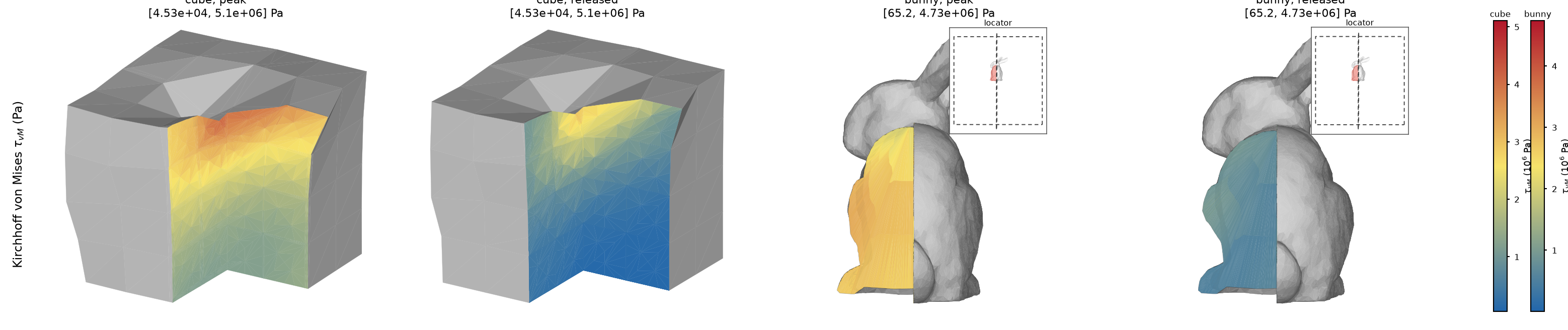}
\caption{Interior state of two of the complete solved bodies, opened after the
solve: the Kirchhoff von Mises stress on the exposed cut faces of the indentation
cube of Figure~\ref{fig:simple-bodies} and of the Bunny character, at peak load and
after complete tool removal. Cutting is post-processing only -- the bodies were
solved uncut and no cut model was re-equilibrated -- and the two planes are fixed in
the rest configuration of the frame each body is drawn in, so only the faces they
expose are drawn. The Bunny is cut on the display surface
Figure~\ref{fig:stanford-states} draws for it, so its cut faces are continuous with
the surface around them: the cube uses its simulation geometry, while the Bunny uses
the same display surface and viewpoint as Figure~\ref{fig:stanford-states}. The
colours are the recovered von Mises field, the element values averaged at the mesh
nodes by volume and read at each drawn point of the cut surface; one colour range is
shared by the peak and the released state of each model, and the ranges and every
number quoted in the text are read from the raw element-constant field. The
supplement repeats these faces in the raw element-constant form and adds the
committed equivalent plastic strain.}
\Description{Interior cutaway of the cube and the Bunny at peak and after
release, showing Kirchhoff von Mises stress on the exposed cut faces.}
\label{fig:interior-cutaways}
\end{figure*}

\begin{figure*}[!t]
\centering
\includegraphics[width=\textwidth]{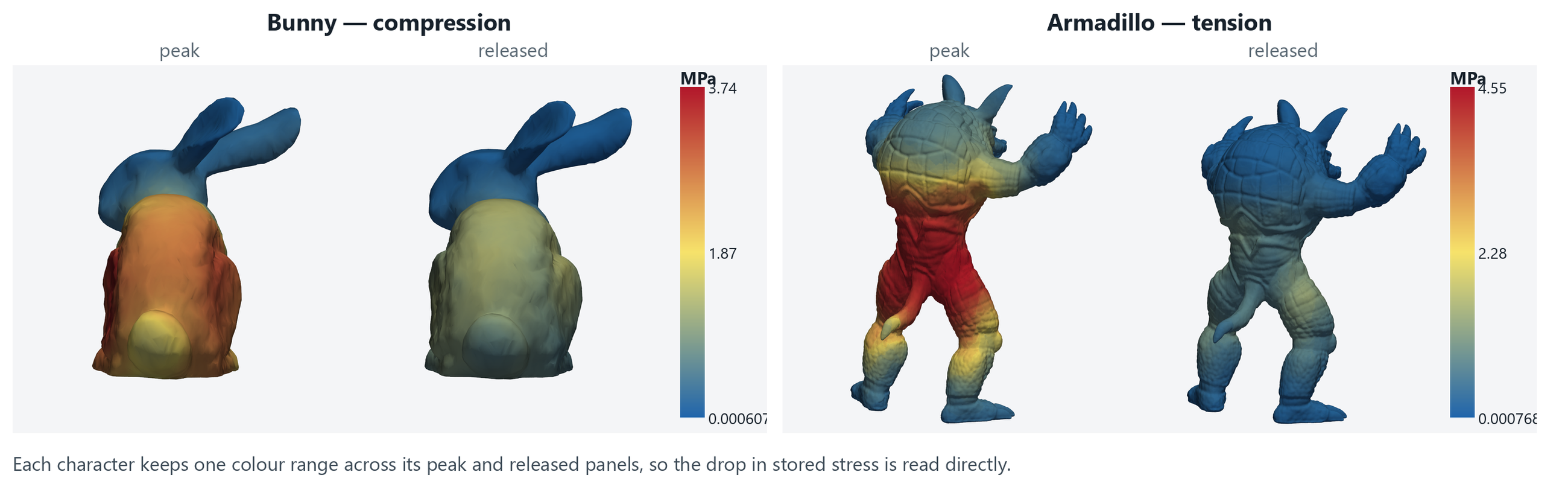}
\caption{Peak and released Kirchhoff von Mises stress on the two Stanford
models. Columns alternate the loaded peak, with the tools or grips active, and
the stress remaining after complete unloading under the stated support
conditions; the pair that belongs to one model shares one colour range, so the
drop from peak to released stress is read directly and not through a rescaling.
The panels show stress fields rather than retained deformation, which the
text quantifies; the plotted colours are a display field, the element values
averaged at the nodes and transferred to the render surface, and the cutaway
figure recovers its cut-face colours by the same rule, read at each drawn point of
the cut surface; the numbers quoted in the text are
read from the raw element-constant field in both cases. The surface drawn here is the display
triangulation of the scanned asset, 12{,}000 triangles for the Bunny and 345{,}944
for the Armadillo, so the fineness of the silhouette is the fineness of that asset
and not a measure of the computed resolution. The Bunny is drawn in the same frame,
from the same viewpoint and on the same triangulation as its panels in
Figure~\ref{fig:interior-cutaways}. The
two models do not share a stress range
because their yield stresses differ.}
\Description{Four panels of peak and released Kirchhoff von Mises stress fields
for the Bunny compression and the Armadillo extension, one colour range per
model.}
\label{fig:stanford-states}
\end{figure*}

\noindent\textbf{Interior and surface stress.}
Figure~\ref{fig:interior-cutaways} opens the same solved bodies at the loaded peak
and after complete tool removal and shows the internal fields that the surface
renderings only imply, with both cut planes fixed in the rest configuration of the
frame each body is drawn in and applied to the deformed state. The two character
panels are drawn in the frame and from the viewpoint of
Figure~\ref{fig:stanford-states}, so the Bunny is read at one orientation across
both figures. Figure~\ref{fig:stanford-states} shows what those
surface renderings carry: the peak and released von Mises field of the two Stanford
models, with one colour range per model. The two figures therefore answer different
questions and are read differently. Both draw a display field recovered from the
element-constant field so that the peak-to-release contrast is visible: the surface
figure recovers it for the rendered bodies, and the cut faces of the cutaway figure
are coloured by the same recovery, the element values averaged at the mesh nodes by
volume and then read at each drawn point of the cut surface. Every
number quoted here is read from the raw element-constant field, not from either
recovered field. The two figures also differ
in what is drawn, not in what was solved, and each body is cut on the surface the
surface figure draws for it. The cube is cut on its computed mesh, \depCubeNodes\ nodes
and \depCubeTetrahedra\ tetrahedra, so each facet on a cut face is one linear
tetrahedron and the field printed on it is evaluated inside that tetrahedron; the cut
surface is printed as pure mapped colour, with no element
outline drawn over it, and the outlined form of the same cut is the supplementary
element-constant record. The Bunny is cut
on the display triangulation of the scanned asset, 12{,}000 triangles, the same surface
Figure~\ref{fig:stanford-states} draws: its cut faces are that surface's own planar
section, they carry no element outlines, and each drawn point of the section is
located in the committed tetrahedral mesh of the deformed body and reads the four
corner values of the tetrahedron that contains it, interpolated linearly there, so the
section is coloured by the computed field at its own positions rather than by a colour
carried in from its edge. The Bunny's computed mesh, 699
nodes and 2{,}502 tetrahedra, is carried by the supplement, which repeats those cut
faces on it in the raw element-constant form, so the front figure's finer silhouette is
a finer display of the same solved state and not a finer computation. The fine
geometric detail of that triangulation is detail of the scanned asset: the computation
does not resolve it and it enters no quantity reported here. Neither figure is an
independent verification of the update and neither is a comparison between methods.
Rendering recovers a smooth field from that element-constant field for display only.
The supplement carries the committed
plastic-strain matrix on the same faces with plane normals and offsets, and repeats
the cut faces in the raw element-constant form.

\noindent\textbf{Computational cost and scope.} The character timings come from an
offline CPU implementation, and the supplement reports the historical seconds and
the available platform metadata, which names the library versions and the
logical processors but not the processor model, so no specific CPU is attached to
those seconds. These timings characterise the current CPU implementation and
exclude validation replay; the character examples verify that SECA remains stable
through finite deformation, contact, unloading and differentiation at character
scale, not real-time performance.

%% file: common/discussion.tex
\section{Discussion}
\subsection{What SECA Establishes}
\emph{SECA regularises the transition while preserving the strict sub-yield
constraint.} The compact gate keeps $G(0)=0$, so a sub-yield trial returns an
exactly zero multiplier, while Equation~\ref{eq:regularity-bias-product} and
Equation~\ref{eq:width-viscosity-decomposition} give its bias and its slope
exactly; committed finite-$\boldsymbol F_p$ history carries that local trade-off
into the released equilibrium.

\emph{This regularity has a measurable solver consequence.} On the shared
material-point path the yield-transition tangent change separates the methods:
over the run whose load phase is a single increment SECA needs less Newton work
and far fewer backtracking halvings than exact return mapping. That coarse
saving is the starting point of a bounded control task, not of a design loop:
Section~\ref{sec:n2-control} differentiates one tool-amplitude control through
the full committed history on a contact scene and drives a two-target loss over
a measured feasible band, while no larger authoring or inverse-design workflow is
evaluated here. The saving is confined to that coarse crossing: on refined
schedules SECA
needs slightly more whole-run Newton work than return mapping, and on the
grip-tension scene the coarse run does not favour SECA.

\emph{The resolved macroscopic response remains closely preserved.} Each
constitutive method converges at a closely comparable rate towards its own
resolved released state, and the controlled cube's peak and released
configurations are quantitatively closely matched across the three. Two
boundaries: it is a same-mesh comparison, since the absolute released depth moves
by roughly a tenth per mesh refinement here, and the cross-method difference in
the released field is of the same order as each method's own temporal
discretisation error. The benefit therefore belongs to the transition tangent and
the solver rather than to the final shape.

\emph{The finite-width bias is controllable, and one normalised setting
generalises across loading modes and material ranges.} Width moves the released
deformation monotonically, and on the reported protocol the accompanying bias is
a few percent rather than an order of magnitude: width tunes residual
deformation, while the imposed load decides it. Only the fixed-$\eta$ layer
isolates the width effect; the per-sweep figures are in
the supplement's sweep records and Section~\ref{sec:ablations}. The same normalised
gate setting carries local indentation, bilateral compression, grip tension on a
doubly connected body, and flexure-dominated bending over a family that varies
modulus, yield stress and hardening together; because $\delta$ and $\eta$ scale
with the material, it is a preset range rather than a single-parameter ablation.

\subsection{Relation to Perzyna Regularisation}
SECA is a compact rate-function choice within the Perzyna lineage: a smooth
Perzyna law using the same $r_\delta$, $\eta$, history update, and time
integrator is locally identical, so the contribution claimed here is the exact
gate analysis and its verified integration through committed history, tool
removal, and released-equilibrium differentiation. Linear-overstress Perzyna does
not remove this slope change on this path, and its global overstress envelope
needs repeated line-search damping at a coarse increment, whereas the compact
activation crosses that increment with the counts recorded in
Section~\ref{sec:micro-numerics}.

\subsection{Scope of the Measurements}
The reported agreement between the three constitutive methods is a matched-mesh
and matched-schedule statement: every comparison shares the settings of the
controlled benchmark, and the error and cost comparison of
Section~\ref{sec:n1-comparison} adds a fourth, purely comparative strict-elastic
Perzyna update under its own calibrated setting. Within each controlled
comparison the conditions are common, but a shared setting is still not a shared
plastic history, and each method's own refined reference is not a cross-method
true error. The path JVP differentiates the actual unprojected residual
on each fixed accepted schedule, the quantity an inverse-design loop consumes,
and the release propagation result applies on a shared stable branch under the
assumptions of Section~\ref{sec:regularity-bias}. The numerical comparison is
against constitutive references, exact return-mapped $J_2$ and linear-overstress
Perzyna, the models that define the competing transition behaviours. The control
task of Section~\ref{sec:n2-control} is one amplitude direction on one contact
scene, over a measured feasible band and on a single partition common to the
whole band, solved for SECA alone: the compared references expose no derivative
at that interface, so the task measures availability and behaviour rather than
an optimisation advantage. The character scenes are offline CPU solves at the
reported timings.

%% file: common/conclusion.tex
\section{Conclusion}
We present SECA, a Strict-Elastic Compact Activation method for differentiable
elastoplastic simulation. SECA prescribes a bounded $C^2$ activation over a finite
positive-overstress interval and integrates it into a Perzyna-type flow law. This
construction preserves exactly zero plastic increments for sub-yield trials while
smoothing the local elastic--plastic transition. For scalar and proportional
isotropic $J_2$ updates, we derive an exact decomposition of the plastic-increment
discrepancy into finite-width and viscosity contributions, together with an
explicit local budget for the transition width. The same construction is
incorporated into finite-deformation simulation with contact, committed plastic
history, complete unloading, and released-equilibrium differentiation.

The controlled benchmarks show closely matched macroscopic loading--release
responses and reduced nonlinear solver work in the tested coarse cube-indentation
setting. Independent full-history replays evaluate the tool-control derivatives,
and a bounded contact task demonstrates their use in released-state control.
Together, these results establish a strict-elastic transition regularisation with
explicit local bias characterisation and demonstrated use in graphics-oriented
elastoplastic simulation.

%% file: common/appendix.tex
\appendix
\section{Proof Details and Higher Regularity}
\label{app:proofs}
For the assumptions in Section~\ref{sec:regularity-bias}, the post-transition
identity follows directly from
\[
\int_0^fD_\delta(z)\,dz
=\int_0^\delta D_\delta(z)\,dz+f-\delta
=f-\int_0^\delta[1-D_\delta(z)]\,dz.
\]
For the implemented quintic,
$D_\delta(f)+D_\delta(\delta-f)=1$ on $(0,\delta)$, proving that the last
integral equals $\delta/2$.

\paragraph{Quintic endpoint system.}
The middle branch of the main paper's activation $D_\delta$ is the degree-five
polynomial
$P(\xi)=c_3\xi^3+c_4\xi^4+c_5\xi^5$ in $\xi=f/\delta$, a form that already
enforces $P(0)=P'(0)=P''(0)=0$. The three remaining conditions $P(1)=1$,
$P'(1)=0$ and $P''(1)=0$ give the linear system
\[
\begin{pmatrix}1&1&1\\3&4&5\\6&12&20\end{pmatrix}
\begin{pmatrix}c_3\\c_4\\c_5\end{pmatrix}
=\begin{pmatrix}1\\0\\0\end{pmatrix},
\qquad\text{so}\qquad c_3=10,\quad c_4=-15,\quad c_5=6 .
\]
Its determinant is $2$, so these coefficients are the only solution in that class
and the window carries no free parameter beyond $\delta$. The same polynomial
satisfies $P(\xi)+P(1-\xi)=1$, the antisymmetry about $(1/2,1/2)$ used above to
obtain the $\delta/2$ offset; consistently, $D_\delta''$ vanishes at
$\xi=1/2$.

The nonzero interior derivatives are
\begin{align}
D_\delta'(f)&=\frac{30}{\delta}\xi^2(1-\xi)^2,\\
D_\delta''(f)&=\frac{60}{\delta^2}\xi(1-\xi)(1-2\xi),\\
D_\delta'''(f)&=\frac{60}{\delta^3}(1-6\xi+6\xi^2).
\end{align}
Thus $D_\delta\in C^2$ but is generally not $C^3$, while
$r_\delta\in C^3$ but is generally not $C^4$. In addition to
Equation~\ref{eq:regularity-bias-product},
\[
\|D_\delta''\|_\infty=\frac{10}{\sqrt3\,\delta^2}.
\]
Differentiating Equation~\ref{eq:gamma-curvature} once more gives
\[
\gamma_{f_{\rm tr}f_{\rm tr}f_{\rm tr}}
=\frac{\alpha D_\delta''(\bar f)}{A^4}
-\frac{3\alpha^2B[D_\delta'(\bar f)]^2}{A^5}.
\]
These are local constitutive regularity statements, not regularity of a
contact-constrained global solution map.

\input{common/regularity_table}

\begin{figure}[t]
  \centering
  \includegraphics[width=\textwidth]{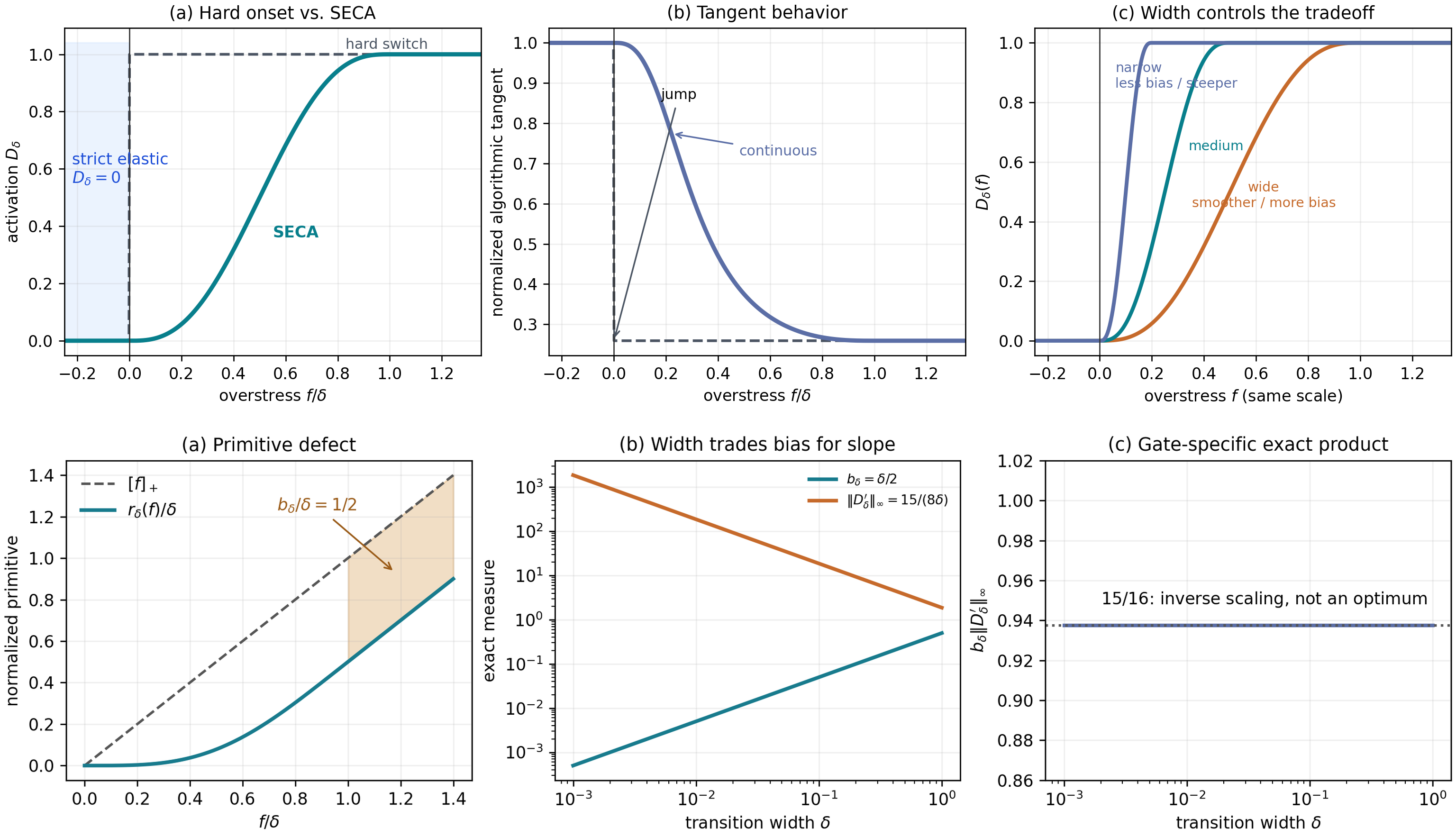}
  \caption{The complete six-panel core-activation figure. Top row: (a) the hard
  switch against compact activation, (b) the implicit SECA tangent across the
  onset against the hard-onset jump, and (c) three transition widths. Bottom
  row: (a) the primitive defect against the positive-part function, (b) the
  bias-versus-slope trade-off as $\delta$ decreases, and (c) the gate-specific
  product, which is exactly $15/16$.}
  \Description{The six panels of the core-activation figure: activation, local
  tangent, three transition widths, primitive defect, width-versus-slope
  trade-off, and the exact product.}
  \label{fig:core-activation-full}
\end{figure}

\paragraph{Complete activation and regularity panels.}
Figure~\ref{fig:core-activation-full} reproduces the full panel set behind the
main paper's core-activation figure, including the two panels that the main text
does not carry. Decreasing the transition width $\delta$ lowers the primitive
defect while increasing the peak activation slope, and their gate-specific
product is exactly $15/16$; that product is
Equation~\ref{eq:regularity-bias-product}, an identity of the implemented
quintic rather than a fitted curve. All six panels are drawn from the closed-form
gate and its primitive alone, at the parameters used in the main text
($E=100$, $H=10$, $\Delta t/\eta=0.04$).

\paragraph{Existence and uniqueness.}
$G(\gamma)$ is continuous and strictly increasing because increasing
$\gamma$ decreases $\bar f$ while $r_\delta$ is nondecreasing.
$G(0)\leq0$ and $G(\gamma)\to+\infty$ as $\gamma\to+\infty$, so the
nonnegative root is unique. For the implemented gate and
$f_{\rm tr}>0$, it satisfies $0<\gamma<f_{\rm tr}/B$.

\paragraph{Dissipation of the flow rule.}
The plastic-work property quoted in the main paper's local-root subsection
follows from the associative direction and the hardening term. With
$\dot{\boldsymbol p}=\dot a\,\boldsymbol n$,
$\boldsymbol n=3\boldsymbol s/(2\sigma_{\rm eq})$ and
$\boldsymbol s:\boldsymbol s=\tfrac23\sigma_{\rm eq}^2$,
\[
\boldsymbol\sigma:\dot{\boldsymbol p}
=\dot a\,\frac{3}{2\sigma_{\rm eq}}\,\boldsymbol s:\boldsymbol s
=\dot a\,\sigma_{\rm eq},
\qquad\text{hence}\qquad
\boldsymbol\sigma:\dot{\boldsymbol p}-Ha\dot a
=(\sigma_{\rm eq}-Ha)\dot a=(\sigma_y+f)\dot a .
\]
On the active branch $\dot a=r_\delta(f)/\eta\geq0$ for $\eta>0$ and
$\sigma_y+f>0$; on $f\leq0$ the gate is exactly zero, so $\dot a=0$ and
$\dot{\boldsymbol p}=\boldsymbol 0$. The local dissipation is therefore
nonnegative, which is what the discrete update approximates increment by
increment; it is a property of the rate form and not an extra constraint imposed
on the solve.

\paragraph{Multiplier derivatives.}
With $A=1+\alpha BD_\delta(\bar f)$ and
$\bar f=f_{\mathrm{tr}}-B\gamma$, and with the previous committed state held
fixed, implicit differentiation of $G(\gamma)=0$ gives
\begin{align}
\gamma_{f_{\mathrm{tr}}}
&=\frac{\alpha D_\delta(\bar f)}{A},&
\left.\gamma_\delta\right|_\eta
&=\frac{\alpha\,\partial_\delta r_\delta(\bar f)}{A},\label{eq:gamma-derivatives-a}\\
\left.\gamma_\eta\right|_{\delta,\Delta t}
&=-\frac{\gamma}{\eta A},&
\gamma_{\Delta t}
&=\frac{\gamma}{\Delta t A}.\label{eq:gamma-derivatives-b}
\end{align}
The width partial differentiates the primitive with respect to its width at fixed
overstress argument,
\begin{equation}
\partial_\delta r_\delta(\bar f)=
\begin{cases}
0,&\bar f\leq0,\\
P(\xi)-\xi D_\delta(\bar f),&0<\bar f<\delta,\\
-\tfrac12,&\bar f\geq\delta,
\end{cases}
\label{eq:primitive-width-partial}
\end{equation}
where $P(\xi)=\tfrac52\xi^4-3\xi^5+\xi^6$ is the accumulation
$\int_0^{\xi}D_\delta(\delta s)\,ds$, so that $r_\delta(\bar f)=\delta P(\xi)$
inside the interval. If an implementation couples $\eta=c\delta$, both the
transition width and the viscosity then move together, and the width derivative
is the total derivative
\begin{equation}
\frac{d\gamma}{d\delta}\bigg|_{\eta=c\delta}
=\frac{\alpha}{A}\left[
\partial_\delta r_\delta(\bar f)-\frac{r_\delta(\bar f)}{\delta}\right].
\end{equation}
The multiplier curvature is
\begin{equation}
\gamma_{f_{\mathrm{tr}}f_{\mathrm{tr}}}
=\frac{\alpha D_\delta'(\bar f)}{A^3},
\label{eq:gamma-curvature}
\end{equation}
and it is the scale a finite-difference check on this update can converge
against. In one dimension, with the previous state fixed, the corresponding
algorithmic modulus is
\begin{equation}
C_{\mathrm{alg}}=E-\frac{E^2\alpha D_\delta}
{1+\alpha(E+H)D_\delta}.
\label{eq:local-tangent}
\end{equation}

\paragraph{Parameter products.}
At fixed previous one-dimensional state,
$f_{\rm tr}=E(\epsilon-p_n)-\sigma_y-Ha_n$ and $B=E+H$. With
$\gamma_B=-\alpha\gamma D_\delta(\bar f)/A$,
\[
\gamma_E=\gamma_f(\epsilon-p_n)+\gamma_B,\qquad
\gamma_{\sigma_y}=-\gamma_f,\qquad
\gamma_H=-a_n\gamma_f+\gamma_B.
\]
The executable checks cover these products together with width, viscosity,
and step duration. The same update supplies the chain when the perturbed
quantity is a material parameter: $\partial B$ enters the denominator through
$B=3\mu+H$, the trial overstress changes both through the elastic trial stress
and through the threshold, and $D_\delta'(\bar f)$ multiplies the resulting
change of the discrete overstress. Those products follow from the update rather
than from a separate constitutive assumption, and they are not themselves
audited here: the derivative checks reported in the main text perturb tool
controls with the material parameters, the gate width and the viscosity held
fixed.

\paragraph{Propagation to released equilibrium.}
Let $\boldsymbol h_n$ collect committed plastic history along a fixed discrete
loading path, and let the two paths compared share their initial committed
history, $\boldsymbol h_0=\widetilde{\boldsymbol h}_0$. Suppose two nearby
updates satisfy
\begin{equation}
\|\boldsymbol h_{n+1}-\widetilde{\boldsymbol h}_{n+1}\|
\leq L_n\|\boldsymbol h_n-\widetilde{\boldsymbol h}_n\|
+M_n|\gamma_n-\widetilde\gamma_n| .
\end{equation}
Iteration yields
\begin{equation}
\|\boldsymbol h_N-\widetilde{\boldsymbol h}_N\|
\leq\sum_{j=0}^{N-1}M_j|\gamma_j-\widetilde\gamma_j|
\prod_{k=j+1}^{N-1}L_k .
\label{eq:history-propagation}
\end{equation}
Here $\boldsymbol h_n$ is the assembled committed state, not one scalar, so a
difference measured at a single material point cannot stand for the history of a
whole body. After complete tool removal, assume
$\boldsymbol R(\boldsymbol x,\boldsymbol h,\boldsymbol0)=\boldsymbol0$
has one continuously connected stable branch between the two histories, with
$\|\boldsymbol R_x^{-1}\|\leq\kappa$ and
$\|\boldsymbol R_h\|\leq L_R$. For an $L_Q$-Lipschitz released observable,
\begin{equation}
|Q(\boldsymbol x_{\mathrm{rel}})-Q(\widetilde{\boldsymbol x}_{\mathrm{rel}})|
\leq L_Q\kappa L_R
\sum_{j=0}^{N-1}M_j|\gamma_j-\widetilde\gamma_j|
\prod_{k=j+1}^{N-1}L_k .
\label{eq:release-propagation}
\end{equation}
Equation~\ref{eq:release-propagation} is stated for trajectories that remain
in one local neighborhood, share the accepted time partition and
contact-feature sequence, and release onto a continuously connected
equilibrium branch with a nonsingular residual Jacobian. Under these
assumptions the bound gives the explicit bridge from multiplier differences
to committed history and released observables on a stable, fixed accepted
graphics path. It is a conditional statement: it does not assert that every
finite-deformation path stays on one branch, and it does not by itself
establish a released-shape error for the cases of the main text, whose
released states are measured directly.

\section{Protocols and Reproducibility}
\label{app:protocols}
\paragraph{Local protocol.}
The scalar derivative set includes elastic, near-onset, transition, plastic,
unloading, and reverse-loading states. Independent bisection solves the
implicit root; centred differences use parameter-scaled steps. A tensor-valued
loading history changes loading direction and compares the finite-width update
with
return-mapped $J_2$ without invoking the proportional theorem.

\paragraph{Contact and derivative protocol.}
The Bunny hierarchy uses non-nested structured interiors with 2,502, 4,992,
and 14,316 tetrahedra. Material, supports, analytic tools, load duration,
30 accepted physical intervals, and the 6,002-vertex observation surface are
shared. Coarse and medium derivatives use two fixed control directions.
Independent plus and minus simulations restart from zero history for every
centred replay. The derivative operator uses the exact unprojected residual;
forward positive-semidefinite projection is excluded.

The tool contact of the main text is the barrier
\begin{equation}
g_{jt}(\boldsymbol x,\boldsymbol c_j)
=\sqrt{d_{\mathrm{PT}}^2(\boldsymbol c_j,\boldsymbol x_t)}-r_j ,
\qquad
B_{\mathrm{tool}}=\sum_{j,t}\kappa_b A_t b(g_{jt}),
\label{eq:contact-barrier}
\end{equation}
with $b(g)=-(g-\hat d)^2\log(g/\hat d)$ for $0<g<\hat d$ and $b(g)=0$ for
$g\geq\hat d$. Here $d_{\mathrm{PT}}$ is the point--triangle distance supplied
by IPC Toolkit, $r_j$ the analytic sphere radius, $A_t$ the boundary triangle
area and $\kappa_b$ the barrier stiffness, so a positive gap below $\hat d$
contributes a finite energy whose first derivative is continuous at $\hat d$.
Relative CCD limits each accepted update for tool and deformable-surface
contact, and the self-contact-capable Bunny path adds IPC boundary collision
candidates, barrier energy and force on top of this tool term.

\paragraph{Denominators.}
Absolute discrepancies are always retained. $L_{\rm ref}$ is the immutable
common-surface bounding-box diagonal and is not a response norm. A
response-relative value names its comparator explicitly. If a response norm
is nonfinite or below $10^{-14}$, the relative value is undefined; no
denominator floor is inserted. Gauge differences between levels are reported
in percentage points.

\paragraph{Tangent-change variant audit.}
The quantity quoted in the results section as $\Delta C_{\mathrm{tr}}$ is the
absolute difference between the secant slope of Kirchhoff stress over a window of
$2\times10^{-4}$ logarithmic strain below yield and the secant slope over the same
window above yield, on the shared material-point path of that section's local
comparison. The estimator reads the integration nodes of the recorded run and not
its output samples. The base protocol is a window of $2\times10^{-4}$ sampled as
twenty steps of $10^{-5}$ on a $2001$-point grid, with
$\Delta t = 1.25\times10^{-3}$~s over $2.5$~s at a strain rate of $0.008$~s$^{-1}$;
there the change is $1.52705\times10^{7}$~Pa for exact return mapping,
$1.58508\times10^{7}$~Pa for Perzyna and $2.44151\times10^{5}$~Pa for SECA, a ratio
of $62.545$. Four protocol changes were measured against that base, each with the
remaining conditions held fixed, and they separate cleanly.
Sampling the same recorded history on $8001$ output points instead of twenty
leaves all three values bit-identical, with a largest shift of $0.0$~Pa, because
the estimator reads integration nodes; drawing density is therefore not a
contributor. Refining the constitutive integration step at fixed physical history,
$2001 \to 4001 \to 8001$ samples with $\Delta t$ halving each time and the duration
held at $2.5$~s, moves return mapping by less than $10^{-3}$~Pa and SECA by a few
per cent, so the ratio moves slightly upward rather than downward, $62.545 \to
65.371 \to 66.859$. The archived refined-grid record is a different kind of change:
it keeps $\Delta t$ and doubles the number of steps, so the physical duration
doubles to $5.0$~s and the strain rate halves to $0.004$~s$^{-1}$. Under that change
the rate-independent return mapping reproduces its value to
$2.49\times10^{-4}$~Pa while SECA's rate-dependent value moves by
$2.14\times10^{5}$~Pa, and the ratio falls to $33.304$ because the loading history
changed and the rate-dependent denominator moved, not because one computation was
sampled differently. The window itself is the first-order driver of the quoted
number: $10^{-4}$, $2\times10^{-4}$ and $4\times10^{-4}$ logarithmic strain, sampled
as ten, twenty and forty steps on the same history, give ratios of $194.116$,
$62.545$ and $10.906$, because the reference methods' secant difference grows with
the window while the compact method's is set by the curvature of its transition.
The quoted factor is therefore only defined together with the window, the
integration resolution and the step it was measured on. One further comparison is
recorded so that it cannot be confused with these: on a frozen history and at the
yield state, the algorithmic tangent
$\mathrm{d}\tau_{11}/\mathrm{d}\epsilon$ is $1.22866\times10^{7}$~Pa for return
mapping, $9.06638\times10^{6}$~Pa for Perzyna and $2.30753\times10^{7}$~Pa for SECA,
with activation values $1$, $1$ and $4.14\times10^{-5}$ at that state, so on a
frozen history SECA's tangent is the largest of the three. That tangent is a
different object from the marched secant change and the two are never quoted as
one another. The headline number is retained at the base protocol with its window,
resolution and step stated, the refined-grid value is reported beside it as a
history-sensitivity study, and neither is called a speed-up; no measurement was
swapped while an older number stayed in place.
\begin{table}[!ht]
\centering
\small
\caption{The recorded variant audit behind the quoted tangent change. Each row
changes one thing against the base protocol and holds the rest fixed; the two
method columns are the absolute secant differences and the last column is their
ratio. The output-sampling row is bit-identical to the base row. The frozen-history
algorithmic tangent is not a row here because it is a different measured object.}
\label{tab:tangent-variants}
\setlength{\tabcolsep}{4pt}
\begin{tabular}{@{}p{0.235\textwidth}p{0.315\textwidth}rrr@{}}
\toprule
Variant & What changes & RM (Pa) & SECA (Pa) & ratio \\
\midrule
base protocol & window $2\times10^{-4}$, $2001$ samples, $\Delta t = 1.25\times10^{-3}$~s, $2.5$~s
  & $1.52705\times10^{7}$ & $2.44151\times10^{5}$ & $62.545$ \\
output sampling only & output points $20 \to 8001$, history unchanged
  & same & same & same \\
integration $\times2$ & $4001$ samples, $\Delta t = 6.25\times10^{-4}$~s, duration fixed
  & $1.52705\times10^{7}$ & $2.33598\times10^{5}$ & $65.371$ \\
integration $\times4$ & $8001$ samples, $\Delta t = 3.125\times10^{-4}$~s, duration fixed
  & $1.52705\times10^{7}$ & $2.28400\times10^{5}$ & $66.859$ \\
archived refined grid & steps doubled at fixed $\Delta t$, duration $5.0$~s, rate $0.004$~s$^{-1}$
  & $1.52705\times10^{7}$ & $4.58519\times10^{5}$ & $33.304$ \\
window $10^{-4}$ & ten steps per half window, same history
  & $8.96039\times10^{6}$ & $4.61600\times10^{4}$ & $194.116$ \\
window $4\times10^{-4}$ & forty steps per half window, same history
  & $1.84256\times10^{7}$ & $1.68947\times10^{6}$ & $10.906$ \\
\bottomrule
\end{tabular}
\end{table}

\paragraph{Released-equilibrium JVP audit.}
The supplement compares the actual-residual algorithmic Jacobian-vector
product (JVP) with centred complete replay on the self-contact-capable Bunny
hierarchy. Each discrepancy is an area-weighted root-mean-square difference
between the algorithmic JVP of the released field and the centred difference of
the two replays, $\lVert\cdot\rVert$ with the reference-area weights renormalised
over the compared vertex set, divided by the root-mean-square of the centred
replay field itself. The whole-object observable $q_{\mathrm{scene}}$ uses all
$6{,}002$ stored observation vertices and the torso observable $q_{\mathrm{obs}}$
uses the $1{,}444$-vertex subset of them; the geometric normalisation divides the
absolute difference by $L_{\mathrm{ref}} = 1.622259039004734$, the bounding-box
diagonal of the reference surface. The step $h \in \{0.01, 0.005, 0.0025\}$ is a
dimensionless increment of the control direction, $c(\epsilon) = c_{\mathrm{base}}
+ \epsilon v$, so the JVP is per unit dimensionless $\epsilon$ and is not divided
by $\lvert v\rvert$; the symmetric control moves both tools equally along the
surface normal and the tangential control moves one tool alone.
Across the stored replay steps that carry a complete field record, the
largest torso-observable field discrepancy relative to replay response is
$0.159\%$, on the medium symmetric-depth control at the middle step, and the
largest whole-object discrepancy is $0.177\%$ response-relative, on the coarse
tangential control at the widest step. The largest geometrically normalised
whole-object discrepancy is $0.0124\%\,L_{\mathrm{ref}}$, on the medium
symmetric-depth control at the middle step, so the two normalisations attain
their maxima on different rows. Linearised equilibrium residuals remain below
$2.35\times10^{-7}$ relative. These bounds describe the audited cases rather than
the hierarchy as a whole: the stored record holds twenty-six relative
discrepancies, eighteen scan values plus eight aliases of the smallest step, and
six of those entries exceed $10^{-3}$; they form five distinct values, because
one value is stored twice, once in the scan and once as the smallest-step result.
One stored unit carries no summary at all, and two of the four
summarised units record a step scan in which one side is incomplete at two and at
one step sizes respectively, so file presence is not replay completeness. On one
medium-mesh tangential control the discrepancy is also not monotone in the step.
Table~\ref{tab:path-jvp-full} lists every stored step of the audited controls,
marks the side that did not complete, and records the aliases of the smallest
step.

\input{common/path_jvp_table_full}

Geometry-normalised differences become smaller
from one mesh pair to the next on this non-nested hierarchy, verifying that
residual deformation and recovery remain physically consistent across these
spatial discretisations.
The Armadillo scene supplies an independent loading mechanism: its complete
17-stage source reproduces the stored observables to $4.77\times10^{-13}$
absolute difference, and symmetric axial and one-grip tangential
actual-residual JVPs, propagated through finite-$\boldsymbol F_p$ history, differ
from complete centred
replay by at most 0.539\% over five released quantities, with a maximum
linearised residual of $7.70\times10^{-8}$.

\paragraph{Material-point sensitivity protocol.}
\label{app:sensitivity-protocol}
This paragraph records the material-point protocol behind the local comparisons of
the main text, not a released-state field comparison. The local comparison quoted
in the local-regularity subsection of the main text is
evaluated in
the scaled prototype configuration recorded with those checks. The largest scalar
stress--strain derivative discrepancy there is $4.50\times10^{-7}$ relative (yield
stress $2.0$, hardening $10.0$, width $0.2$, $\eta/\delta=0.2$, unit loading
duration), and the exact width/rate decomposition closes to
$3.47\times10^{-18}$; both sides of that identity are floating-point evaluations
of the same expression, so the residual measures round-off alone rather than
agreement between two derivations. Repeating the six
states at the evaluation point of that section separates the protocols: the
algorithmic tangent and its closed form agree to $8.4\times10^{-16}$ relative, the
centred difference at $h=10^{-6}$ to $9.3\times10^{-11}$, and only the near-onset
state deviates at $h=10^{-4}$ ($1.6\times10^{-5}$) because a step of that size
leaves the elastic branch. On the nonproportional width sweep quoted in the main
text, $0.2\to0.05\to0.005$ gives $3.480\%$, $0.890\%$ and $0.089\%$ maximum
yield-normalised stress discrepancy. In the same configuration and at fixed width
$0.2$ and physical loading duration, 20--2000 scalar load steps produce a final
stress of $2.321412$ to the shown precision, while changing the duration changes
the response, as expected for a viscous law. Onset is reached at the same strain
($0.07222222222222202$) by all three methods, and at the switch point the
one-sided relative gaps are $9.35\times10^{-1}$ for return mapping,
$6.07\times10^{-1}$ for Perzyna, and $3.23\times10^{-10}$ for SECA, whose
automatic-differentiation tangent lies inside the one-sided interval.
The tables below are the complete pre-registered records behind the released-state
sensitivity subsection of the main text: the response grid in
Table~\ref{tab:sensitivity-curve} and the full step scan at the three control
points in Table~\ref{tab:sensitivity-derivative}. Every difference in the second
table is a difference of two independently re-integrated
load--hold--unload--release histories, not a perturbation of a frozen state; a
dagger marks a response row on which at least one method needed substep refinement
of an increment before the residual tolerance was met, and the flags in the second
table record contact-set changes, substep refinement in a rerun, unresolved reruns,
and steps that cross the onset. Within the response grid, $9$ of the $63$ rows
recorded one increment that did not meet the residual tolerance, on $4$ of the $21$
grid points; each was rerun with the increment itself subdivided, which leaves the
physical loading path and the total simulated duration unchanged, and every one then
met the tolerance. Two of the $27$ difference rows were formed from a perturbed
rerun whose minus side needed the same in-place subdivision, both at the plastic
control and the widest step, and those rows are not used as fixed-partition
comparisons.

\begin{table}[htbp]
\centering
\footnotesize
\setlength{\tabcolsep}{4pt}
\caption{Released-state response of the controlled cube indentation over the
pre-registered tool-depth grid. $Q_{\mathrm{rel}}$ is the mean released
displacement of a fixed rest-space region of the indented face; the volume
fraction is the plastically active share of the mesh. A dagger marks a grid point
at which at least one method needed substep refinement of an increment before the
residual tolerance was met.}
\label{tab:sensitivity-curve}
\input{common/appendix_g05_curve}
\end{table}

\begin{table}[htbp]
\centering
\footnotesize
\setlength{\tabcolsep}{4pt}
\caption{Central-difference derivative of the released response at the three
pre-registered control points, for all three step sizes of the scan. Each row is
one comparison of complete reruns; the flags record a changed released contact
set, substep refinement of an increment in a rerun, and a step that crosses the
onset.}
\label{tab:sensitivity-derivative}
\input{common/appendix_g05_derivative}
\end{table}

\paragraph{Simple bodies and parameter sweeps.}
The four-body comparison and the four parameter sweeps are produced inside
\path{revision_supervisor_round2}, which leaves the frozen cube and character
evidence untouched and writes new file stems only.
\path{extra_scenes.py} defines the grip-tension and cantilever scenes on the
same solver, Newton tolerance, and Armijo line search as the controlled cube
benchmark, so the only difference between scenes is the loading.
\path{extra_benchmark.py} runs one scene and writes a \path{_summary.json}, a
per-step \path{_records.csv}, and a \path{_states.npz};
\path{extra_studies.py} runs one sweep and writes a \path{study_*.json}.
The four-body and ablation figures are rendered by \path{dep_final_figures.py}
from those retained state packages alone, and every number quoted in this paper
is a \LaTeX{} macro that \path{make_numbers.py} generates from the same files, so
no value is transcribed by hand. \path{build_final_delivery.py} runs the chain
and refuses to finalise a build while any macro still lacks a retained source.

The delivered code and logs regenerate the reported figures, tables and numbers
from the delivered records; runs marked there as reused were not repeated in this
revision.

\section{Solver Work and Increment Sensitivity}
\label{app:increments}
Table~\ref{tab:increments} reports the total solver work of the shared
indentation path against the number of load increments. The one- and
two-increment rows are the ones that expose the cost of crossing the transition
inside a single solve; the finer rows show the two counts converging as the
transition is resolved. The counts are totals over the whole load, hold, and
release path, so each row sums the work of every solve in that schedule.

\begin{table}[t]
\centering
\small
\caption{Total solver work against the number of load increments on the shared
indentation path, regenerated in the environment recorded in the release notes.
One load increment means a single Newton solve must cross the entire
elastic--plastic transition; $16$ and $32$ resolve it. Each entry is a sum over
every solved stage of that run, so a Newton count is the whole run's Newton work
and a line-search count is a sum of backtracking halvings rather than a number of
evaluations. Diagonal-damping fallbacks occurred at $1$, $4$ and $8$ increments and
nowhere else: $3$, $7$ and $9$ for return mapping, $4$, $6$ and $9$ for Perzyna, and
$0$, $8$ and $10$ for SECA, so the coarse row is not a comparison between two
identically globalised iterations.}
\label{tab:increments}
\setlength{\tabcolsep}{4pt}
\begin{tabular}{rrrrrrr}
\toprule
& \multicolumn{2}{c}{exact RM} & \multicolumn{2}{c}{Perzyna} & \multicolumn{2}{c}{SECA} \\
\cmidrule(lr){2-3}\cmidrule(lr){4-5}\cmidrule(lr){6-7}
increments & Newton & line search & Newton & line search & Newton & line search \\
\midrule
$1$  & $36$  & $117$ & $27$ & $37$ & $\mathbf{19}$ & $\mathbf{8}$ \\
$2$  & $20$  & $2$   & $23$ & $3$  & $22$ & $3$ \\
$4$  & $55$  & $52$  & $57$ & $47$ & $60$ & $44$ \\
$8$  & $97$  & $76$  & $97$ & $66$ & $99$ & $69$ \\
$16$ & $170$ & $212$ & $179$ & $226$ & $184$ & $283$ \\
$32$ & $244$ & $21$  & $256$ & $21$ & $258$ & $23$ \\
\bottomrule
\end{tabular}
\end{table}

The one-increment row is the schedule quoted in the main text, and its load stage
is worth separating from those totals: that stage, which must cross the whole
transition inside a single solve, needs \depNewtonCrossingRM\ Newton iterations
under return mapping and \depNewtonCrossingDEP\ under SECA, with
\depLineSearchCrossingRM\ and \depLineSearchCrossingDEP\ backtracking halvings
respectively, so on that stage alone SECA needs \depNewtonSavedCrossing\% fewer
Newton iterations and \depLineSearchSavedCrossing\% fewer halvings. The stage
carries all \depLineSearchCrossingRM\ return-mapping halvings but only
\depLineSearchCrossingDEP\ of the \depLineSearchDEP\ SECA ones, so the stage and
whole-run readings are different splits of the same run, and the halves are not
interchangeable.

\section{Parameter Study Detail}
\label{app:parameter-detail}
\begin{figure*}[!t]
\centering
\includegraphics[width=0.94\textwidth]{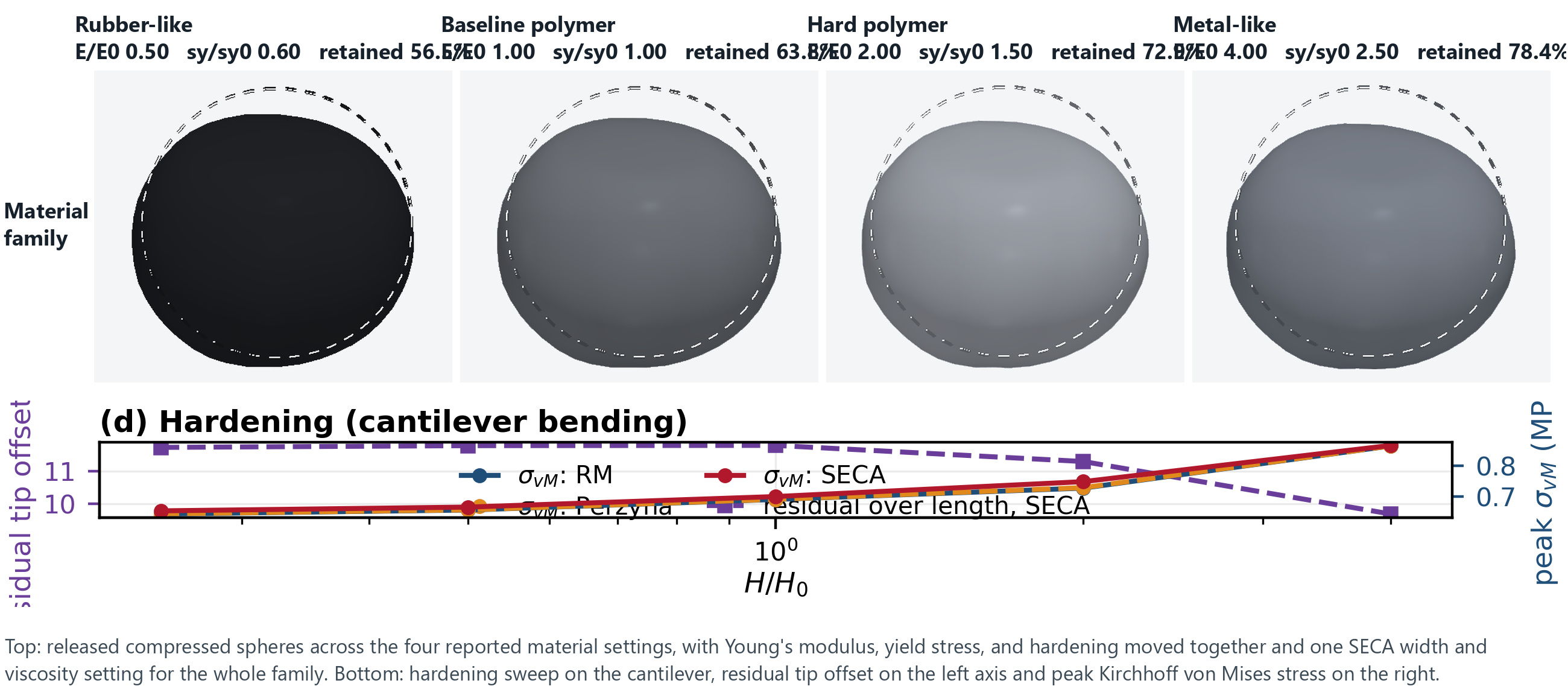}
\caption{Material family and hardening on fixed protocols. Top: released
compressed spheres across the family, with Young's modulus, yield stress, and
hardening moved together under one normalised SECA gate setting
($\delta/\sigma_y$ and $\eta/\delta$ fixed, so the absolute width and viscosity
scale with the material); four of the five presets reach a converged release and
are the ones reported. Bottom: hardening sweep on the cantilever, with the
residual tip offset on the left axis and the peak Kirchhoff von Mises stress of
all three constitutive methods on the right axis.}
\Description{Released compressed spheres in four material appearances above a
twin-axis hardening plot of residual tip offset and peak Kirchhoff stress.}
\label{fig:supplement-studies}
\end{figure*}

Figure~\ref{fig:supplement-studies} reports the two parameter sweeps whose
released shapes are not carried by the main parameter-control figure: the
material family on the compressed sphere, and the hardening sweep on the
cantilever. The family moves Young's modulus over
$E/E_0\in[\depFamilyYoungMin,\depFamilyYoungMax]$ and the yield stress over
$\sigma_y/\sigma_{y,0}\in[\depFamilyYieldMin,\depFamilyYieldMax]$ with one SECA
width and one viscosity setting, and its retained fraction rises from
\depFamilyTrendMin\% to \depFamilyTrendMax\%. The hardening sweep moves
$H/H_0\in[\depHardLo,\depHardHi]$, the residual tip offset from \depHardFirst\% to
\depHardLast\%, and the peak Kirchhoff von Mises stress by \depHardStressRel\% over
the same range. Both keep the mesh, boundary condition, load path, time step,
tolerance, and line search of the controlled cube benchmark; only the named
quantity changes.

\section{Discretisation Sensitivity}
\label{app:discretization-sensitivity}
All three fixed forward levels complete the same 30-interval
load--withdrawal--hold schedule on non-nested meshes. For the common surface
field $q_{\rm scene}$, coarse--medium and medium--fine discrepancies are
0.887\% and 0.530\% of $L_{\rm ref}$, but 63.25\% and 51.85\% relative to
the corresponding finer response. The common $q_{\rm obs}$ gauge changes from
14.071\% to 12.611\% and 10.159\% compression, giving differences of 1.460
and 2.452 percentage points. Geometry-normalised discrepancies decrease
from one mesh pair to the next, verifying that residual deformation and
recovery remain physically consistent across these spatial discretisations.

\begin{figure}[t]
\centering
\includegraphics[width=\linewidth]{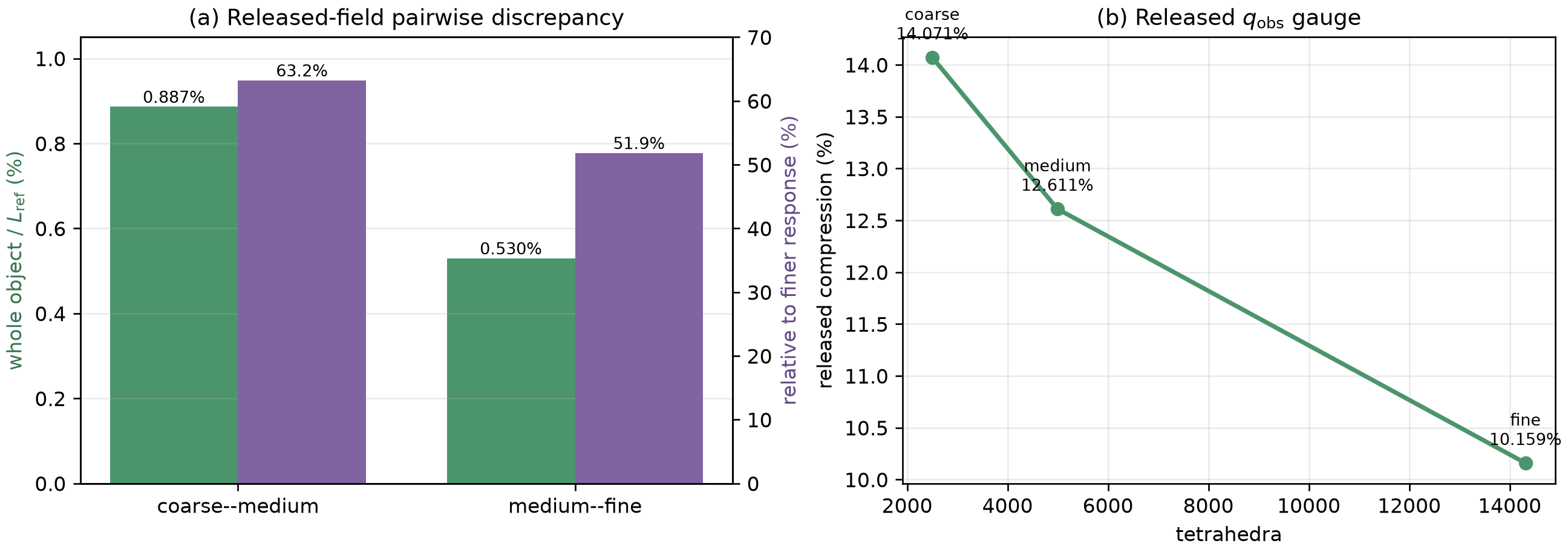}
\caption{Three-level forward discretisation sensitivity. (a) The same released
surface-field discrepancy is shown relative to $L_{\rm ref}$ and relative to
the finer response. (b) The fixed observation-region gauge at each
non-nested level. The finer level is a comparator, not continuum truth.}
\Description{Pairwise released-field discrepancies under two normalisations
and released gauge compression on coarse, medium, and fine discretisations.}
\label{fig:discretization-sensitivity}
\end{figure}

\paragraph{Performance.}
The production profile excludes validation summaries, hierarchy construction,
centred replay, cross-level comparison, and artefact compression. The retained
profile reports $1039.96$~s for the forward solve of the $14{,}316$-tetrahedron
Bunny fine simulation and $890.76$~s for its algorithmic-gradient path
($1376$ residual JVPs). The finite
Bunny run uses Python~3.12.10, NumPy~2.5.3,
SciPy~1.18.1, PyTorch~2.14 CPU, and IPC Toolkit~1.6.0 on Windows~11 with $24$
logical processors. The retained profile states the platform, the library
versions and the timing scope but not the processor model, so the reported
seconds belong to that profile rather than to a named CPU.

\begin{figure}[t]
\centering
\includegraphics[width=\linewidth]{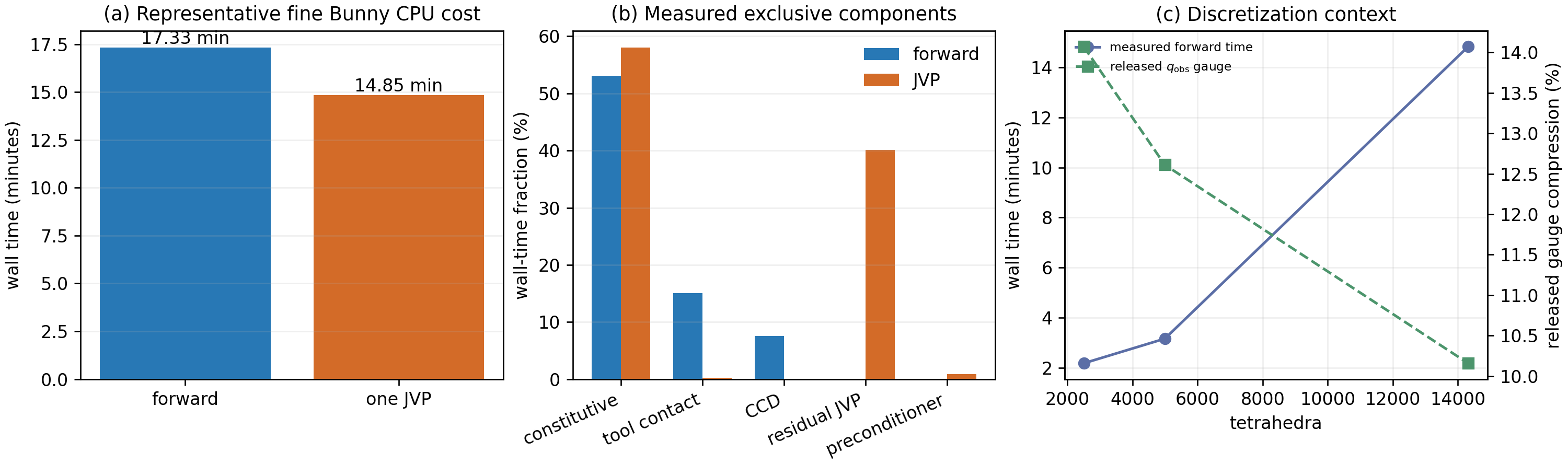}
\caption{Complete CPU profile and discretisation context. Production forward
and one algorithmic JVP exclude validation replay. Component fractions expose
the measured bottlenecks; the hierarchy panel reports recorded runtime and
the released gauge at each spatial discretisation.}
\Description{Production timing, measured component fractions, and mesh-level
runtime and released-gauge values for the Bunny CPU prototype.}
\label{fig:performance-profile}
\end{figure}

\paragraph{What this round did not re-run.}
The evidence kept here is retained record, and four of its boundaries are stated
rather than smoothed over. First, three legacy modules cannot be collected in the
current environment because the contact library they import is absent
(\path{importlib.util.find_spec("ipctk")} returns \path{None}); their earlier
artefacts are retained, no result in this paper is recomputed from them, and no
unit test of theirs is claimed to pass here. Second, the presets that the methods
check marks \textsc{analysis-only} are analytic readings of the update rather
than executed runs, and they are not claimed as numerical results. Third, the
production seconds are read from the retained profile, which is not re-measured
in this round and which records wall and CPU time per component but no processor
model. Fourth, the path-derivative audit covers the coarse and medium levels of
the Bunny hierarchy; the fine level carries no stored replay summary, so the
audit is quoted for the levels that were replayed and not for the hierarchy as a
whole.

\section{Simple Released-Shape Protocols}
\label{app:simple-scenes}
\begin{figure*}[t]
\centering
\includegraphics[width=0.8\textwidth]{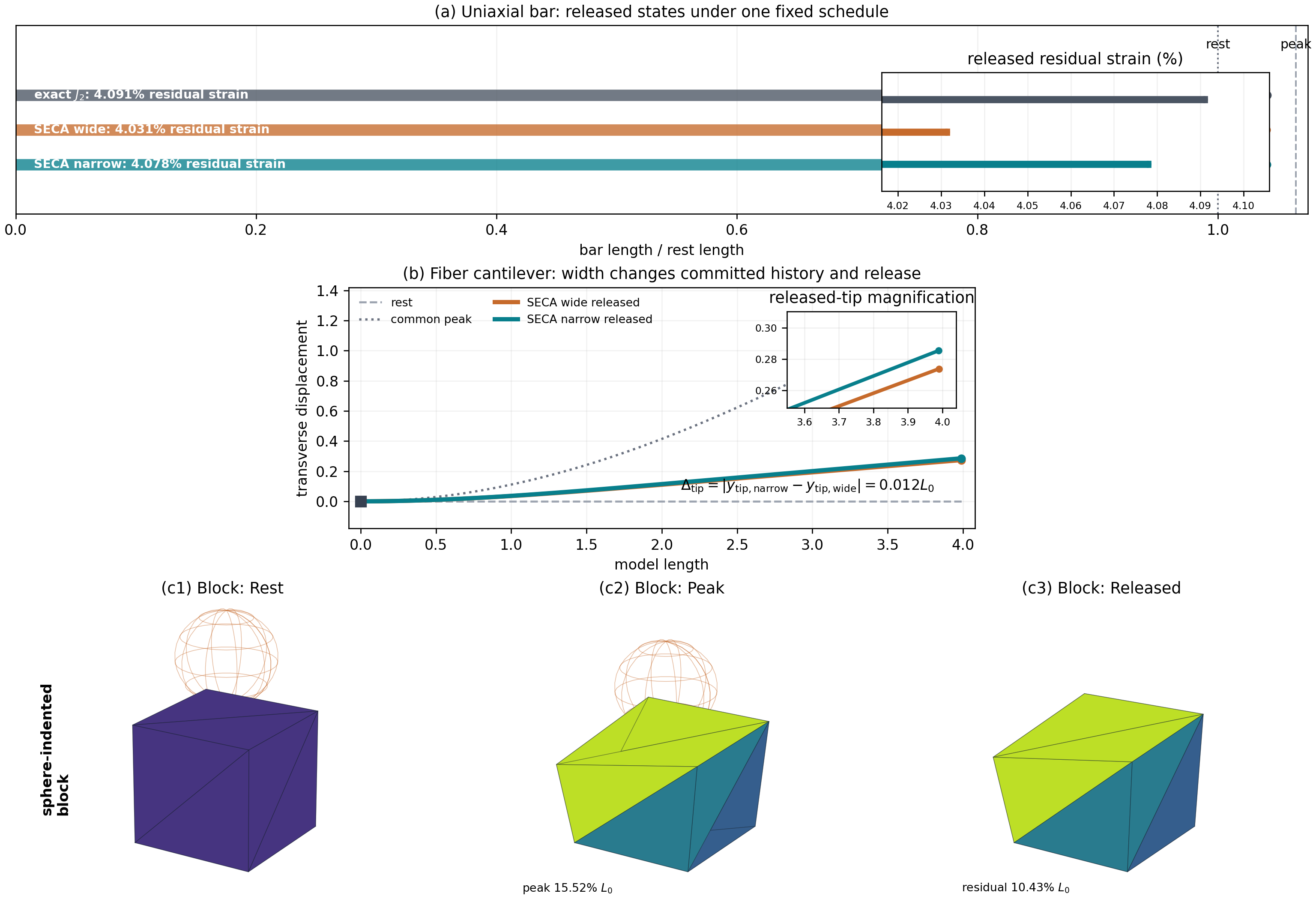}
\caption{Transition width changes committed history and therefore released
shape. (a) Exact-$J_2$, SECA-wide, and SECA-narrow bars share one peak strain;
the inset exposes their released residual strains. (b) Wide and narrow SECA
cantilevers share one imposed peak curvature but release to different
centrelines and tip positions. (c) The tetrahedral block records normalised
committed history through analytic-sphere indentation, withdrawal, and
release. All geometry is shown at unit displacement scale. Referenced from
Section~\ref{sec:results}.}
\Description{Magnified released bar tips, overlaid released cantilever
centrelines, and rest, peak, and released sphere-indented block states make
the transition-width to plastic-history to released-shape chain explicit.}
\label{fig:recovery-consequence}
\end{figure*}

The three explanatory scenes in Figure~\ref{fig:recovery-consequence} are
generated by
\path{experiments/dep/care_simple_scenes.py}. The script writes one
\path{metrics.json} record and one \path{states.npz} array package. It does
not invoke the character solver or reuse a rendered deformation.

\paragraph{Uniaxial bar.}
The material parameters are $E=100$, $\nu=0.3$, $\sigma_y=2$, and $H=10$.
The imposed strain rises monotonically from zero to 0.065 in 80 steps over
one time unit. Both SECA cases use the same physical
$\eta=0.01$, $\Delta t=0.0125$, and duration; only
$\delta\in\{0.4,0.05\}$ changes. The implementation API accepts
$\eta/\delta$, so the two numerical arguments are 0.025 and 0.2 while their
product with width remains the same physical viscosity. The exact-$J_2$
reference uses the repository's rate-independent scalar return map.
After loading, the free zero-traction equilibrium is
$\epsilon_{\rm rel}=p$, which gives exactly zero stress without an additional
plastic increment.

\paragraph{Fibre cantilever.}
The disclosed Euler--Bernoulli model has length 4, height 0.4, breadth 0.25,
65 sections, and 41 trapezoidally weighted fibres. Its prescribed loading is
\[
\kappa(x,t)=s(t)\,0.24(1-x/4),
\]
where $s$ rises from zero to one in the same 80-step SECA protocol. Each
fibre commits its own scalar plastic state after every step. With history
fixed, the released curvature at each section is the unique linear solution
of zero section moment,
\[
\kappa_{\rm rel}(x)
=\frac{\sum_j A_jp_j(x)y_j}{\sum_jA_jy_j^2}.
\]
An independent SECA update at this curvature confirms nonpositive released
overstress and zero further plastic increment for every fibre. This
reduced-order scene visualises the release mechanism; it is not used as
finite-deformation character validation.

\paragraph{Sphere-indented block.}
The block contains eight nodes and six tetrahedra with its four bottom nodes
fixed. It uses $E=100$, $\nu=0.3$, $\sigma_y=3$, $H=12$,
$\delta=0.075$, and $\eta=0.015$. An analytic sphere of radius 0.38 moves
from centre height 1.50 to 1.25 in ten equal stages and then returns in ten
stages. Point--triangle distance minus radius defines the gap; the activation
distance is 0.075. Each equilibrium trial starts from the previous committed
state, and only a converged feasible stage commits history. The released
barrier is inactive, its reaction is zero, and the recorded residual norm is
$2.35\times10^{-14}$.

\paragraph{Measured outcomes.}
Exact $J_2$, wide ($\delta=0.4$), and narrow ($\delta=0.05$) activation retain
respectively 4.091\%, 4.031\%, and 4.078\% residual strain in the bar after a
6.5\% peak strain, so the narrow gate approaches the hard reference more closely
without changing the viscosity or the schedule. The reduced-order cantilever
releases to residual tip displacements $0.274L_0$ and $0.286L_0$ for the wide and
narrow gates, with moment residuals below $2\times10^{-18}$ and no fibre resuming
plastic flow during the release check. The tetrahedral block converges all 21
load--withdrawal stages with positive gap and $\min\det\boldsymbol F=0.9038$, and
a peak indentation of $0.155L_0$ releases to a residual $0.104L_0$.

\begin{table}[t]
\centering
\small
\caption{Primary evidence records; the study tables above name the protocol
behind each quoted number, and the release package carries the exhaustive
file-level coverage list in its experiment and figure manifests. Hashes and
complete per-stage arrays are stored beside each record.}
\label{tab:evidence-ledger}
\begin{tabular}{p{0.28\linewidth}p{0.64\linewidth}}
\toprule
Purpose & Retained record\\
\midrule
Theory & \texttt{theory/regularity\_bias\_derivation.md}\\
Literature boundary & \texttt{theory/novelty\_matrix.md}\\
Simple released scenes & \texttt{experiments/output/care\_simple\_scenes/}\\
Character forward paths & \texttt{frozen/metrics.json}; \texttt{frozen/F14\_b02\_highres\_manifest.json}\\
Three-level forward & \texttt{self\_contact\_results/hierarchy/fixed/summary.json}\\
Two-level JVP/replay & \texttt{self\_contact\_results/hierarchy/derivatives/}\\
Metric definitions & \texttt{b02\_spatial/metric\_dictionary.json}\\
Production timing & \texttt{performance/profile.json}\\
\bottomrule
\end{tabular}
\end{table}

The arithmetic-only figure commands use \texttt{python -m} with modules
\path{D_EP.figures.build_paper_analysis_figures} and
\path{D_EP.figures.build_b02_sensitivity_figure}.
They read retained JSON and images and do not execute a physical solve.
The two manuscripts use \texttt{python -m} with
\path{D_EP.paper.build}.

\begin{figure*}[t]
\centering
\includegraphics[width=0.9\textwidth]{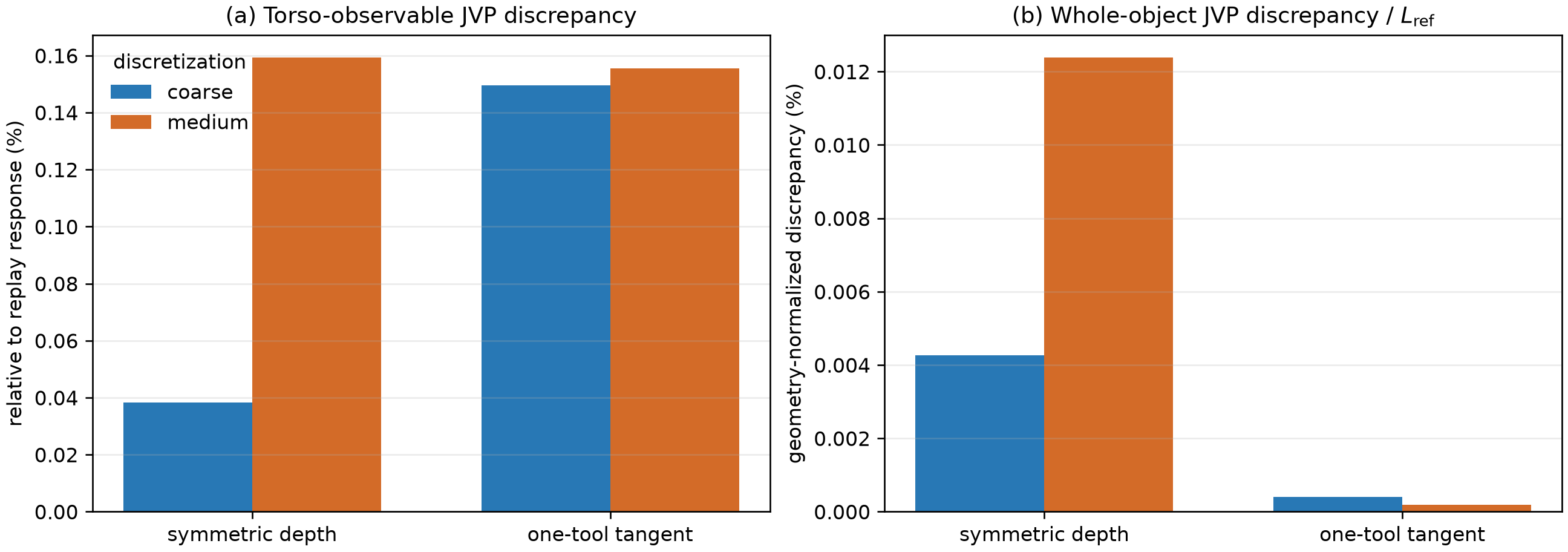}
\caption{Released-equilibrium sensitivity validation. Coarse and medium
actual-residual JVPs are checked against independent complete centred replay
for symmetric-depth and one-tool tangential controls. The two panels report
response-relative and geometry-normalised discrepancies, confirming that
residual deformation and recovery remain physically consistent across these
spatial discretisations. Referenced from Section~\ref{sec:results}.}
\Description{Two panels compare actual-residual algorithmic products with
independent complete replay under response-relative and geometry-normalised
metrics.}
\label{fig:contact-sensitivity-audit}
\end{figure*}

\section{Material-Family Convergence Record}
\label{app:family-record}
Each preset moves the modulus, the yield stress and the hardening modulus together,
and the gate setting is normalised ($\delta/\sigma_y = 0.05$, $\eta/\delta = 0.2$~s),
so the absolute width and viscosity scale with $\sigma_y$. The table reports every
preset that was run, including the one whose release stage did not converge; the
family result in the main text is quoted over the converged subset only.
\begin{table}[!ht]
\centering
\small
\caption{Material-family presets as run. ``Release'' states whether the final
tool-free stage of that preset reached the residual tolerance, and ``failed'' counts
the stages that did not converge anywhere in the schedule; the retained fraction
and the peak stress of the converged presets are quoted in the main text and drawn
in the study-detail figure of the supplement.}
\label{tab:family-record}
\setlength{\tabcolsep}{3pt}
\begin{tabular}{lrrrrrrrc}
\toprule
preset & $E$ (Pa) & $\nu$ & $\sigma_y$ (Pa) & $H$ (Pa) & $\delta$ (Pa) & $\eta$ (Pa\,s) & rel. & failed \\
\midrule
Foam-like & $7.5\times10^{6}$ & 0.30 & $0.88\times10^{6}$ & $0.72\times10^{6}$ & $0.438\times10^{5}$ & $0.88\times10^{4}$ & no & 19 \\
Rubber-like & $15.0\times10^{6}$ & 0.30 & $1.50\times10^{6}$ & $1.44\times10^{6}$ & $0.750\times10^{5}$ & $1.50\times10^{4}$ & yes & 1 \\
Baseline polymer & $30.0\times10^{6}$ & 0.30 & $2.50\times10^{6}$ & $2.40\times10^{6}$ & $1.250\times10^{5}$ & $2.50\times10^{4}$ & yes & 3 \\
Hard polymer & $60.0\times10^{6}$ & 0.30 & $3.75\times10^{6}$ & $3.60\times10^{6}$ & $1.875\times10^{5}$ & $3.75\times10^{4}$ & yes & 2 \\
Metal-like & $120.0\times10^{6}$ & 0.30 & $6.25\times10^{6}$ & $4.80\times10^{6}$ & $3.125\times10^{5}$ & $6.25\times10^{4}$ & yes & 2 \\
\bottomrule
\end{tabular}
\end{table}

\section{Noncompact-Smoothing Diagnostic}
\label{app:noncompact-diagnostic}
The second difficulty in the introduction is that a noncompact smooth activation can
produce plastic flow below the intended threshold. The tables measure that claim for one
explicitly named family instead of asserting it: a softplus overstress law
$r(f) = s\log(1+e^{f/s})$ with sharpness $s = 5\times10^{4}$~Pa, at the same material
point, the same viscosity ratio $\eta/\delta = 0.2$~s and the same time step
$\Delta t = 1.282\times10^{-2}$~s as the SECA material-point protocol. Within each
experiment the two laws share the step, the modulus and the tolerances; only the rate
function differs, so this is a rate-function ablation and not a reproduction of an
external implementation. The softplus path never reaches the compact branch that returns
exactly zero: the residual was evaluated $31\,943$ times at negative overstress in the
recorded run, and the compact primitive and the library kernel are not called from it.

The first table separates two quantities that are easy to conflate. Evaluating the
softplus law once at the trial overstress, $\gamma = \alpha r(f_{\mathrm{tr}})$ with
$\alpha = \Delta t/\eta$, reproduces the sub-yield column reported earlier. The root of
the implicit step, $\gamma = \alpha r(f_{\mathrm{tr}} - B\gamma)$ with $B = 3\mu + H$,
does not: at $f_{\mathrm{tr}} = +0.125\times10^{6}$~Pa the explicit value is
$6.613\times10^{-2}$ and the root is $5.350\times10^{-3}$, a factor of $12.4$, and at
$f_{\mathrm{tr}} = 0$ the two differ by a factor of $6.2$. Both columns are printed so
the distinction is checkable; each root is bracketed to a relative width of
$9.8\times10^{-16}$ with a maximum relative residual of $8.1\times10^{-16}$. The compact
column is the printed primitive solved implicitly, exactly zero for every
$f_{\mathrm{tr}} \le 0$, and the library scalar kernel, which evaluates the same primitive
explicitly at the trial overstress over modulus $E + H$, also returns exactly zero there.
\begin{table}[!ht]
\centering
\small
\caption{Sub-yield behaviour of the named softplus law against the compact SECA gate at
the same step and modulus. The explicit column is one evaluation of the rate function at
the trial overstress, the implicit column is the root of the same backward-Euler step, and
the compact column is the printed primitive solved the same way. For
$f_{\mathrm{tr}} \le 0$ the compact law returns exactly zero, so states that should be
recoverable are recoverable, while the softplus law returns a positive multiplier at every
overstress whose exponential is representable. One named family at one sharpness is
measured here.}
\label{tab:noncompact}
\input{common/appendix_g06_softplus}
\end{table}

The second table carries the consequence through a hold and a release, which is where an
unpreserved state shows up. Holding a constant strain whose initial trial overstress is
$10^{6}$, $5\times10^{5}$ and $10^{5}$~Pa below the threshold for $200$ increments, the
compact law keeps the plastic strain and the released stress at exactly zero, while the
softplus law accumulates $1.057\times10^{-8}$, $2.168\times10^{-4}$ and
$9.407\times10^{-3}$ respectively. The per-increment multiplier falls during the hold,
from $1.381\times10^{-3}$ to $7.814\times10^{-6}$ at the shallowest level, but it never
reaches zero, so the accumulation is what survives the release: at that level the released
plastic strain is $11.8\%$ of the applied strain and the released stress is
$-2.822\times10^{5}$~Pa rather than zero. The leak stops being representable only when the
double-precision exponential underflows, at
$f_{\mathrm{tr}} \approx -3.71\times10^{7}$~Pa or $742$ sharpnesses below the threshold,
which is a property of floating-point arithmetic and not of the model. This is one named
family at one sharpness, and the size of the sub-yield accumulation depends on the hold
protocol; the measured claim is that the compact gate returns exactly zero where a
noncompact smooth law returns a positive multiplier that accumulates.
\begin{table}[!ht]
\centering
\small
\caption{The same two laws through a $200$-increment hold at constant strain below the
threshold, followed by a release to zero strain. $\gamma$ first and last are the first and
last multipliers of the hold, $\alpha$ is the accumulated multiplier, and the released
plastic strain is measured after the release increment. The compact law preserves the
state exactly; the softplus law accumulates plastic strain that survives the release.}
\label{tab:softplus-hold}
\input{common/appendix_g06_hold}
\end{table}

\section{Grip-Tension Solver Work}
\label{app:tension-work}
The indentation table above is the controlled comparison. The same frozen matrix was
also run on the finite-area grip-tension scene of the main paper with
the identical solver settings, and the outcome there is the opposite at the coarse end:
with one load increment SECA needs \emph{more} Newton iterations than return mapping.
The table reports both scenes so the admission in the main text is checkable.
The Armadillo grip schedule holds the grip stiffness at its full value through the
nine loading stages that bring the control to its peak, then keeps the control at
that peak while ramping the stiffness linearly to exactly zero over the eight
removal stages; the final stage is therefore a grip-free equilibrium on the
unchanged supports, and no stage weakens or removes a support.
\begin{table}[!ht]
\centering
\small
\caption{Whole-run Newton iterations and backtracking halvings on the grip-tension scene
(and, for comparison, the indentation scene), regenerated in the recorded environment.
The indentation rows repeat the controlled table above so that the two scenes can be
read side by side.}
\label{tab:tension-work}
\setlength{\tabcolsep}{4pt}
\begin{tabular}{rlrrrrrr}
\toprule
& & \multicolumn{2}{c}{RM} & \multicolumn{2}{c}{Perzyna} & \multicolumn{2}{c}{SECA} \\
\cmidrule(lr){3-4}\cmidrule(lr){5-6}\cmidrule(lr){7-8}
scene & increments & Newton & halvings & Newton & halvings & Newton & halvings \\
\midrule
grip tension & 1 & 27 & 31 & 63 & 229 & 40 & 28 \\
grip tension & 2 & 24 & 2 & 27 & 4 & 28 & 5 \\
grip tension & 4 & 56 & 58 & 39 & 4 & 51 & 46 \\
grip tension & 8 & 64 & 8 & 64 & 4 & 66 & 5 \\
grip tension & 16 & 129 & 58 & 118 & 13 & 221 & 753 \\
grip tension & 32 & 214 & 3 & 246 & 179 & 213 & 18 \\
indentation & 1 & 36 & 117 & 27 & 37 & 19 & 8 \\
indentation & 2 & 20 & 2 & 23 & 3 & 22 & 3 \\
indentation & 4 & 55 & 52 & 57 & 47 & 60 & 44 \\
indentation & 8 & 97 & 76 & 97 & 66 & 99 & 69 \\
indentation & 16 & 170 & 212 & 179 & 226 & 184 & 283 \\
indentation & 32 & 244 & 21 & 256 & 21 & 258 & 23 \\
\bottomrule
\end{tabular}
\end{table}

\section{Interior Cutaway Record}
\label{app:interior-cutaways}

The interior figure of the main paper carries the representative stress row only.
This section records the complete matrix behind it: the same two cut faces of the
indentation cube and of the Bunny character, at peak load and after complete tool
removal, for the Kirchhoff von Mises stress and for the committed equivalent
plastic strain $a$. Each body is cut on the surface the main paper's surface figure
draws for it: the cube on its computed mesh, the Bunny on the display triangulation
of the scanned asset.

\paragraph{Cut-plane convention.}
Both planes are fixed in the rest configuration at the body centre, expressed in
the frame that body is drawn in, and are then applied to the deformed state. Plane 1
has normal $n=[1,0,0]$ and keeps
$n\cdot x\le x_c$; plane 2 has normal $n=[0,-1,0]$ and keeps $n\cdot x\le -y_c$.
The kept body is therefore three quarters of the solid, and the exposed faces are
the planar sections the two planes cut open: on the cube they are element faces of
the computed mesh, on the Bunny they are the section of the display triangulation,
and in neither case are the element interfaces inside the kept pieces drawn, which
are counted separately in the rendering record. The two
characters are drawn in the frame and from the viewpoint of the surface figure of
the main paper, so a character keeps one orientation across both figures; rotating
a state into that frame is a rigid rotation of the whole body and leaves every
element-constant value unchanged.

\paragraph{What the fields are.}
Cutting is post-processing only: the bodies were solved uncut, no cut model was
re-equilibrated, and no field was recomputed on the cut faces. The colours on the
cut faces are a display field: the element-constant values are averaged at the mesh
nodes with the reference element volumes as weights, and that nodal field is then
read on the drawn surface: on the cube each drawn facet is evaluated inside the linear
tetrahedron that owns it, and on the Bunny each drawn point of the planar section is
located in the committed tetrahedral mesh of the deformed body and reads the four
corner values of the tetrahedron that contains it, interpolated linearly there, so the
section is coloured by the computed field at its own positions. This is the same recovery
the surface figure of the main paper uses before transferring its field to the
display triangulation, so the two figures colour their surfaces by one rule. Every
number reported in the main text is read from the raw element-constant field, and the
second matrix below repeats these faces in that raw form, one constant per drawn
facet. The outer shell is drawn in neutral grey, and only the faces the planes expose
are drawn: the element interfaces inside the kept body are not part of the cut
surface, and on the Bunny, whose cut is taken on the display triangulation, the
silhouette therefore matches the surface figure of the main paper. Each model keeps
one colour
range across its peak and released panels, separately for the two quantities, so
the change between peak and release is read without rescaling. A mostly blue
released face is an observation about that state, not a failure of the update.

\begin{figure}[t]
  \centering
  \includegraphics[width=\textwidth]{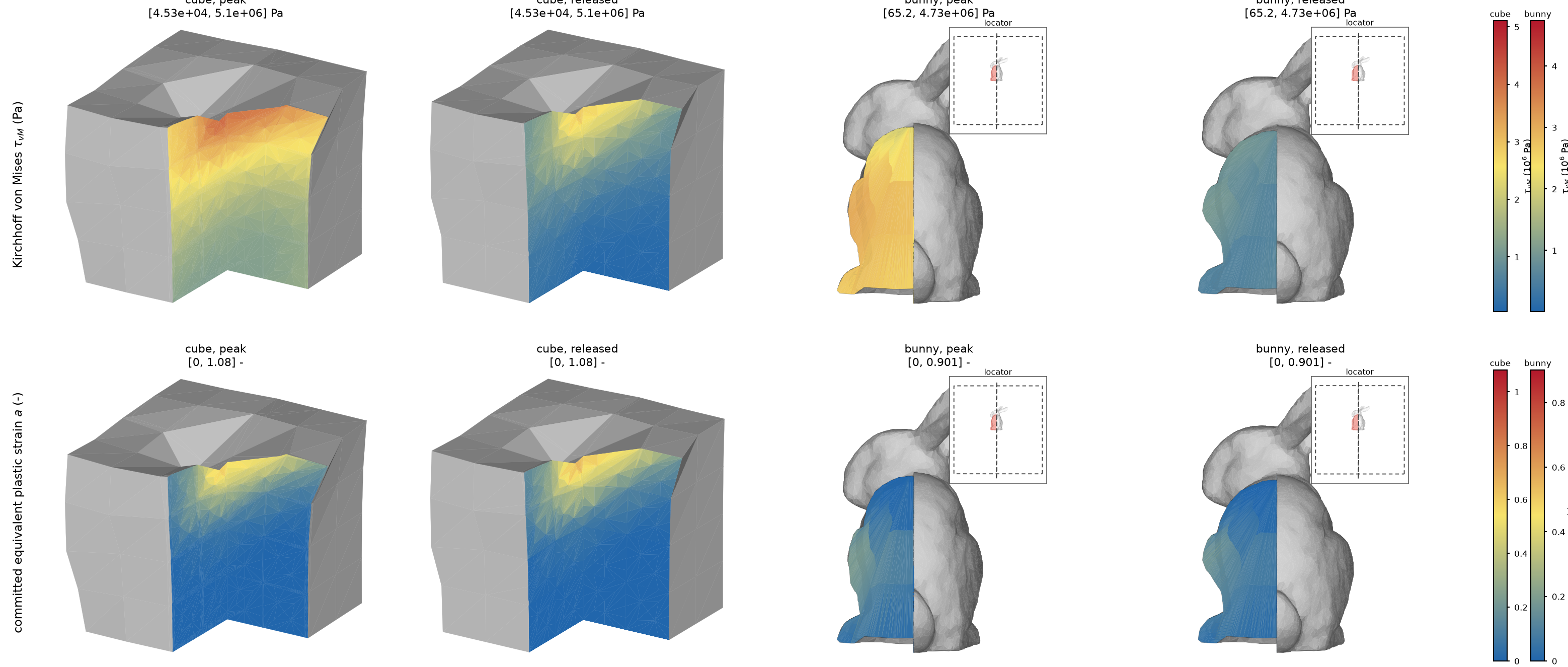}
  \caption{The complete interior-cutaway matrix. The upper row is the Kirchhoff
  von Mises stress on the exposed cut faces of the indentation cube and the Bunny
  character, at peak load and after complete tool removal; the lower row is the
  committed equivalent plastic strain $a$ on the same faces and states. The cut
  planes are fixed in the rest configuration of the frame each body is drawn in, at
  the body centre, as stated above, and the faces are coloured by the recovered
  nodal field described above, as in the main figure. The cube is cut on its
  computed mesh, printed here as pure mapped colour without the element outlines, which
  are carried by the raw element-constant record below; the Bunny is cut on the
  display triangulation of the scanned asset, so its cut faces carry no element
  outlines either and its silhouette is the one of the main paper's surface figure.}
  \Description{Two rows of cutaway panels: Kirchhoff von Mises stress and
  committed equivalent plastic strain on the exposed cut faces of the cube and the
  Bunny, at peak load and after complete tool removal.}
  \label{fig:interior-cutaways-full}
\end{figure}

\begin{figure}[t]
  \centering
  \includegraphics[width=\textwidth]{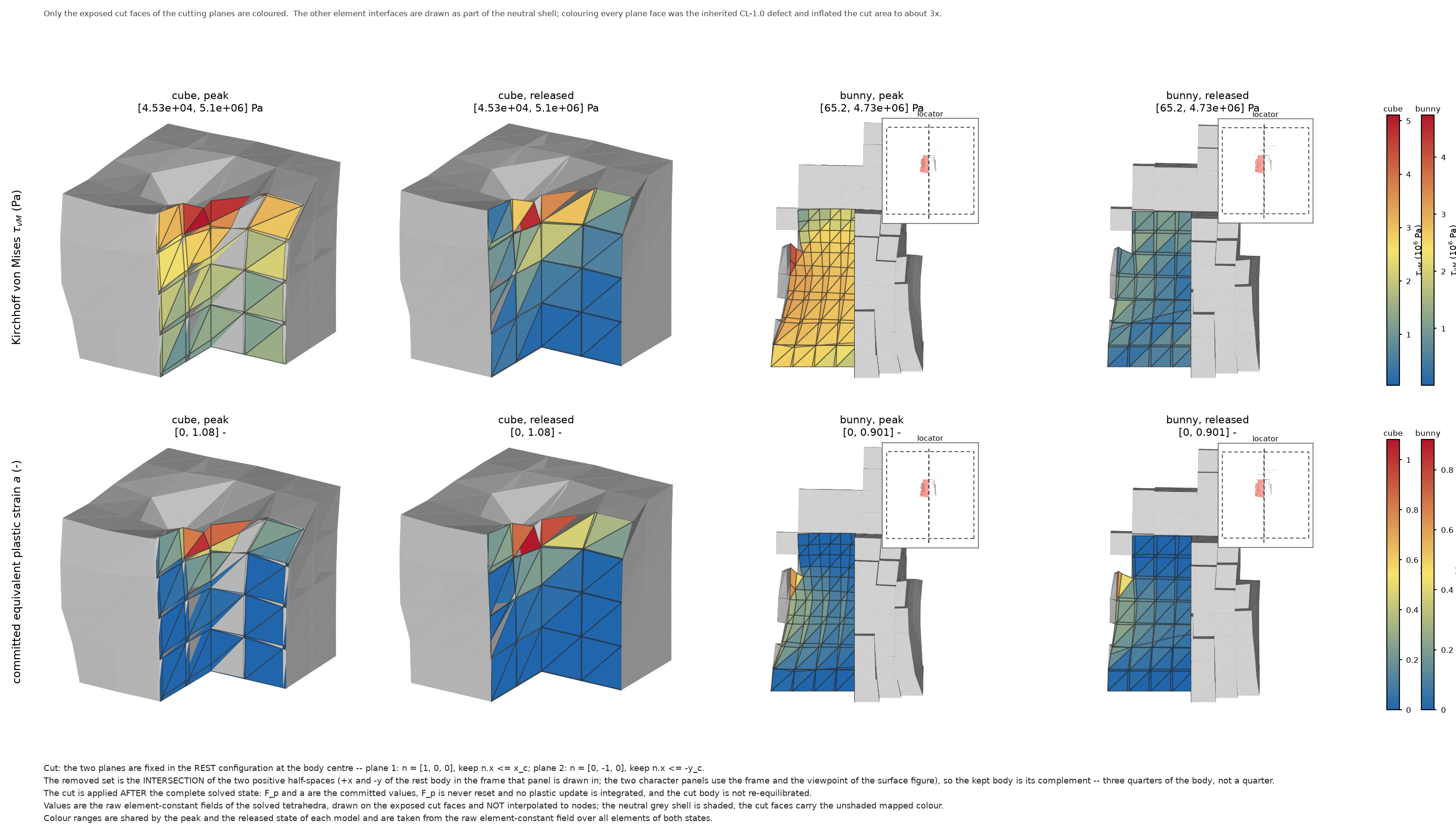}
  \caption{The same cutaway matrix in raw element-constant form, on the computed mesh
  of each body: \depCubeNodes\ nodes and \depCubeTetrahedra\ tetrahedra for the cube
  and 699 nodes and 2{,}502 tetrahedra for the Bunny. Each drawn facet
  carries the constant value of the linear tetrahedron that owns it, without any
  nodal recovery, and this is the field every number reported in the main text is
  read from. On the cube these are the same element faces the matrix above and the
  main figure draw, coloured there by the recovered field so that the stress varies
  continuously across a cut face; on the Bunny this matrix is the computed-mesh form
  of the same cutaway, while the matrix above and the main figure take its cut on the
  display triangulation instead. The two
  drawings show one solved state, and the difference between them is the display
  rule and, for the Bunny, the surface the cut is taken on. The cut planes are those
  of the matrix above.}
  \Description{Two rows of cutaway panels, as in the previous figure, with the raw
  element-constant value on each facet.}
  \label{fig:interior-cutaways-element}
\end{figure}

\section{Independent Proxy Calibration Record}
\label{app:proxy-calibration}

\paragraph{What this record is, and what it is not.}
This section documents, as an independent exploration, the small inverse problem in
which one tool depth is recovered from a target released dent: a differentiable
proxy fits the depth in a first stage and the exact return-mapping update refines it
in a second stage. It is not a validation of the SECA constitutive update, it is
not a test of the actual-residual path derivative audited in the main text, and it
is not evidence of a relative advantage of the printed method. The target is
generated with the exact update, no other law was run to the same target from the
same starting point with the same budget, and no optimisation advantage is claimed
for the printed method. The two derivatives are different objects and neither is
evidence for the other: the proxy gradient is the reverse-mode derivative of that
proxy's own unrolled released-shape solve, while the audited path derivative is the
algorithmic derivative of the finite-strain residual on one fixed accepted schedule.

\paragraph{Scene, objective and optimiser.}
The scene is the smaller single-tool contact body: a $6\times6\times1$ tetrahedral
plate whose indented face carries $49$ nodes, one tool of radius $0.23$ held at the
centre of that face, and a load--withdraw--release schedule over the load fractions
$0.55$, $1.0$ and $0$ with the tool active. The target released dent
$q_{\mathrm{target}}$ is generated with the exact return-mapping update at a tool
depth of $2.10$, and its released shape has scaled norm $5.217$. Only the depth is
optimised, with the footprint held at the centre of the face, inside the bounds
$[0.30,2.50]$ and from a starting value of $1.700$ that reproduces the target to a
relative error of $0.864$. The objective is
$e(\theta)=\lVert q(\theta)-q_{\mathrm{target}}\rVert /
\max\!\bigl(\lVert q_{\mathrm{target}}\rVert,10^{-12}\bigr)$ on the released shape
of the indented face, with the two regularisation weights of the optimiser set to
zero.

\paragraph{Stage A: depth fit on a differentiable proxy.}
Stage~A fits the depth on a differentiable proxy, and its gradient is the
reverse-mode derivative of that proxy's own unrolled released-shape solve, not a
difference quotient and not the algorithmic path derivative audited in the main
text. Adam with a learning rate of $0.05$ runs for at most $80$ iterations and stops
once the relative error reaches $0.02$ at yield scale one; the configuration also
carries continuation scales, but they are consumed only by the continuation route,
so this run solved a single stage. The tolerance was met at iteration $50$, leaving
$0.0198$, with $50$ backward and $52$ forward evaluations.

\paragraph{Stage B: refinement on the exact update.}
Stage~B refines that depth on the exact return mapping, which is a different model:
Adam at a learning rate of $0.025$ continues from the proxy depth and, if the exact
error still exceeds the tolerance, a bounded L-BFGS-B search with at most $80$
iterations and $400$ evaluations. Two further iterations moved the depth from
$2.0769$ to $2.1019$ and left a relative error of $4.8\times10^{-3}$ measured with
the exact update, so the bounded local search reported zero evaluations. Stage~B
therefore changes the optimised parameter and supplies the final depth and the final
error; neither belongs to stage~A alone.

\paragraph{What the proxy route is.}
The fitted route of stage~A is the compact-history candidate of the retained
differentiable stack, not the printed method, and the two are not one
implementation. Its plastic ratio is a softplus smoothing of the strain ratio in
yield units, positive for every positive strain ratio ($7.1\times10^{-5}$ at half
the nominal yield ratio and $1.6\times10^{-3}$ at nine tenths of it), and its
history is a running maximum envelope of that ratio, whereas the printed update is
the implicit scalar root of the compact primitive and its activation, which are
exactly zero below yield. Its scene is a small-strain linear tetrahedral body of one
assembled constant stiffness, solved through a differentiable Cholesky
factorisation, not the finite-strain Hencky residual of the main text. Its tool is a
softplus penalty evaluated a fixed number of times per stage, not the feasible-set
logarithmic barrier, and the released shape that both stages measure comes from a
linear solve driven only by the committed plastic state, so the objective is
contact-free even though every loading stage keeps the tool active. Its gradient is
a reverse-mode derivative through that unrolled implementation, while the path
derivative audited in the main text is the algorithmic derivative of the
finite-strain residual on one fixed accepted schedule. The released calibration
record lists the file and line of each of these statements.

\paragraph{Timing provenance.}
The two stages were timed separately on one thread by the recorded re-run of this
calibration: $6.04$\,s for stage~A, $0.26$\,s for stage~B and $6.63$\,s for the
run, the remainder being the target solve and the starting solve, of which
$5.18$\,s and $0.18$\,s respectively were spent inside the two optimiser calls;
along the optimised path the smallest recorded determinant of $\boldsymbol F_p$ is
$0.9438$ and the equilibrium residuals stay below $5\times10^{-16}$. Those three
numbers describe that one run. The whole run recorded with the sealed generator
took $6.3$\,s and covers the same two stages without the target and starting
solves, so the two timing sets are reported separately from their own records and
are not spliced together.

\begin{figure*}[t]
\centering
\includegraphics[width=\textwidth]{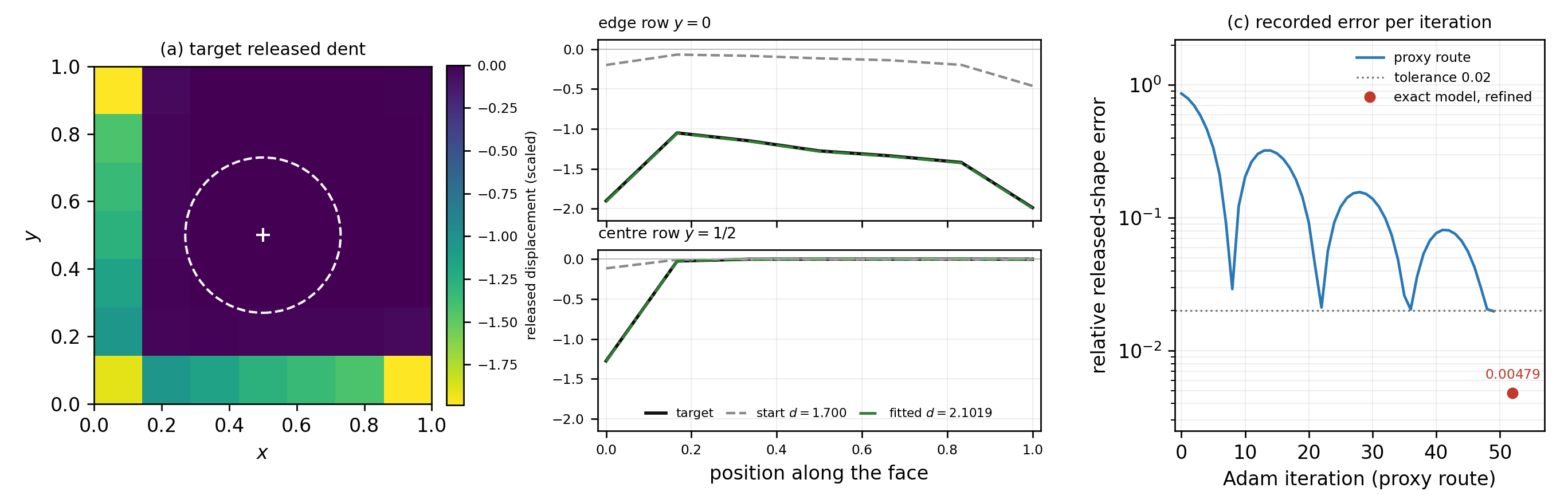}
\caption{The recorded depth-only calibration. (a)~The target released dent on the
indented face of the $6\times6\times1$ tetrahedral plate, normalised by the
output scale of the scene, with the tool footprint of radius $0.23$ at the centre
of the face. (b)~The same target against the starting depth $1.700$ evaluated on
the proxy route and the fitted depth $2.1019$ evaluated on the exact update, on
the edge row and on the centre row of the $7\times7$ node face. (c)~The relative
released-shape error recorded at every Adam iteration of the proxy stage, the
$2\%$ stopping tolerance, and the error of the refined depth measured with the
exact update, which is a different model and is therefore drawn as a separate
point.}
\Description{Three panels: the target released dent on the indented face with the
tool footprint, profiles of the target against the starting and the fitted depth
on two rows of that face, and the recorded error per optimisation iteration with
the exact-model error of the refined depth.}
\label{fig:calibration-record}
\end{figure*}

\begin{table}[t]
\centering
\small
\caption{Every tenth iteration of the recorded proxy-stage log of the
depth-only calibration, and the final recorded iteration of that stage; the depth
column is the single optimised control, which the exact-update stage refines from
that final value. The complete per-iteration log, together with the refinement
log of the second stage, is part of the delivered record for this run.}
\label{tab:calibration-log}
\begin{tabular}{rrrrr}
\toprule
iteration & depth & relative error & $\lVert\nabla e\rVert$ & $\min\det\boldsymbol F_p$ \\
\midrule
$0$  & $1.7000000000$ & $8.639600\times10^{-1}$ & $9.5675\times10^{-1}$ & $0.958441$ \\
$5$  & $1.9498149063$ & $3.379175\times10^{-1}$ & $8.6808\times10^{-1}$ & $0.948428$ \\
$10$ & $2.1558335841$ & $2.031049\times10^{-1}$ & $5.3561\times10^{-1}$ & $0.941050$ \\
$20$ & $2.1135760489$ & $9.243211\times10^{-2}$ & $2.3886\times10^{-1}$ & $0.942540$ \\
$30$ & $2.0265390820$ & $1.395174\times10^{-1}$ & $3.6134\times10^{-1}$ & $0.945637$ \\
$40$ & $2.1074731791$ & $7.673617\times10^{-2}$ & $1.9624\times10^{-1}$ & $0.942756$ \\
$49$ & $2.0769171007$ & $1.983029\times10^{-2}$ & $1.6088\times10^{-2}$ & $0.943838$ \\
\bottomrule
\end{tabular}
\end{table}

%% file: common/regularity_table.tex
\begin{table*}[t]
\centering
\small
\caption{Regularity claims are attached to explicit mathematical objects.
The discrete selector and adaptive partition are intentionally outside the
fixed-protocol derivative claim.}
\label{tab:regularity-objects}
\setlength{\tabcolsep}{4pt}
\begin{tabular}{p{0.16\textwidth}p{0.17\textwidth}p{0.25\textwidth}p{0.33\textwidth}}
\toprule
Object & Variable & Held fixed & Established scope\\
\midrule
$D_\delta(f)$ & overstress $f$ & $\delta>0$ &
Bounded and globally $C^2$; exactly zero for $f\leq0$.\\
$r_\delta(f)$ & overstress $f$ & $\delta>0$ &
$C^3$ primitive of $D_\delta$ and zero on the elastic branch.\\
$\gamma(f_{\rm tr})$ & trial overstress & previous state,
$\Delta t,\delta,\eta>0$ &
The implicit root is locally $C^2$ because
$G_\gamma=1+(\Delta t/\eta)K_\gamma D_\delta\geq1$.\\
$\boldsymbol P(\boldsymbol F)$ & deformation gradient & previous committed
state and one local trial branch &
Differentiable where the polar/logarithmic strain and trial direction are
regular; the fixed-state energy derivative and discrete-update tangent are
distinct.\\
$\boldsymbol R(\boldsymbol x,\boldsymbol\theta)$ & nodal coordinates and tool
controls &
previous state, contact feature branch, positive gaps &
Piecewise differentiable material--contact residual; no claim crosses a
closest-feature change or a singular structural tangent.\\
$\boldsymbol x_{\rm rel}(\boldsymbol\theta)$ & prescribed tool controls & material,
parameters, physical partition, converged solution branch &
Implicitly differentiable if every stage tangent is nonsingular and the
stated contact branch remains feasible.\\
Selector / adaptive execution & candidate index or partition & --- &
Discrete changes do not inherit the regularity of the six fixed objects
above; fixed-partition and adaptive replay are evaluated separately.\\
\bottomrule
\end{tabular}
\end{table*}

%% file: common/path_jvp_table_full.tex
% Generated by SECA_CL2.0/audit/v12_jvp_table.py -- do not edit by hand.
% Source: SECA_CL2.0/derived/g12_derivative_evidence.json (read-only)
%         SECA_CL2.0/derived/g05_retained_jvp.json (read-only)
\begin{table}[!ht]
\centering
\footnotesize
\setlength{\tabcolsep}{4pt}
\caption{Every stored replay step of the audited path Jacobian--vector
product, with the two observables of the grouped table in the main paper and the
completeness of the centred replay. $h$ is the dimensionless step of the
control direction; a dash marks a replay that did not complete on one side,
and values above the pre-set target $10^{-3}$ are marked $^{\dagger}$ and
retained. The stored record also holds the smallest step as eight aliases of
the nine scan values, so its twenty-six entries are eighteen printed values
plus those aliases.}
\label{tab:path-jvp-full}
\begin{tabular}{llrrrl}
\hline
Mesh (tetrahedra) & Audited control & $h$ & $q_{\mathrm{scene}}$ & $q_{\mathrm{obs}}$ & Centred replay \\
\hline
coarse (2{,}502) & symmetric sphere depth & 0.01 & -- & -- & incomplete (minus) \\
 & symmetric sphere depth & 0.005 & -- & -- & incomplete (minus) \\
 & symmetric sphere depth & 0.0025 & $4.76\times10^{-4}$ & $3.83\times10^{-4}$ & complete \\
 & one-sphere tangential translation & 0.01 & $1.77\times10^{-3}$$^{\dagger}$ & $1.50\times10^{-3}$$^{\dagger}$ & complete \\
 & one-sphere tangential translation & 0.005 & $2.78\times10^{-4}$ & $3.19\times10^{-4}$ & complete \\
 & one-sphere tangential translation & 0.0025 & $4.73\times10^{-4}$ & $5.75\times10^{-4}$ & complete \\
\hline
medium (4{,}992) & symmetric sphere depth & 0.01 & -- & -- & incomplete (plus) \\
 & symmetric sphere depth & 0.005 & $1.48\times10^{-3}$$^{\dagger}$ & $1.59\times10^{-3}$$^{\dagger}$ & complete \\
 & symmetric sphere depth & 0.0025 & $4.95\times10^{-4}$ & $5.89\times10^{-4}$ & complete \\
 & one-sphere tangential translation & 0.01 & $3.74\times10^{-4}$ & $3.82\times10^{-4}$ & complete \\
 & one-sphere tangential translation & 0.005 & $1.62\times10^{-4}$ & $1.70\times10^{-4}$ & complete \\
 & one-sphere tangential translation & 0.0025 & $9.82\times10^{-4}$ & $1.56\times10^{-3}$$^{\dagger}$ & complete \\
\hline
\multicolumn{6}{l}{fine (14{,}316): no stored summary in the retained record} \\
\hline
\end{tabular}
\end{table}

%% file: common/appendix_g05_curve.tex
% generated by SECA_CL2.0/audit/g12_appendix_rows.py -- do not edit by hand
\begin{tabular}{lrrrrrr}
\toprule
$\theta$ & \multicolumn{3}{c}{released $Q_{\mathrm{rel}}$} & \multicolumn{3}{c}{plastically active volume fraction} \\
\cmidrule(lr){2-4}\cmidrule(lr){5-7}
 & RM & Perzyna & SECA & RM & Perzyna & SECA \\
\midrule
0.1000 & $-8.069\times10^{-12}$ & $-8.069\times10^{-12}$ & $-8.069\times10^{-12}$ & 0.00000 & 0.00000 & 0.00000 \\
0.1643 & $-5.072\times10^{-12}$ & $-5.072\times10^{-12}$ & $-5.072\times10^{-12}$ & 0.00000 & 0.00000 & 0.00000 \\
0.2286 & $-1.899\times10^{-13}$ & $-1.899\times10^{-13}$ & $-1.899\times10^{-13}$ & 0.00000 & 0.00000 & 0.00000 \\
0.2929 & $-4.026\times10^{-4}$ & $-4.026\times10^{-4}$ & $-3.804\times10^{-4}$ & 0.02083 & 0.02083 & 0.02083 \\
0.3571 & $-1.737\times10^{-3}$ & $-1.737\times10^{-3}$ & $-1.706\times10^{-3}$ & 0.03125 & 0.03125 & 0.03125 \\
0.4214 & $-3.021\times10^{-3}$ & $-3.019\times10^{-3}$ & $-2.975\times10^{-3}$ & 0.03125 & 0.03125 & 0.03125 \\
0.4857 & $-4.498\times10^{-3}$ & $-4.498\times10^{-3}$ & $-4.466\times10^{-3}$ & 0.03125 & 0.03125 & 0.03125 \\
0.5500 & $-5.949\times10^{-3}$ & $-5.950\times10^{-3}$ & $-5.920\times10^{-3}$ & 0.04167 & 0.04167 & 0.04167 \\
0.5833 & $-6.817\times10^{-3}$ & $-6.819\times10^{-3}$ & $-6.780\times10^{-3}$ & 0.05729 & 0.05729 & 0.05729 \\
0.6167 & $-7.778\times10^{-3}$ & $-7.781\times10^{-3}$ & $-7.737\times10^{-3}$ & 0.06250 & 0.06250 & 0.06250 \\
0.6500 & $-8.794\times10^{-3}$ & $-8.797\times10^{-3}$ & $-8.750\times10^{-3}$ & 0.06771 & 0.06771 & 0.06771 \\
0.6833$^{\dagger}$ & $-9.878\times10^{-3}$ & $-9.882\times10^{-3}$ & $-9.831\times10^{-3}$ & 0.07292 & 0.07292 & 0.07292 \\
0.7167$^{\dagger}$ & $-1.096\times10^{-2}$ & $-1.097\times10^{-2}$ & $-1.091\times10^{-2}$ & 0.07292 & 0.07292 & 0.07292 \\
0.7500 & $-1.206\times10^{-2}$ & $-1.207\times10^{-2}$ & $-1.202\times10^{-2}$ & 0.07292 & 0.07292 & 0.07292 \\
0.7833 & $-1.322\times10^{-2}$ & $-1.323\times10^{-2}$ & $-1.318\times10^{-2}$ & 0.08333 & 0.08333 & 0.08073 \\
0.8167 & $-1.469\times10^{-2}$ & $-1.470\times10^{-2}$ & $-1.462\times10^{-2}$ & 0.10677 & 0.10677 & 0.10677 \\
0.8500 & $-1.663\times10^{-2}$ & $-1.664\times10^{-2}$ & $-1.655\times10^{-2}$ & 0.11719 & 0.11719 & 0.11719 \\
0.8875$^{\dagger}$ & $-1.906\times10^{-2}$ & $-1.908\times10^{-2}$ & $-1.894\times10^{-2}$ & 0.14323 & 0.14323 & 0.14323 \\
0.9250$^{\dagger}$ & $-2.181\times10^{-2}$ & $-2.182\times10^{-2}$ & $-2.168\times10^{-2}$ & 0.16927 & 0.17188 & 0.17708 \\
0.9625 & $-2.466\times10^{-2}$ & $-2.468\times10^{-2}$ & $-2.452\times10^{-2}$ & 0.19531 & 0.19531 & 0.19792 \\
1.0000 & $-2.770\times10^{-2}$ & $-2.773\times10^{-2}$ & $-2.754\times10^{-2}$ & 0.22135 & 0.22135 & 0.22656 \\
\bottomrule
\end{tabular}
% direct non-converged increments: 9; repaired by substep refinement: 9; unresolved: 0
% grid points carrying a dagger: 4

%% file: common/appendix_g05_derivative.tex
% generated by SECA_CL2.0/audit/g12_appendix_rows.py -- do not edit by hand
\begin{tabular}{llrrrr}
\toprule
control & $h$ & RM & Perzyna & SECA & flags \\
\midrule
elastic ($\theta=0.1643$) & 0.005 & $-1.126\times10^{-9}$ & $-1.126\times10^{-9}$ & $-1.126\times10^{-9}$ & -- \\
elastic ($\theta=0.1643$) & 0.02 & $5.288\times10^{-14}$ & $5.288\times10^{-14}$ & $5.288\times10^{-14}$ & -- \\
elastic ($\theta=0.1643$) & 0.05 & $1.908\times10^{-12}$ & $1.908\times10^{-12}$ & $1.908\times10^{-12}$ & -- \\
\midrule
near\_onset ($\theta=0.2929$) & 0.005 & $-1.792\times10^{-2}$ & $-1.792\times10^{-2}$ & $-1.705\times10^{-2}$ & -- \\
near\_onset ($\theta=0.2929$) & 0.02 & $-1.656\times10^{-2}$ & $-1.656\times10^{-2}$ & $-1.628\times10^{-2}$ & -- \\
near\_onset ($\theta=0.2929$) & 0.05 & $-1.414\times10^{-2}$ & $-1.413\times10^{-2}$ & $-1.383\times10^{-2}$ & onset crossed \\
\midrule
plastic ($\theta=0.7500$) & 0.005 & $-3.347\times10^{-2}$ & $-3.350\times10^{-2}$ & $-3.354\times10^{-2}$ & -- \\
plastic ($\theta=0.7500$) & 0.02 & $-3.371\times10^{-2}$ & $-3.373\times10^{-2}$ & $-3.377\times10^{-2}$ & -- \\
plastic ($\theta=0.7500$) & 0.05 & $-3.458\times10^{-2}$ & $-3.474\times10^{-2}$ & $-3.450\times10^{-2}$ & substep refinement in a rerun \\
\bottomrule
\end{tabular}

%% file: common/appendix_g06_softplus.tex
\begin{tabular}{rrrr}
\hline
$f_{\mathrm{tr}}$ (Pa) & softplus, explicit & softplus, implicit & compact (SECA) \\
\hline
$-2000000$ & $1.089\times 10^{-19}$ & $1.089\times 10^{-19}$ & $0$ \\
$-1000000$ & $5.285\times 10^{-11}$ & $5.285\times 10^{-11}$ & $0$ \\
$-500000$ & $1.164\times 10^{-6}$ & $1.163\times 10^{-6}$ & $0$ \\
$-100000$ & $3.255\times 10^{-3}$ & $1.296\times 10^{-3}$ & $0$ \\
$0$ & $1.777\times 10^{-2}$ & $2.878\times 10^{-3}$ & $0$ \\
$125000$ & $6.613\times 10^{-2}$ & $5.350\times 10^{-3}$ & $2.081\times 10^{-3}$ \\
\hline
\end{tabular}

%% file: common/appendix_g06_hold.tex
\begin{tabular}{lrrrrr}
\hline
law & hold $f_{\mathrm{tr}}$ (Pa) & $\gamma$ first & $\gamma$ last & $\alpha$ after 200 steps & released $\varepsilon^{p}$ \\
\hline
compact & $-1000000$ & $0$ & $0$ & $0$ & $0$ \\
compact & $-500000$ & $0$ & $0$ & $0$ & $0$ \\
compact & $-100000$ & $0$ & $0$ & $0$ & $0$ \\
softplus & $-1000000$ & $5.285\times 10^{-11}$ & $5.285\times 10^{-11}$ & $1.057\times 10^{-8}$ & $1.057\times 10^{-8}$ \\
softplus & $-500000$ & $1.163\times 10^{-6}$ & $1.012\times 10^{-6}$ & $2.168\times 10^{-4}$ & $2.168\times 10^{-4}$ \\
softplus & $-100000$ & $1.381\times 10^{-3}$ & $7.814\times 10^{-6}$ & $9.407\times 10^{-3}$ & $9.407\times 10^{-3}$ \\
\hline
\end{tabular}

%% file: common/appendix_moves_a.tex
% Material moved out of the main-text sections that were trimmed for the CGF
% submission revision (the local method, regularity analysis, finite-deformation
% increment, contact and path derivative, and discussion sections).
% The main text keeps the definitions and formulas, the measurement objects, the
% error definitions, the coverage and completion counts, the negative and null
% results, and the limitation of every claim; the detail below is the part that
% moved here, so that no measurement, threshold, failure statement or scope
% qualification is lost with the shorter main text.
% This supplement compiles on its own, so references to main-paper objects use
% either the textual stand-ins declared in the supplement's preamble or a
% descriptive name; references to this supplement's own sections and equations
% are ordinary \ref/\eqref calls.
% One moved display, the post-transition identity of the activation, is already
% displayed in Section~\ref{app:proofs} and is therefore pointed at instead of
% being repeated here; its unused main-text label was eq:compact-offset.

\makeatletter
\@ifundefined{c@localalgorithm}{\newcounter{localalgorithm}}{}
\makeatother

\section{Moved Detail: Local Method in the Main Paper}
\label{app:moves-local-method}

\paragraph{Quintic endpoint system and the two properties of the gate.}
The main paper's activation is the lowest-degree polynomial matching
$P(0)=0$, $P(1)=1$ with vanishing first and second derivatives at both ends;
the endpoint system and the interior derivatives it implies are solved in
Section~\ref{app:proofs}. Two properties were separated in the trimmed text.
\emph{Exactly inert below yield} means $D_\delta$, and therefore the
multiplier, is identically zero on $f\leq0$, not merely small. \emph{Finite
window} means the non-trivial variation is confined to $0<f<\delta$:
$D_\delta'$ is nonzero exactly on that interval, so its support is the closed
interval $[0,\delta]$, whereas $D_\delta$ itself is not of compact support,
because it equals one for every $f\geq\delta$.

\paragraph{Consistency of the candidate state.}
Under the radial-correction assumption the trial state is corrected along a
fixed direction, so $f^{n+1}=f_{\mathrm{tr}}-(3\mu+H)\gamma$ is exactly the
argument of the primitive in the main paper's multiplier equation; the scalar
equation solved by the update is therefore the consistency condition of the
candidate state, and no separate variable is introduced.

\paragraph{Dissipation of the flow rule.}
Whenever flow is active the local dissipation
$\boldsymbol\sigma:\dot{\boldsymbol p}-Ha\dot a=(\sigma_y+f)\dot a$ is
nonnegative, and both internal rates vanish at $f\leq0$. The derivation is in
Section~\ref{app:proofs}; the property is a check on the flow rule and not an
input to the solver.

\paragraph{Scope of the audited local derivative.}
The audited multiplier derivative holds the committed history, the material
parameters, $\delta$, $\eta$ and $\Delta t$ fixed; it is not the stress tangent
of the three-dimensional problem, which also carries the derivative of the flow
direction and of the elastic stress update. When those quantities are
perturbed, the fixed-$B$ factor is replaced by the derivative chain of the main
paper's regularity analysis, which is written out in Section~\ref{app:proofs}
next to the checks that cover it. The one-dimensional reduction joins the
elastic modulus continuously at onset for fixed $\delta>0$, and is not a
substitute for the assembled three-dimensional Jacobian. Material derivatives
follow from the same update but are not audited here; at fixed $\eta$ and
$\Delta t$, $\delta\to0$ approaches the linear-overstress root, not
rate-independent $J_2$, which is obtained only when the viscosity also
vanishes.

\section{Moved Detail: Regularity and Bias Analysis in the Main Paper}
\label{app:moves-regularity}

\paragraph{Activation class and the defect identity.}
The post-transition offset was stated for any integrable
$D_\delta:\mathbb R\rightarrow[0,1]$ satisfying $D_\delta(f)=0$ for $f\leq0$
and $D_\delta(f)=1$ for $f\geq\delta$. With the defect
$b_\delta=\int_0^\delta[1-D_\delta(z)]\,dz$, measured in the same units as $f$,
splitting the primitive at $\delta$ gives, for every $f\geq\delta$,
\[
r_\delta(f)=\int_0^\delta D_\delta(z)\,dz+f-\delta=f-b_\delta .
\]
No smoothness is needed for this identity: it depends on exact saturation above
$\delta$ and therefore does not hold in this form for logistic or softplus
tails. The same identity is displayed in Section~\ref{app:proofs}, where the
antisymmetry of the implemented quintic is used to evaluate the defect.

\paragraph{Peak slope and the constant product.}
For the main paper's quintic, symmetry
$D_\delta(f)+D_\delta(\delta-f)=1$ inside the transition yields the defect, and
$D_\delta'$, whose maximum sits at $\xi=1/2$, yields the peak slope. The product
is an exact property of the implemented quintic, not a fitted curve: narrowing
the transition reduces its primitive defect linearly while increasing its
maximum slope inversely.

\paragraph{The four multiplier derivatives.}
Implicit differentiation of the multiplier equation with the previous committed
state held fixed was written with $A=1+\alpha BD_\delta(\bar f)$ and
$\bar f=f_{\mathrm{tr}}-B\gamma$. It gives the partial derivatives of the
multiplier with respect to the trial overstress, the transition width, the
viscosity and the time increment, together with the fixed-argument width
partial of the primitive, the total derivative when $\eta=c\delta$ couples the
two, and the multiplier curvature; all of these are written out in
Section~\ref{app:proofs}. The width partial is not a new material parameter but
the derivative of the same primitive with respect to the same width, taken with
the overstress argument held fixed.

\paragraph{Limits and the non-uniform convergence of derivatives.}
Above the transition the width term equals $\delta/[2(B+\beta)]$. Holding
$\beta>0$ fixed while removing width reaches the linear-overstress root; it does
not reach rate-independent $J_2$, which additionally requires $\beta\to0$. If
$\eta=c\delta$ at fixed $\Delta t$, both terms are $O(\delta)$. Multiplier
values converge in this limit, but derivatives do not converge uniformly at
yield, because the right derivative changes from zero at every positive width to
a nonzero hard-onset value.

\paragraph{What the multiplier bound does not certify.}
The bound on the multiplier difference is a bound on the local multiplier only:
its constants are not measured against any three-dimensional released geometry,
so they are not a numerical certificate for the released results of
Section~\ref{sec:results}, which are measured directly.

\paragraph{Propagation to released equilibrium.}
The propagation assumptions, the iteration that carries the local difference to
a released observable, and the constants it uses are stated in
Section~\ref{app:proofs}: the Lipschitz constant $L_n$ of the history update,
the per-step contribution $M_n|\gamma_n-\widetilde\gamma_n|$ of the multiplier
difference, the residual bounds $\|\boldsymbol R_x^{-1}\|\leq\kappa$ and
$\|\boldsymbol R_h\|\leq L_R$, and the $L_Q$-Lipschitz released observable,
combined in Equations~\eqref{eq:history-propagation} and
\eqref{eq:release-propagation}. Here $\boldsymbol h_n$ is the assembled
committed state, not one scalar, so a difference measured at a single material
point cannot stand for the history of a whole body. The bound therefore holds
for trajectories that remain in one local neighbourhood, share the accepted time
partition and contact-feature sequence, and release onto a continuously
connected equilibrium branch with a nonsingular residual Jacobian; it does not
assert that every finite-deformation path stays on one branch, and it does not
by itself establish a released-shape error for the measured cases of
Section~\ref{sec:results}.

\section{Moved Detail: Finite-Deformation Increment in the Main Paper}
\label{app:moves-finite}

\paragraph{One accepted increment.}
The main text describes this procedure in prose; the numbered statement is kept
here.
\begin{center}
\noindent\fbox{\parbox{\dimexpr\columnwidth-2\fboxsep-2\fboxrule}{%
\refstepcounter{localalgorithm}%
\textbf{Algorithm \thelocalalgorithm\ (one accepted increment).}\label{alg:global}\\
\emph{Input:} accepted state $(\boldsymbol x^n,\boldsymbol F_{p}^n,a^n)$, the
control at this increment, the time increment $\Delta t$.\\
1.\quad Newton loop: form the element trial deformation from the current iterate
$\boldsymbol x$.\\
2.\quad call the selected local evolution equation of the main paper's
local-update subsection for every element with the \emph{committed} history
$(\boldsymbol F_p^n,a^n)$; the SECA branch is the main paper's local algorithm.\\
3.\quad assemble the material residual of the main paper plus the support and
barrier/contact terms of its contact subsection.\\
4.\quad if the relative free-dof residual is below the tolerance of this case,
exit the loop.\\
5.\quad assemble the tangent, solve the linear system, and globalise with the
line search; a rejected trial discards its candidate history.\\
6.\quad \emph{commit} $(\boldsymbol F_p^{n+1},a^{n+1})$ only when the increment
is feasible and converged; record the residual, the iteration counts, the
contact state, and $\min\det\boldsymbol F$.}}
\end{center}
\vspace{2pt}

\paragraph{Order of operations.}
Two properties of that order are load bearing. The candidate history is not
committed before the increment is accepted, so a rejected trial leaves the
committed state untouched. And the derivatives of that whole sequence---the flow
direction, the exponential update, and the dependence of the elastic state and
the stress on the trial deformation---define the residual the solver iterates,
so assembling a different Newton matrix changes the iteration and not the
residual that defines the solution.

\paragraph{Volume preservation of the plastic state.}
The flow direction of the main paper's finite update is trace free but not
unit-Frobenius, so $\det\boldsymbol F_p=1$ holds in exact arithmetic from an
initially isochoric state.

\paragraph{Newton fallbacks.}
In the controlled cube and simple-body experiments the iteration matrix is the
consistent material tangent of the main paper's finite update plus the
analytically differentiated tool term. Where the plain Newton direction is not a
descent direction or the linear solve fails, the solver falls back to a
diagonally damped system and, if necessary, to a least-squares direction; those
fallbacks are counted together with the iteration counts reported in the main
paper.

\paragraph{Scope of the radial reduction.}
For the isotropic Hencky model the radial direction is coaxial with the trial
elastic stretch, so the updated equivalent strain obeys
$e_{\mathrm{eq},n+1}=e_{\mathrm{eq,tr}}-\gamma$ and the yield function reduces
to $f_{n+1}=f_{\mathrm{tr}}-(3\mu+H)\gamma$; that proportional reduction is not
asserted for anisotropic elasticity or a noncoaxial flow rule. The finite update
reuses the scalar root in that form, under the stated radial-correction
condition, and is not a general projection of an arbitrary finite-strain model
onto the small-strain update.

\paragraph{Supports, tools and the grip schedule.}
A support is a prescribed-displacement condition on a fixed node set: those
displacements are part of the data in every state, so the free degrees of
freedom are their complement and no stage reduces a support. A tool or grip is
the opposite kind of object: it enters the residual as a penalty term with a
recorded stiffness---the barrier stiffness $\kappa_b$ and the per-pair contact
area $A_t$ defined in Section~\ref{app:protocols}---and removing it is part of
the schedule rather than a change of the supports. In the Armadillo character
the grip stiffness is held at its full value through the loading stages and then
ramped linearly to exactly zero at fixed control, so the last state is a
grip-free equilibrium on the unchanged supports rather than a body with a
weakened grip; the stage counts and the stiffness schedule across the stages are
recorded in Section~\ref{app:tension-work}.

\section{Moved Detail: Contact, Globalisation and Path Derivative in the Main Paper}
\label{app:moves-contact}

\paragraph{Barrier expressions.}
For analytic sphere $j$ and boundary triangle $t$ the tool gap is measured from
the point--triangle distance and the sphere radius, and positive gaps below the
barrier distance $\hat d$ contribute the logarithmic barrier energy with the
standard $-(g-\hat d)^2\log(g/\hat d)$ profile, which vanishes for
$g\geq\hat d$ and is strictly positive below it. The barrier energy, the
point--triangle distance, the per-pair contact area and the barrier stiffness
are written out in Section~\ref{app:protocols} and
Equation~\eqref{eq:contact-barrier}.

\paragraph{Hierarchy diagnostics of the controlled path.}
The controlled hierarchy uses the same material, supports, tools, and
30-interval physical-time schedule on 2,502, 4,992, and 14,316 tetrahedra. All
three fixed-schedule forward paths complete. The fine path activates
self-contact at three stages, reaches minimum self distance
$7.64115\times10^{-4}L_0$, maintains positive tool gap, and has
$\min\det\boldsymbol F=0.672440$. The largest relative force and moment
imbalances over the hierarchy are $1.60\times10^{-5}$ and $2.93\times10^{-5}$.

\paragraph{Linear solve.}
The path system is solved by GMRES with a frozen elastic factor as
preconditioner. Neither the positive-semidefinite forward matrix nor a perturbed
forward solve is substituted for $R_u$: the projected forward matrix is an
optimisation device for globalisation, while the residual and the path
derivative use the exact unprojected tool and self-contact Hessian actions.

\paragraph{Operator mapping.}
The spaces are fixed by the discretisation: $\boldsymbol u_k$ collects the free
nodal displacements of increment $k$, so the supports are eliminated rather than
differentiated; $\boldsymbol z_k$ collects $(\boldsymbol F_{p,k},a_k)$ over all
elements of the mesh; and $\boldsymbol\theta$ collects the audited controls, the
tool parameters that the trajectory is generated from. The operators $R_u$,
$R_z$ and $R_\theta$ map directions in those three spaces to residuals of the
assembled dimension, and $\Phi_u$, $\Phi_z$ and $\Phi_\theta$ map the same
directions to an increment of the committed history, so the path
Jacobian-vector product is the forward propagation of one directional
derivative along the accepted sequence.

\section{Moved Wording: Related Work and Discussion in the Main Paper}
\label{app:moves-wording}

The remaining differences between the trimmed related-work and discussion
sections and their earlier full wording are editorial: every claim, citation and
boundary is retained in the main text, including that exact return mapping is
used as the reference implementation rather than as a straw man, that a smoothed
level-set envelope and a one-sided compact activation are different objects,
that contact is an enabling component rather than the analysed object, and that
the reported agreement is a matched-mesh and matched-schedule statement. One
statement was shortened rather than reworded and is recorded here.

\begin{itemize}
\item The cross-method difference in the released field is of the same order as
each constitutive method's own temporal discretisation error, so
``comparable'' means comparable at the reported discretisation and not
identical.
\end{itemize}

%% file: common/appendix_moves_b.tex
% appendix_moves_b.tex --- detail moved out of the main text when its Results were
% re-partitioned for the CGF submission (the macroscopic, parameter and
% complex-model question).  Cut-and-move only: every sentence, enumeration and
% record pointer that left the main text reappears here with its numbers, its
% scene and its settings intact, and nothing was recomputed.  The main text keeps
% the measured outcomes, the failure and non-convergence facts, the limitation of
% each claim and the interior/surface-stress statements; the arrays, per-preset
% detail, protocol enumerations and record pointers live here.

\section{Controlled Cube Indentation: Moved Detail}
\label{sec:b-cube-detail}

Moved from the main text's subsection on the macroscopic, parameter and
complex-model response. On the unit cube of that controlled benchmark, with all
three constitutive methods on one tetrahedral mesh, one tool path, one physical
duration, one Newton tolerance, one Armijo line search and one committed-history
rule, the removed framing sentence asked whether the regularised transition
leaves the resolved macroscopic load--release response intact once the
material-point behaviour is embedded in a full solve, since that behaviour only
predicts structure if the transition survives assembly, contact and the Newton
loop. The removed scope sentence recorded that the advantage of the method lies
in that regularity and in the nonlinear solver work rather than in the released
shape. The peak and released dent of that cube run, $18.589\%$ to $18.591\%$ and
$9.602\%$ to $9.642\%$ of the body height, are carried by the caption of the cube
figure in the main text, and the per-mesh ladder quoted in the main text is the
one measured on that same run: refining from \depCubeTetrahedra\ to $24{,}576$
tetrahedra raises the released dent from $9.602\%$ to $11.946\%$ for SECA, from
$9.642\%$ to $11.958\%$ for exact return mapping and from $9.642\%$ to $12.005\%$
for Perzyna, while the spread between the three constitutive methods stays within
$0.06$ percentage points at every level, which is why an absolute indentation
figure is quoted together with its mesh level.

\section{Parameter Studies: Moved Detail}
\label{sec:b-parameter-detail}

Moved from the main text's parameter-control paragraphs, which sweep one named
quantity at a time on a fixed mesh, boundary condition, load path, time step,
tolerance and line search while the constitutive method is held fixed.

\subsection*{Sweep taxonomy}

The removed enumeration distinguished three kinds of sweep: single-quantity
sweeps that move the loading depth, the yield threshold, or the transition width
alone; a coupled sweep that moves the transition width and the viscosity together
at fixed $\eta/\delta=0.2$ beside a fixed-$\eta$ layer, which is the pair a user
would tune; and a material-family preset that moves four parameters at once.

\subsection*{Loading magnitude}

The removed record pointer noted that the internal record of the shallowest run
is what backs the reading of a released dent of zero: its stored states carry a
maximum accumulated plastic strain of exactly zero and a multiplier of exactly
zero at every increment, rather than a small one, and the sub-yield table of this
supplement shows the same compact column evaluated directly at negative
overstress. The removed figure remark recorded that panel~(a) of the
parameter-control figure shows the released curve separating from the peak curve
as depth grows, which is the qualitative signature of plasticity rather than of a
stiff elastic response. The swept values are $d/h$ of \depLoadDepthMin\ to
\depLoadDepthMax\ in \depLoadCount\ steps, and the released dent moves from
\depLoadDentMin\% to \depLoadDentMax\%.

\subsection*{Strength: threshold, hardening and material family}

The removed sentence stated that the vanishing released deformation at
$\sigma_y/\sigma_{y,0}=2$ is the direct numerical confirmation of the
strict-elastic property claimed in the main text's analysis of the constitutive
method, where the stored states of those runs report a maximum accumulated
plastic strain of exactly zero rather than a small number. On the cantilever, the removed
explanation recorded why the retained length responds more weakly than the stress
level: the tip displacement is prescribed and the plastic strain field is largely
fixed by the kinematics of the beam, so it is the threshold, not the hardening,
that decides retention. The hardening sweep moves $H/H_0$ over
\depHardLo--\depHardHi, with a peak Kirchhoff von Mises stress of
\depHardStressFirst\ to \depHardStressLast\ MPa and a residual tip offset of
\depHardFirst\% to \depHardLast\%.

For the material family, the removed pointer recorded that $E$, $\nu$,
$\sigma_y$, $H$, $\delta$ and $\eta$ are listed here for every preset that was
run, under one normalised gate setting ($\delta/\sigma_y = 0.05$ and
$\eta/\delta = 0.2$~s, so the absolute width and viscosity scale with
$\sigma_y$), and that the family renders, the hardening sweep and the per-preset
convergence record are carried by the parameter-study and material-family
sections of this supplement. The retained fraction rises monotonically from
\depFamilyTrendMin\% to \depFamilyTrendMax\% over the four converged presets
while their peak stress spans \depFamilyStressMin\ to \depFamilyStressMax\ MPa,
and the fifth (softest) preset does not converge its final tool-free stage, so it
is excluded from the quoted range and kept in the material-family record rather
than dropped. The removed scope sentence stated what that joint preset does and
does not show: one normalised gate setting carries the family across the reported
span without retuning, not that any one of the four parameters drives the trend.

\subsection*{Transition width}

The removed opening sentence identified the transition width as the parameter
specific to this work, because it is the quantity that trades regularity for
bias. The sweep is run in two layers on the gripped torus: in the coupled layer
$\eta/\delta=0.2$ is held, so width and viscosity move together, and in the
fixed-$\eta$ layer only the width moves; over
$\delta/\sigma_y\in[\depWidthCoupledLo,\depWidthCoupledHi]$ the released
major-axis extension moves monotonically from \depWidthCoupledFirst\% to
\depWidthCoupledLast\% in the coupled layer and from \depWidthFixedFirst\% to
\depWidthFixedLast\% in the fixed-$\eta$ layer. The removed derivation pointer
recorded that the two exact relations plotted in panel~(d) of the core-activation
figure of the main text and stated in
Equation~\ref{eq:regularity-bias-product},
$\|D_\delta'\|_\infty=15/(8\delta)$ and $b_\delta=\delta/2$, are properties of
the local mapping, so the released-shape change measured here is a structural
consequence of that choice on this protocol. The removed closing sentence
recorded the result for the protocols that were run: over the swept ranges the
finite-width bias of the compact transition behaves as a control rather than as
scatter, the two width layers separate, and an application that selects a width
can see the released-shape consequence it is choosing.

\section{Diverse Loading Conditions: Moved Detail}
\label{sec:b-simple-bodies-detail}

Moved from the main text's four-geometry experiment, which runs the same three
constitutive methods on local indentation of a block, bilateral platen
compression of a sphere, finite-area grip tension of a genus-one torus and
cantilever bending of a slender beam, with one material, one mesh per body, one
Newton tolerance, one Armijo line search, and complete removal of every tool or
grip before the released state is measured. The retained fractions of that one
SECA setting are \depCubeRetainedPct\%, \depSphereRetainedPct\%,
\depTorusRetainedPct\%\ and \depBeamRetainedPct\%, and the released-state table
of the main text lists the measured peak, released and retained fractions with
the smallest deformation determinant on each path.

The removed purpose sentence recorded that the experiment is not intended to
distinguish SECA from return mapping visually; it tests whether the same
transition model remains usable under qualitatively different loading
mechanisms, and character-scale deformation is an important graphics test while
evidence from one loading condition would not establish broader applicability.

The removed paragraph explained why the two cases that are not simply connected
or not singly loaded carry the most information. The torus is the only case that
combines tension with a doubly connected body, so it is the one case in which the
plastic zone can run around a ring instead of through a simply connected section;
its grips are an area-weighted stiffness on the patch nodes rather than a
prescribed displacement on a single layer, so the patch is free to shear and
contract. The beam is the one mode in which the two faces carry opposite stress
signs, so a material that responded differently in tension and compression would
show it here. In both cases the released state is a genuine residual deformation,
and the retained fraction remains monotone in the imposed amplitude. The removed
record pointer added that this supplement records where each path's states come
from together with the reduced-order release check that accompanies it, and the
protocol detail of the released-shape scenes is in the simple released-shape
protocol section below.

\section{Character-Scale Models: Moved Detail}
\label{sec:b-character-detail}

Moved from the main text's two character paths and their cost paragraph. Both
characters are loaded to a peak, held, and then released by removing the tools
completely; the reported quantity is the released equilibrium, not the pose under
load.

\subsection*{Bunny compression}

The displayed path uses 699 nodes, 2,502 tetrahedra and a 968-triangle physical
boundary, with two opposed analytic spheres executing the
load--withdrawal--hold schedule while a torso gauge is measured on fixed
rest-space side patches. At dimensionless depth 0.85 the gauge reaches
\depBunnyPeak\% peak compression and retains \depBunnyRetained\% after complete
withdrawal, for \depBunnyRecovery\ percentage points of recovery. The removed
verification numbers of that run are: every accepted state has positive
sphere--triangle gap, with a minimum of $0.021513L_0$; the minimum
$\det\boldsymbol F$ is $0.797316$; the maximum $|\det\boldsymbol F_p-1|$ is
$3.553\times10^{-15}$; the maximum relative free residual is
$1.993\times10^{-5}$; and the two peak reaction magnitudes, $0.401825F_0$ and
$0.401800F_0$, agree to $6.2\times10^{-5}$ relative difference. The removed
pointer noted that the peak and released stress fields of this body are the first
pair of panels in the peak-versus-released stress figure of the main text.

\subsection*{Armadillo tension}

The second character uses the same finite-$\boldsymbol F_p$ update on a 731-node,
2,280-tetrahedron mesh, with three pelvic supports and two rest-area-weighted
finite grip patches whose targets are held at peak before their stiffness is
reduced to zero. With symmetric axial control 1.00, all 17 equilibrium stages
converge; the gauge extends \depArmadilloPeak\% at peak and retains
\depArmadilloRetained\% after complete grip removal, corresponding to
\depArmadilloRecovery\ percentage points of recovery. The removed verification
numbers of that run are: the minimum $\det\boldsymbol F$ is $0.8875$; the largest
departure of $\det\boldsymbol F_p$ from one over the $17$ recorded stages is
$1.8\times10^{-15}$; and a further four-unit zero-grip hold produces no measured
gauge or history change at the recorded tolerance. The removed pointer noted that
the released panels of the same figure show the corresponding fields.

\subsection*{Interior and surface stress}

The interior cutaway opens the same solved bodies and the indentation cube at both
the loaded peak and after complete tool removal, with both cut planes fixed in
the rest configuration of the frame each body is drawn in, at the body centre, and
applied to the deformed state; the two characters keep the frame and the viewpoint
of the surface figure, so their orientation is the same in both figures. As
the main text states, it is a view of bodies that the surface and quantitative
evidence already verifies, not an independent verification of the update and not
a comparison between methods. Values on the cut faces are a display field recovered
from the raw element-constant fields of the linear tetrahedra: the element values are
averaged at the mesh nodes with the reference element volumes as weights and that
nodal field is read on the drawn surface, evaluated inside the element that owns
each drawn facet on the cube and, on the Bunny, located in the committed tetrahedral
mesh of the deformed body and interpolated linearly at each drawn point of the planar
section, which is the same recovery the surface panels of the character figures use
before transferring it to the drawn boundary. Each body is cut on the surface the
surface figure draws for it. Every number quoted in the main text is read
from the raw element-constant field, and this supplement also draws the same faces
with it, on the computed mesh of each body, so the Bunny's cut faces are recorded
there both on that mesh and on the display triangulation. This supplement carries the
committed equivalent plastic strain on the same faces, with the plane normals and
offsets.

\subsection*{Timing provenance}

The removed timing sentences recorded that the timings come from an offline CPU
implementation: this supplement reports the historical seconds for the forward
solve of the 14,316-tetrahedron Bunny fine simulation and for its
algorithmic-gradient path (1376 residual JVPs), with the platform metadata
available for them. That record states the platform, the library versions and
$24$ logical processors but not the processor model, so no specific CPU is
attached to those seconds. These timings characterise the current CPU
implementation and exclude validation replay; the character examples verify that
the method remains stable through finite deformation, contact, unloading and
differentiation at character scale, not real-time performance.

\subsection{Sweep values and released-state summaries}
\label{app:sweep-tables}

The main text carries one table of method properties, one of measured cost and
accuracy, and one of the derivative audit; the two record tables that used to sit
beside them are kept here. Table~\ref{tab:ablations} lists, for every parameter
row, the swept values and the observable that is read from the released state,
and Table~\ref{tab:simple-bodies} collects the measured peak, released and
retained fractions with the smallest deformation determinant on each path.

\begin{table*}[!t]
\centering
\small
\caption{Parameter studies. Each row moves one named quantity, one named pair, or
one preset on a fixed
protocol and names the swept values, the observable read from the released state, and
where the row is reported; the one pair is the coupled width row, where the
transition width and the viscosity move together at fixed $\eta/\delta$, beside the
fixed-$\eta$ width row that moves the width alone; the two width rows span a factor of $24$ in
$\delta/\sigma_y$. Loading depth is an external control, so the prescribed tool
command is not the measured dent, and hardening is reported through the peak
Kirchhoff von Mises stress. The material-family row moves four parameters at once and
is a joint preset rather than a one-parameter ablation; the supplement lists $E$,
$\nu$, $\sigma_y$, $H$, $\delta$ and $\eta$ for every preset that was run, and the
span quoted here covers the four presets whose release converged. The numbers in
this table are SECA; the threshold row of the main paper's parameter-control figure is the
row that carries all three constitutive methods.}
\label{tab:ablations}
\setlength{\tabcolsep}{4pt}
\begin{tabular}{@{}p{0.115\textwidth}p{0.185\textwidth}p{0.235\textwidth}p{0.215\textwidth}p{0.155\textwidth}@{}}
\toprule
Study & Symbol or preset & Swept values & Observable & Reported in \\
\midrule
Loading depth & tool command $d/h$ (control)
  & \depLoadDepthMin--\depLoadDepthMax & released dent depth (\%)
  & main paper, loading-depth sweep \\[3pt]
Material family & joint preset of $(E,\nu,\sigma_y,H)$
  & $E/E_0 = 0.25$--$4.00$; the four converged presets span
  \depFamilyYoungMin--\depFamilyYoungMax
  & retained fraction (\%); peak $\sigma_{vM}$ (MPa)
  & supplement, family record and renders \\[3pt]
Yield threshold & $\sigma_y/\sigma_{y,0}$ & \depYieldLo--\depYieldHi
  & released compression (\%) & main paper, yield sweep \\[3pt]
Hardening & $H/H_0$ & \depHardLo--\depHardHi & peak $\sigma_{vM}$ (MPa)
  & supplement, sweep figure \\[3pt]
Width (coupled) & $\delta/\sigma_y$ at fixed $\eta/\delta$
  & \depWidthCoupledLo--\depWidthCoupledHi & released major extension (\%)
  & main paper, width sweeps \\[3pt]
Width (fixed $\eta$) & $\delta/\sigma_y$ at fixed $\eta$
  & \depWidthFixedLo--\depWidthFixedHi & released major extension (\%)
  & main paper, width sweeps \\
\bottomrule
\end{tabular}
\end{table*}

\begin{table}[!t]
\centering
\small
\caption{Released-state summary for the controlled geometries of
the main paper's controlled-geometry figure, in percent. ``Peak'' and ``released'' are the
measured deformation at the loaded peak and after complete tool or grip removal,
both as a fraction of the relevant rest dimension (indentation depth over block
height, platen stroke over sphere diameter, grip extension over rest grip span,
tip deflection over beam length). ``Retained'' is the ratio of released to peak
deformation of the same row, and $\min\det F$ is the smallest deformation
determinant on that row's path. Each released state is taken after the
converged steps of that run.}
\label{tab:simple-bodies}
\setlength{\tabcolsep}{3pt}
\begin{tabular}{lrrrr}
\toprule
Body & Peak (\%) & Released (\%) & Retained (\%) & $\min\det F$ \\
\midrule
Block indentation & \depCubePeakPct & \depCubeReleasedPct
  & \depCubeRetainedPct & \depCubeMinimumDetF \\
Sphere compression & \depSpherePeakPct & \depSphereReleasedPct
  & \depSphereRetainedPct & \depSphereMinimumDetF \\
Torus tension & \depTorusPeakPct & \depTorusReleasedPct
  & \depTorusRetainedPct & \depTorusMinimumDetF \\
Beam bending & \depBeamPeakPct & \depBeamReleasedPct
  & \depBeamRetainedPct & \depBeamMinimumDetF \\
\bottomrule
\end{tabular}
\end{table}

\subsection{Controlled cube-indentation render record}
\label{app:cube-render}

The render sheet of the controlled cube indentation is the record behind the
numbers quoted in the main text for that benchmark, so it is kept here in full.
The measurements it illustrates are reported in the Results section of the main
paper.

\begin{figure*}[!t]
\centering
\includegraphics[width=\textwidth]{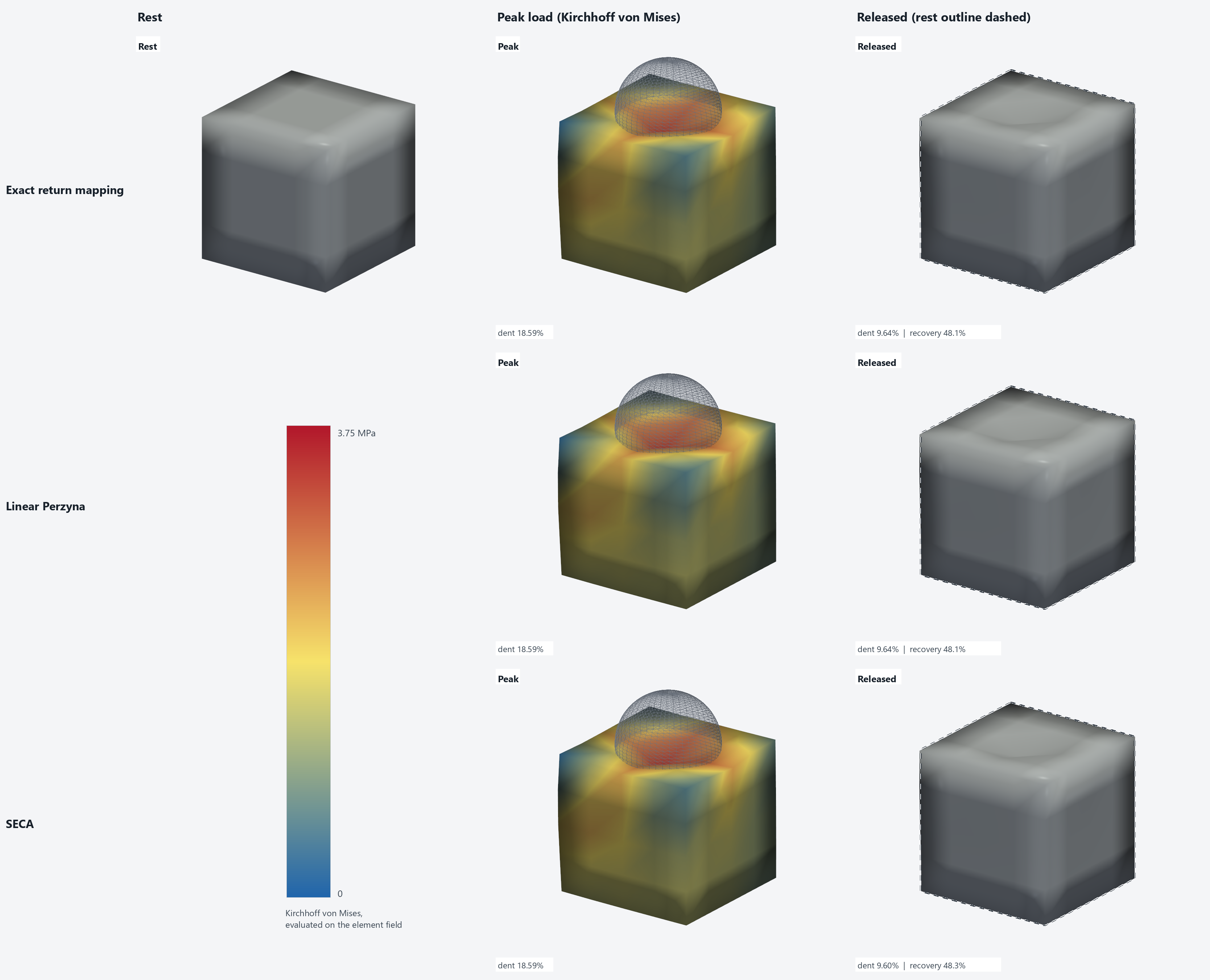}
\caption{Controlled spherical indentation of the unit cube on one mesh and one
tool path. Rows are the three constitutive methods, columns are the rest state,
the peak state coloured by Kirchhoff von Mises stress, and the released state
after complete tool
removal with the undeformed silhouette overlaid as a dashed contour. The rest
state is shared by all three rows, with one mesh, one tool path, one time step
and one tolerance and with zero stress and zero plastic history, so it is drawn
once beside them together with the colour scale the peak panels share, whose
upper limit of $3.75\,$MPa is a display ceiling read off the redrawn peak-state
element fields and not a stress quoted in the text; only the
plastic multiplier differs. Peak dent is
$18.589\%$ to $18.591\%$ of the body height for all three constitutive methods and the released
dent is $9.602\%$ to $9.642\%$.}
\Description{Three rows of rest, peak-stress, and released cube states compare
exact return mapping, Perzyna, and SECA on one indentation protocol.}
\label{fig:cube-indentation}
\end{figure*}

%% file: common/appendix_moves_main.tex
% Supplement file for the CGF rewrite round: everything removed from the three new Results
% subsections (6.1 Experimental Setup and Local Properties, 6.2 Controlled Computational
% Benefit, 6.3 Released-State Derivative and Minimal Control Task).
% Nothing here is new evidence: every number is the number the main text carried before the
% split, kept together with the scene, the stored record and the settings it was measured under.

\section{Records Moved from the Experimental Setup and Local Properties Results}
\label{app:moves-local}

\paragraph{Order of the measurements.}
The setup subsection used to list the order in which the measurements follow: the local
regularity and material-point comparison first, then the released-state sensitivity, the
controlled cube benchmark, the parameter studies, the simple bodies and the character examples.
The list is kept here, unchanged, because the main agent rewires those subsection labels in this
round; it names no measured quantity.

\paragraph{Local evaluation point and the protocol records behind it.}
The local comparison quoted in the main text is evaluated in the scaled prototype configuration
with yield stress $2.0$, hardening $10.0$, width $0.2$, $\eta/\delta=0.2$ and unit loading
duration. This is a pointer to values this supplement already records in its material-point
sensitivity protocol, repeated here so that the moved statement stays complete: the largest
scalar stress--strain derivative discrepancy $4.50\times10^{-7}$ relative in that
configuration, the agreement of the algorithmic tangent with its closed form
$8.4\times10^{-16}$ relative over the same six states, the centred difference at $h=10^{-6}$ of
$9.3\times10^{-11}$, the near-onset deviation of $1.6\times10^{-5}$ at $h=10^{-4}$ because a
step of that size leaves the elastic branch, the exact width/rate decomposition residual
$3.47\times10^{-18}$ (round-off alone, both sides being floating-point evaluations of the same
expression), the nonproportional width sweep $0.2\to0.05\to0.005$ giving $3.480\%$, $0.890\%$
and $0.089\%$ maximum yield-normalised stress discrepancy, and the load-step and duration checks
of the same configuration ($20$--$2000$ scalar load steps reproduce a final stress of $2.321412$
at fixed width $0.2$ and physical duration, while changing the duration changes the response,
as expected for a viscous law). Nothing in that record is withdrawn.

\paragraph{Settings record kept from the setup subsection.}
The setup text also stated that the character solver of the character-scale examples uses its own
tolerance and its own iteration matrix, described with those examples, and that the accounting of
the diagonal-damping and least-squares fallbacks is as described in the finite-deformation solver
section. Both statements are conditions of the reported runs and are kept here; the main text
keeps the settings that are common to the controlled comparisons and states that a common
setting is not a common plastic history.

\paragraph{Tangent-change figure and its protocol variants.}
The main text drew the quoted $\Delta C_{\mathrm{tr}}$ values together with the protocol variants
in the following figure, which moves here with its caption unchanged. Its record is the one the
main text cites: refining the output sampling alone leaves all three values unchanged, whereas the
archived refined-grid record doubles the number of steps at fixed $\Delta t$, which doubles the
physical duration and halves the strain rate, so the ratio falls because the loading history
changed rather than because one computation was sampled differently.

\begin{figure*}[t]
\centering
\includegraphics[width=\textwidth]{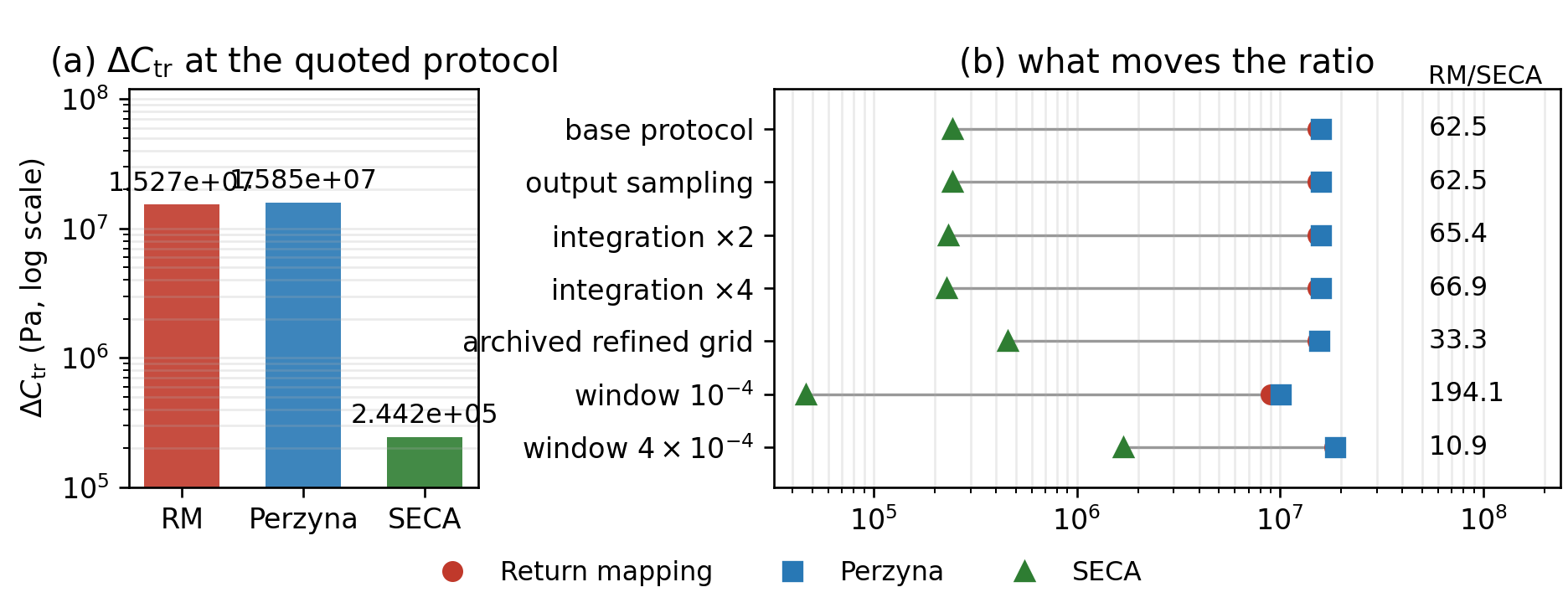}
\caption{$\Delta C_{\mathrm{tr}}$ at the quoted protocol and what moves the ratio.
(a)~The three secant-slope changes at the quoted protocol, a window of
$2\times10^{-4}$ logarithmic strain sampled as twenty steps of $10^{-5}$ on a
$2001$-point grid.
(b)~The same three quantities under each protocol variant, on the common
logarithmic axis of $\Delta C_{\mathrm{tr}}$; the resulting
return-mapping/SECA ratio is printed in the right-hand column, so the ratio is not
read from the quantity axis. The variants are integration-step refinement at fixed
history ($\times2$, $\times4$), the archived refined grid (doubled physical
duration), window changes ($10^{-4}$ and $4\times10^{-4}$ logarithmic strain), and
output-sampling refinement alone ($8001$ samples).}
\Description{Two panels: a bar chart of the three tangent-change magnitudes at the
quoted protocol, and a dot plot of the same magnitudes under seven protocol
variants with the resulting ratios.}
\label{fig:tangent-jump-metric}
\end{figure*}

\paragraph{Onset alignment at $32$ load increments.}
The strict-elasticity check on the controlled cube benchmark recorded that at $32$ load increments
the retained run records report a maximum accumulated plastic strain of exactly zero through the
first eight increments and a first non-zero value at the ninth, with the same plastic volume
fraction ($0.015625$ of the mesh) for all three constitutive methods. The main text keeps the
outcome, that the three methods enter the plastic branch at the same load step with the same
volume fraction and the compact onset therefore does not shift the apparent yield point.

\section{Records Moved from the Controlled Computational Benefit Results}
\label{app:moves-cost}

\paragraph{Micro-numeric derivative protocol and its per-state records.}
The derivative comparisons of the material-point comparison used the micro-numeric protocol:
they differentiate the accumulated equivalent plastic strain of a complete load--unload cycle
with respect to its peak strain control, and each side of a comparison is integrated
independently from the initial state. Under that protocol the compact update is consistent with
independent centred differences at the sub-yield, near-yield and yield controls, and at the first
plastic control it agrees to $1.4\times10^{-9}$ relative at $h=10^{-4}$, a value that the same
protocol rerun at the evaluation point used throughout the results reproduces to its last printed
digit, with both sides of the difference integrated as complete load--unload histories. Where
the return-mapping derivative vanishes identically below the onset, the comparison reports the
floored relative error of exactly $1$, an artefact of normalising by a vanishing difference: the
absolute deviation of that row is exactly zero, and the retained per-state records mark such
rows as carrying no meaningful relative error. Because the transition has a finite curvature
scale rather than a branch switch, its centred-difference discrepancy contracts at second order
in the difference step, from $2.8\times10^{-3}$ at $h=10^{-3}$ to $2.7\times10^{-9}$ at
$h=10^{-6}$: the finite difference converges to the analytic tangent rather than to a one-sided
limit of a kink, which is the property a derivative-based optimiser consumes.

\paragraph{Stage split of the whole-run solver counts.}
The whole-run totals of the controlled cube indentation with one load increment are sums over
every solved stage, and their split is not uniform. The load stage that crosses the transition
needs $32$ Newton iterations under return mapping and $10$ under SECA, with $117$ and $7$
backtracking halvings respectively, so on that stage alone SECA needs $69\%$ fewer Newton
iterations and $94\%$ fewer halvings. That stage carries all $117$ return-mapping halvings but
only $7$ of the $8$ SECA ones, so the stage and whole-run readings are different splits of the
same run and the halves are not interchangeable. The same numbers are already recorded in the
solver-work section of this supplement, under the same whole-run convention; the main text keeps
the split statement without the stage-level counts.

% Moved out of the main text in the CGF rewrite round: the micro-numeric
% protocol comparison of the controlled computational benefit subsection.
% Its main-text section pointers were rewritten as prose so that the
% supplement never references a main-text label.
\begin{table}[!t]
\centering
\small
\caption{Comparison of the three constitutive methods on the same protocol, split into
the quantities that belong to the material point and the quantities that belong to the
whole structural run.
\emph{Material point} (the main paper's material-point comparison, one uniaxial path
loaded across the transition and unloaded back to zero total strain):
$\Delta C_{\mathrm{tr}}$ is the yield-transition tangent change;
$a$ is the accumulated equivalent plastic strain at the zero-total-strain endpoint,
the history variable of the whole path rather than the multiplier of one
increment; and $\tau_{\mathrm{res}}=\tau_{11}$ is the signed axial Kirchhoff stress
component there, negative on this tension path.
\emph{Structural} (the run of the main paper's controlled cube-indentation run, counted in the
text below, where $N$ counts the load
increments of the schedule): every count is a whole-run sum over
every solved stage of the same run, whose load phase is a single increment that must cross
the whole transition, so \emph{Newton} means whole-run Newton iterations and
\emph{line search} means whole-run Armijo backtracking halvings, not line-search
evaluations; $e_{\mathrm{rel}}$ is the released-state error of that run against its own
$32$-increment reference. Arrows mark the preferred direction of the
cross-method quantities; rows without a preferred direction carry none. Bold marks
the smaller value on those same rows and nothing more: the rows with a single load
increment are the only schedules in which SECA's counts are the lower ones, the
text states the schedules where that ordering reverses, and the row that uses each
method's own reference carries neither an arrow nor bold.}
\label{tab:micro-numerics}
\setlength{\tabcolsep}{4pt}
\begin{tabular}{lrrr}
\toprule
Quantity & exact RM & Perzyna & SECA \\
\midrule
\multicolumn{4}{l}{\emph{Material point, single transition}}\\
$\Delta C_{\mathrm{tr}}$ (Pa) $\downarrow$ & \depTangJumpRM & \depTangJumpPerzyna & \textbf{\depTangJumpDEP} \\
$a$ (--) & \depMpAlphaRM & \depMpAlphaPerzyna & \depMpAlphaDEP \\
$\tau_{\mathrm{res}}$ (MPa) & \depMpResidualRM & \depMpResidualPerzyna & \depMpResidualDEP \\
\midrule
\multicolumn{4}{l}{\emph{Structural, controlled cube indentation, whole run ($N=1$)}}\\
Newton $\downarrow$ & \depNewtonRM & \depNewtonPerzyna & \textbf{\depNewtonDEP} \\
line search $\downarrow$ & \depLineSearchRM & \depLineSearchPerzyna & \textbf{\depLineSearchDEP} \\
$e_{\mathrm{rel}}(1)$ $\downarrow$ & \depCoarseErrRM & \depCoarseErrPerzyna & \depCoarseErrDEP \\
$e_{\mathrm{rel}}(32)$ (own reference) & \depFineErrRM & \depFineErrPerzyna & \depFineErrDEP \\
\bottomrule
\end{tabular}
\end{table}

\section{Records Moved from the Released-State Derivative Results}
\label{app:moves-derivative}

\paragraph{Complete-rerun difference against the bracketing-grid estimate.}
The derivative of the released observable is compared with an independent estimate taken from
the two bracketing grid points, that is, from a different discretisation of the same continuous
path. At the plastic control the complete-rerun difference at $h=5\times10^{-3}$ gives
$-0.033466$, $-0.033495$ and $-0.033544$ for the three methods, and the bracketing-grid estimate
gives $-0.033812$, $-0.033818$ and $-0.033959$: relative differences of $1.0\%$, $1.0\%$ and
$1.2\%$. At the near-onset control the two discretisations differ by $13.7\%$, $13.7\%$ and
$17.3\%$, because the bracketing interval spans a region in which the tangent changes rapidly
and the widest scanned step crosses the onset there. At the elastic control both estimates are
at the level of round-off ($7.6\times10^{-11}$ and $-1.1\times10^{-9}$ on a response of order
$10^{-12}$) and their ratio carries no information.

\paragraph{Rerun and subdivision rule for the response grid.}
A response row whose first attempt did not meet the residual tolerance was rerun with that
increment subdivided in place, which leaves the physical loading path and the total simulated
duration unchanged, so a marked row ends on a converged state sequence. Such a row is not used
as a fixed-partition comparison, which requires both sides to share one accepted partition. The
response grid carries $63$ rows, three methods at each of its $21$ grid points; nine of those
rows, on $4$ of the grid points, recorded one increment that did not meet the residual tolerance
and are marked with a dagger in the response table of the released-shape protocols of this
supplement, and two of the $27$ difference rows were formed from such a rerun and are not used
as fixed-partition comparisons.

\paragraph{Normalisation detail of the discrepancy values.}
The audit text spelled the two observables out: the discrepancy is reported as $q_{\mathrm{scene}}$
over all $6{,}002$ stored observation vertices and $q_{\mathrm{obs}}$ over the $1{,}444$-vertex
torso subset of those vertices. Both counts are also printed in the caption of the JVP table that
stays in the main text; the main text keeps the normalisation and the upper bound in words.

\paragraph{Linearised equilibrium residual bound.}
The audit text also stated that linearised equilibrium residuals stay below
$2.35\times10^{-7}$ relative on every unit; the same bound is printed in the caption of the JVP
table that stays in the main text, and it is repeated here so the moved sentence keeps its number.

\paragraph{Released-state response figure.}
The response sweep of the controlled cube indentation was drawn in the main text by the
following figure; it moves here with its caption unchanged. Its labels are kept so that no other
file needs editing.

\begin{figure*}[t]
\centering
\includegraphics[width=\textwidth]{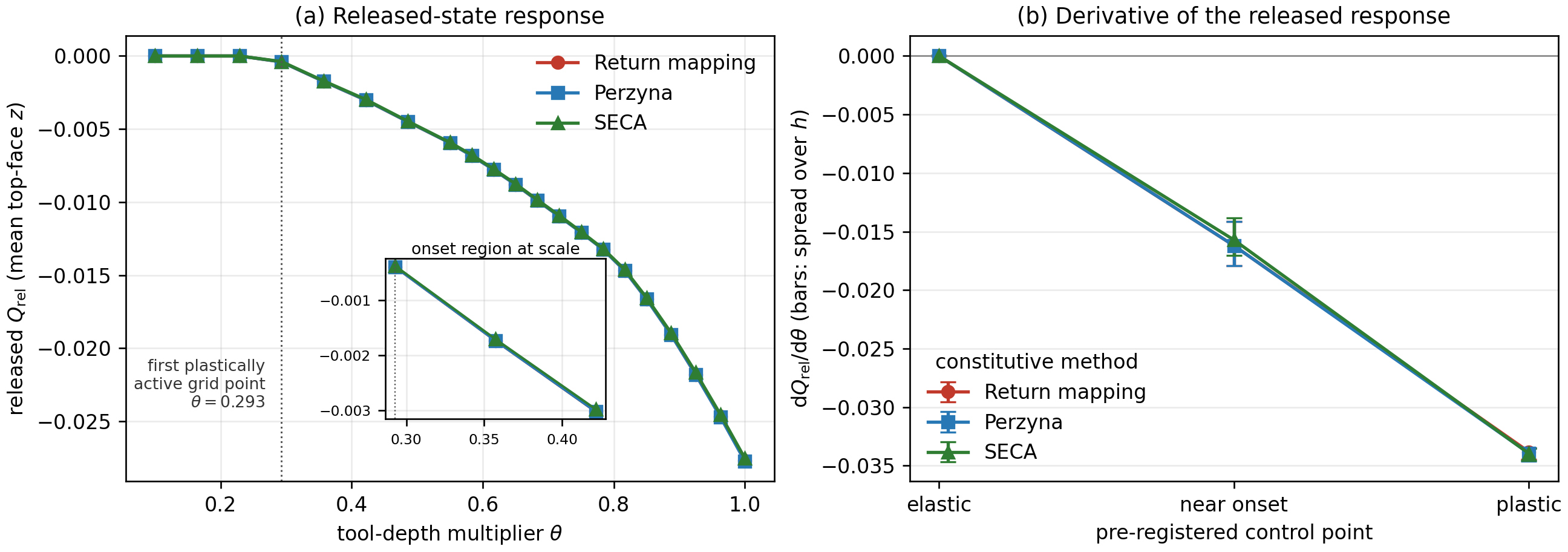}
\caption{Released-state sensitivity of the controlled cube indentation.
(a)~Mean released displacement over a fixed rest-space region of the indented
face against the tool-depth multiplier, over the pre-registered grid; the dotted
line marks the first plastically active grid point and the inset shows the onset
region at scale. (b)~Central-difference derivative of the same observable at the
three pre-registered control points, with bars spanning the three step sizes of
the scan; each difference is a complete load--hold--unload--release rerun, and the
widest step at the near-onset control crosses the onset, so that entry is a secant
across the transition rather than a derivative.}
\label{fig:released-sensitivity}
\end{figure*}

%% file: common/appendix_n1.tex
% Protocol, calibration ladder and full records of the cross-method error and cost
% comparison (N1).  This file is included by the supplement only; the main paper keeps the
% summary table and the error--cost figure.
\section{Protocol and Full Records of the Cross-Method Error and Cost Comparison}

The comparison in the main paper's controlled computational benefit results section asks a
narrow question with a protocol that was frozen before the first three-dimensional solve:
four constitutive updates --- exact return mapping, SECA, linear-overstress Perzyna and a
strict-elastic quartic-overstress Perzyna neighbour --- were run on two coarse scenes, cube
indentation and grip tension, at three loading-increment levels and, for the three
regularised updates, at three calibrated accuracy grades. All four updates enter through the
same constitutive interface of the same solver, which is what makes the comparison
controlled, but the local cost per evaluation is not assumed to be equal: exact return
mapping and linear-overstress Perzyna evaluate their multiplier in closed form, while the
compact activation and the quartic neighbour each solve an implicit scalar equation by a
$64$-step bisection on a bracketed interval. The fourth update is defined by its rate
function and its own viscosity,
\begin{equation}
r_{\mathrm{pow}}(f) = \sigma_\star\left(\frac{[f]_+}{\sigma_\star}\right)^{4},
\qquad \dot a = \frac{r_{\mathrm{pow}}(f)}{\eta_{\mathrm{pow}}},
\qquad \sigma_\star = \sigma_y = 2.5\times10^{6}\,\mathrm{Pa},
\end{equation}
with the implicit increment
\begin{equation}
\gamma = \frac{\Delta t}{\eta_{\mathrm{pow}}}\, r_{\mathrm{pow}}(f_{\mathrm{tr}} - B\gamma),
\qquad
\gamma \in \Bigl[0,\ \min\Bigl(\frac{f_{\mathrm{tr}}}{B},\
\frac{\Delta t}{\eta_{\mathrm{pow}}}\, r_{\mathrm{pow}}(f_{\mathrm{tr}})\Bigr)\Bigr],
\end{equation}
solved on that bracket and differentiated through the implicit-function theorem. It is strictly elastic below yield
and leaves the elastic branch with vanishing slope, but its rate is unbounded, so unlike the
compact activation it has no finite saturation interval. Its three calibrated viscosities are
$\eta_{\mathrm{pow}} = 3.96\times10^{-7}$, $6.28\times10^{-6}$ and $9.95\times10^{-5}\,\mathrm{Pa\,s}$;
the corresponding compact-activation widths are $\delta = 1250$, $2494$ and $4976\,\mathrm{Pa}$
at $\eta/\delta = 0.2\,\mathrm{s}$, and the linear-overstress viscosities are
$\eta_{\mathrm{lin}} = 396.2$, $790.6$ and $1577.4\,\mathrm{Pa\,s}$. The global cost
differences reported here therefore come from the number and shape of the Newton iterations
each update induces as well as from its own local work.
\subsection*{Measured object and observation sets}

The error of a run is the weighted relative deviation of its released configuration from
the reference configuration,
\begin{equation}
e_{\mathrm{rel}} = \frac{1}{L_0}\sqrt{\frac{\sum_i w_i \lVert x_i - x_i^{\mathrm{ref}}\rVert^2}{\sum_i w_i}},
\end{equation}
with uniform weights $w_i = 1$, observation nodes fixed in advance, and $L_0$ the largest
extent of the rest bounding box. No alignment, registration or post-hoc node selection is
performed: the nodes are compared in their material coordinates. Cube indentation observes
the indented interior surface nodes of the block; grip tension observes every node of the
specimen. The error is measured on the released state, after the full
load--hold--unload--release history, because that is the state the method is asked to
reproduce for an inverse or design problem.

The reference for each scene is the same scene solved with exact return mapping on a finer
loading partition of the same mesh. Refinement continues while the change between adjacent
levels stays above a threshold of $10^{-4}\,L_0$ fixed before the runs, and stops at a
pre-registered level cap; the levels, the measured adjacent differences and the decision are
recorded per scene in \texttt{data/n1\_scenes.json} in the submission package, and the level
used is named in the main paper. On the cube scene the adjacent difference falls below the
threshold within the cap and stops there. On the grip-tension scene the adjacent difference
has not fallen below the threshold when the cap is reached, so the finest computed level is
used as the reference and its remaining adjacent difference is quoted in the main paper as the
reference uncertainty of that scene: it is larger than the cube's, and still below the
smallest error budget of the comparison. The threshold was not relaxed after the runs, and
every computed level is kept in the record rather than discarded.

Cost is the wall time of one complete load--hold--unload--release solve on a single thread
in the same environment, measured as the median of three independent repeats of the whole
history; minimum and maximum of the repeats are recorded with the median. Timings from
earlier rounds that ran on unidentified hardware are not mixed into these numbers. Newton
iterations, line-search halvings and the number of increments whose iteration stopped without
reaching the residual tolerance are recorded per run as well, together with whether the run is
ranked in the comparison or kept as a diagnostic.

\subsection*{Calibration of the accuracy knobs}

Each regularised update has exactly one accuracy knob: the activation width $\delta$ of the
compact activation (with the relaxation ratio $\eta/\delta$ held at its frozen value), the
viscosity of the linear-overstress update, and the viscosity of the quartic-overstress
update (with its reference stress fixed at the yield stress and its exponent fixed at four).
The three knobs were calibrated on a material-point history before any three-dimensional
solve, using one common rule: starting from the shared reference flow scale of the frozen
evaluation point, each knob walks the same uniform downward ladder
$\mathrm{knob}_j = \mathrm{knob}_0 \cdot 10^{-0.1 j}$, and the grade is the candidate closest
to the target offset on a logarithmic measure. The calibration measure is the maximum
relative deviation of the accumulated plastic multiplier from the exact return-mapping
solution over the entire strain history, not the endpoint deviation: an endpoint measure is
not usable here, because a hold at constant strain relaxes the linear-overstress update back
to the yield surface, so its endpoint deviation falls to the level of solver noise for every
viscosity and would rank the grades arbitrarily. The ladder, the realised parameters of each
selected grade, and the material-point offsets are recorded in the accompanying data file;
the three grades used in the comparison are the same for all four updates.

\subsection*{Meaning of a missed increment}

An increment is marked missed when its Newton iteration stops without reaching the residual
tolerance: the iteration limit is reached, no descent direction is found, the line search
fails to produce a decrease, or the trial deformation gradient becomes near-singular. A missed
increment produces no accepted state under the commit rule of the local update. The comparison
driver does not roll such an increment back: it records the missed increment and continues the
history from the last iterate for diagnosis only, so the run still completes and still yields a
released configuration and a wall time, but the increments that follow --- including the
release --- start from a state the acceptance rule does not admit. Five cube records of this study contain one missed
increment each, and in every one of them it is the second of the ten recorded increments, the
second loading step, with a final relative residual between $8.0\times10^{-3}$ and
$1.2\times10^{-2}$ after $44$ to $60$ Newton iterations and $544$ to $884$ line-search
halvings; the eight later increments of those runs report convergence from that state. Those
five records are therefore diagnostics rather than released solutions of converged history:
they keep their rows in the tables marked $^{\dagger}$, carry no error value and no time to
solution, and are excluded from the error--cost figure of the main paper and from every
budget count, as that figure's caption states. The length of each run is preserved in the
record as a diagnostic quantity and is not quoted as a time to solution. Runs that could not
complete the history at all are kept with their refusal text and are likewise absent from the
figure. Table~\ref{tab:n1-diagnostics} locates the missed increment of each record.

\input{common/n1_diagnostics_table}

\input{common/n1_cost_table_full}

\subsection*{Grip-tension reference ladder and refused schedules}

The grip-tension reference is a rate-independent return-mapping solve on the same mesh at five
increment levels. Every level completed the full history without missing the residual tolerance
in any increment, and the
deformation gradient stayed well away from the guard at all of them
($\min\det\boldsymbol F \ge 0.6836$):

\begin{table}[!t]
\centering
\footnotesize
\setlength{\tabcolsep}{4pt}
\begin{tabular}{@{}rrrrr@{}}
\toprule
$n$ & steps & Newton & wall (s) & adjacent difference ($L_0$) \\
\midrule
32  & 70  & 429  & 384.04  & $7.973\times10^{-4}$ \\
64  & 140 & 537  & 469.13  & $4.206\times10^{-4}$ \\
128 & 280 & 1038 & 912.63  & $2.480\times10^{-4}$ \\
256 & 560 & 1983 & 1773.05 & $1.738\times10^{-4}$ \\
512 & 1120 & 3706 & 3224.49 & --- \\
\bottomrule
\end{tabular}
\caption{The grip-tension reference ladder. The adjacent difference is the weighted
release-shape difference between two neighbouring levels, computed on the same observation set
as the error of the comparison. All five levels are kept; the threshold of $10^{-4}\,L_0$ is
not reached before the pre-registered level cap, so the $512$-increment level is the reference
and its remaining difference, $1.738\times10^{-4}\,L_0$, is that scene's reference
uncertainty.}
\label{tab:n1-tension-reference}
\end{table}

The scene is larger than the cube one ($L_0 = 0.96$ against a unit cube, $1250$ observed
nodes) and the tensile history is the one the solver finds hardest: at one and at four
increments the update is refused outright by the positive-determinant guard of the
finite-strain material update,
\texttt{ValueError: F and Fp must have positive determinant}, both for the reference and for
the regularised updates. Each refusal is stored as a record with its own error text and no
error value rather than being retried with different settings, so the grip-tension panels of
the figure and the table carry no point at those two increment counts, and the refusal count
is reported next to the completed runs. The runs at sixteen increments use the same three
repeats and the same pre-registered error definition, and are entered in the full table above
as they complete; the record file is the authority for which of them exist.

%% file: common/n1_diagnostics_table.tex
% generated by SECA_CL2.0/audit/cgf_n1_report.py
\begin{table}[!t]
\centering
\footnotesize
\setlength{\tabcolsep}{4pt}
\begin{tabular}{@{}llrrrrr@{}}
\toprule
Method & grade & increment & phase & residual & Newton & halvings \\
\midrule
SECA & 5.00e-04 & 1 of 10 (load) & 1.16e-02 & 60 & 745 \\
SECA & 2.00e-03 & 1 of 10 (load) & 8.01e-03 & 59 & 884 \\
Perzyna, linear & 1.00e-03 & 1 of 10 (load) & 8.02e-03 & 54 & 615 \\
Perzyna, linear & 2.00e-03 & 1 of 10 (load) & 8.03e-03 & 44 & 544 \\
Perzyna, quartic & 1.00e-03 & 1 of 10 (load) & 7.97e-03 & 60 & 748 \\
\bottomrule
\end{tabular}
\caption{Every \emph{Cube indentation} record whose committed history contains an increment that did not reach the residual tolerance. The increment column gives the position of the first such increment in the recorded step sequence and its phase, the residual is the final relative residual of that increment's Newton iteration, and Newton and halvings are that increment's own counts. The increments after it converged, but they started from the iterate it left behind, which is why these records are diagnostics and not released solutions of the acceptance rule.}
\label{tab:n1-diagnostics}
\end{table}

%% file: common/n1_cost_table_full.tex
% generated by SECA_CL2.0/audit/cgf_n1_report.py
\begin{table}[!t]
\centering
\footnotesize
\setlength{\tabcolsep}{3.5pt}
\begin{tabular}{@{}llrrrrrl@{}}
\toprule
Method & grade & $n$ & $e_{\mathrm{rel}}$ (\%) & wall (s) & Newton & missed & status \\
\midrule
\multicolumn{8}{l}{\emph{Cube indentation}} \\
Return mapping (ref.) & --- & 1 & 0.1700 & 6.30 & 36 & 0 & ranked \\
Return mapping (ref.) & --- & 4 & 0.0669 & 8.46 & 55 & 0 & ranked \\
Return mapping (ref.) & --- & 16 & 0.0144 & 27.64 & 170 & 0 & ranked \\
Return mapping (ref.) & --- & 32 & --- & --- & 244 & 0 & diagnostic \\
Return mapping (ref.) & --- & 64 & --- & --- & 411 & 0 & diagnostic \\
SECA & 5.00e-04 & 1 & 0.1699 & 3.80 & 22 & 0 & ranked \\
SECA & 1.00e-03 & 1 & 0.1698 & 4.13 & 24 & 0 & ranked \\
SECA & 2.00e-03 & 1 & 0.1696 & 6.50 & 33 & 0 & ranked \\
SECA & 5.00e-04 & 4 & --- & --- & 96 & 1 & diagnostic \\
SECA & 1.00e-03 & 4 & 0.0668 & 10.01 & 56 & 0 & ranked \\
SECA & 2.00e-03 & 4 & --- & --- & 99 & 1 & diagnostic \\
SECA & 5.00e-04 & 16 & 0.0144 & 32.17 & 176 & 0 & ranked \\
SECA & 1.00e-03 & 16 & 0.0144 & 32.74 & 176 & 0 & ranked \\
SECA & 2.00e-03 & 16 & 0.0143 & 32.02 & 176 & 0 & ranked \\
Perzyna, linear & 5.00e-04 & 1 & 0.1700 & 4.56 & 28 & 0 & ranked \\
Perzyna, linear & 1.00e-03 & 1 & 0.1700 & 3.24 & 21 & 0 & ranked \\
Perzyna, linear & 2.00e-03 & 1 & 0.1699 & 3.47 & 23 & 0 & ranked \\
Perzyna, linear & 5.00e-04 & 4 & 0.0669 & 9.48 & 61 & 0 & ranked \\
Perzyna, linear & 1.00e-03 & 4 & --- & --- & 94 & 1 & diagnostic \\
Perzyna, linear & 2.00e-03 & 4 & --- & --- & 85 & 1 & diagnostic \\
Perzyna, linear & 5.00e-04 & 16 & 0.0144 & 27.70 & 172 & 0 & ranked \\
Perzyna, linear & 1.00e-03 & 16 & 0.0144 & 27.82 & 174 & 0 & ranked \\
Perzyna, linear & 2.00e-03 & 16 & 0.0144 & 27.70 & 174 & 0 & ranked \\
Perzyna, quartic & 5.00e-04 & 1 & 0.1698 & 4.03 & 25 & 0 & ranked \\
Perzyna, quartic & 1.00e-03 & 1 & 0.1696 & 4.10 & 24 & 0 & ranked \\
Perzyna, quartic & 2.00e-03 & 1 & 0.1692 & 4.48 & 26 & 0 & ranked \\
Perzyna, quartic & 5.00e-04 & 4 & 0.0668 & 9.49 & 58 & 0 & ranked \\
Perzyna, quartic & 1.00e-03 & 4 & --- & --- & 97 & 1 & diagnostic \\
Perzyna, quartic & 2.00e-03 & 4 & 0.0661 & 10.07 & 58 & 0 & ranked \\
Perzyna, quartic & 5.00e-04 & 16 & 0.0144 & 31.41 & 176 & 0 & ranked \\
Perzyna, quartic & 1.00e-03 & 16 & 0.0142 & 31.05 & 177 & 0 & ranked \\
Perzyna, quartic & 2.00e-03 & 16 & 0.0140 & 31.68 & 177 & 0 & ranked \\
\midrule
\multicolumn{8}{l}{\emph{Grip tension}} \\
Return mapping (ref.) & --- & 16 & 0.3153 & 150.56 & 167 & 0 & ranked \\
Return mapping (ref.) & --- & 32 & --- & --- & 429 & 0 & diagnostic \\
Return mapping (ref.) & --- & 64 & --- & --- & 537 & 0 & diagnostic \\
Return mapping (ref.) & --- & 128 & --- & --- & 1038 & 0 & diagnostic \\
Return mapping (ref.) & --- & 256 & --- & --- & 1983 & 0 & diagnostic \\
Return mapping (ref.) & --- & 512 & --- & --- & 3706 & 0 & diagnostic \\
SECA & 5.00e-04 & 16 & 0.3141 & 157.83 & 169 & 0 & ranked \\
SECA & 1.00e-03 & 16 & 0.3129 & 159.99 & 172 & 0 & ranked \\
SECA & 2.00e-03 & 16 & 0.3104 & 195.25 & 199 & 0 & ranked \\
Perzyna, linear & 5.00e-04 & 16 & 0.3133 & 161.29 & 176 & 0 & ranked \\
Perzyna, linear & 1.00e-03 & 16 & 0.3113 & 163.06 & 177 & 0 & ranked \\
Perzyna, linear & 2.00e-03 & 16 & 0.3074 & 157.00 & 173 & 0 & ranked \\
Perzyna, quartic & 5.00e-04 & 16 & 0.3149 & 158.99 & 172 & 0 & ranked \\
Perzyna, quartic & 1.00e-03 & 16 & 0.3143 & 174.87 & 183 & 0 & ranked \\
Perzyna, quartic & 2.00e-03 & 16 & 0.3129 & 145.42 & 159 & 0 & ranked \\
\bottomrule
\end{tabular}
\caption{Every recorded setting of the cross-method comparison: both scenes, all three calibrated grades, all increment levels and every repeat-median cost, including the reference levels that carry no error value of their own because they define it. Return mapping is the rate-independent reference model; \texttt{---} marks its single uncalibrated series. \emph{Ranked} records commit only increments that reached the residual tolerance and are the records the comparison uses; \emph{diagnostic} records advanced from the last iterate of an increment that missed it, so no error and no time to solution are quoted for them. The measured run lengths and the located increments are in the diagnostics table.}
\label{tab:n1-cost-full}
\end{table}

%% file: common/appendix_n2.tex
% Protocol and full records of the minimal control task (N2).  Included by the supplement
% only; the main paper keeps the summary of the derivative cross-check and of what the
% control task does and does not support.
\section{Protocol and Records of the Minimal Control Task}

The main paper reports two questions about the released-state path derivative of the SECA
chain: whether differentiating the stored history reproduces what a full re-solve would give,
and whether that derivative can drive a small contact control problem. This section states
the measured object, the partition on which every run of the study was performed, the frozen
targets, and every evaluation that did not complete, including the ones that failed for
reasons intrinsic to the derivative rather than to the solver.
\subsection*{Scene, control and measured quantity}

The scene is the coarse symmetric Bunny indentation used elsewhere in this work: two tools
press the specimen from opposite sides along $z$. The control $\theta$ is the common $z$
amplitude of both tools, so the control vector is $[0,0,\theta,0,0,\theta]$ and the direction
used for the derivative is $[0,0,1,0,0,1]$; the chain rule from the amplitude to the tool
motion is therefore exact and no interpolation of the control enters the derivative. The
measured quantity is the released gauge compression of the specimen, and the secondary
recorded quantity is the released shape loss. Both are evaluated on the full
load--hold--unload--release history. The objective used by the control task is
\begin{equation}
L(\theta) = \frac{1}{2}\left(\frac{Q(\theta) - Q^\star}{L_0}\right)^2 ,
\end{equation}
with $Q$ the released gauge compression, $Q^\star$ a target value and $L_0$ the rest extent,
so the derivative of the objective is the path derivative of the released state multiplied by
a scalar. The path derivative itself is a full-history Jacobian-vector product along the
stored states; no perturbed forward solve is called to obtain it.

\subsection*{One common partition for the whole band}

The derivative is defined on a fixed partition of the loading history, and a partition is
only usable where the solver accepts every interval of it. The retained partition of the
frozen history contains 26 intervals. With the iteration cap used when that partition was
constructed, only the control $\theta = 0.85$ completed; every neighbouring control stopped
with \texttt{predictor\_path\_}\allowbreak\texttt{requires\_substep:}\allowbreak%
\texttt{relative\_ccd}, because the amplitude change
moves contact onset into an interval that the fixed partition may not subdivide. Raising the
iteration cap, and not the tolerance, makes all five exploration controls complete on the
same 26 intervals, so the study uses one common partition for every run and no control is
given a partition of its own. The accepted band is therefore a measured set:
\begin{center}
\begin{tabular}{@{}lrrr@{}}
\hline
$\theta$ & released gauge compression & released shape loss & path derivative \\
\hline
0.84150 & 0.123544 & $1.6209\times10^{-4}$ & 0.424883 \\
0.84745 & 0.126076 & $1.7126\times10^{-4}$ & 0.426395 \\
0.85000 & 0.127165 & $1.7533\times10^{-4}$ & 0.427458 \\
0.85255 & 0.128256 & $1.7949\times10^{-4}$ & 0.428086 \\
0.85850 & 0.130808 & $1.8956\times10^{-4}$ & 0.429826 \\
\hline
\end{tabular}
\end{center}
Two attempts to widen the band by refining the partition were made and failed, and both are
kept in the record: a uniform doubling of every interval of the retained partition accepted
$\theta = 0.85$ only and additionally produced a Newton iteration-limit failure at
$\theta = 0.70$, and a local refinement of the intervals around contact onset accepted
$\theta = 0.85$ only. The band is also not symmetric about the control used to build the
partition, which is a limitation of this study rather than a property of the method.

\subsection*{Cross-check of the derivative against full re-solves}

For two controls inside the band, the path derivative was compared with a central difference
of the complete measured quantity, using full re-solves at $\theta \pm h$ for four step
sizes $h = \{8.5\times10^{-3}, 2.55\times10^{-3}, 8.5\times10^{-4}, 2.55\times10^{-4}\}$.
All perturbed runs completed on the common partition (26 intervals). The result is reported
in full for both controls:
\begin{center}
\begin{tabular}{@{}lrrrr@{}}
\hline
$\theta$ & $h$ & central difference & path derivative & relative difference \\
\hline
0.85000 & $8.5\times10^{-3}$  & 0.427281 & 0.427458 & $4.13\times10^{-4}$ \\
0.85000 & $2.55\times10^{-3}$ & 0.427318 & 0.427458 & $3.27\times10^{-4}$ \\
0.85000 & $8.5\times10^{-4}$  & 0.427368 & 0.427458 & $2.11\times10^{-4}$ \\
0.85000 & $2.55\times10^{-4}$ & 0.427406 & 0.427458 & $1.22\times10^{-4}$ \\
0.85255 & $8.5\times10^{-3}$  & 0.428033 & 0.428086 & $1.22\times10^{-4}$ \\
0.85255 & $2.55\times10^{-3}$ & 0.428115 & 0.428086 & $6.85\times10^{-5}$ \\
0.85255 & $8.5\times10^{-4}$  & 0.428087 & 0.428086 & $2.68\times10^{-6}$ \\
0.85255 & $2.55\times10^{-4}$ & 0.428119 & 0.428086 & $7.80\times10^{-5}$ \\
\hline
\end{tabular}
\end{center}
At $\theta = 0.85$ the difference decreases with the step size, as a truncation error of the
difference quotient should, from $4.13\times10^{-4}$ at the widest step to
$1.22\times10^{-4}$ at the finest. At $\theta = 0.85255$ it is at most
$1.22\times10^{-4}$ at every step size, the same level as the finest step at the first
control, so the derivative agrees with full re-solves at both sampled controls. An earlier
evaluation of this same check left a difference of about $4.5\times10^{-3}$ at the second
control that did not decrease with the step size; that difference came from the contact term
of the derivative being assembled at the control that built the partition instead of at the
control of the run, and the derivative was corrected. The pre-correction and the corrected
derivative are both kept in the record, per control.

\subsection*{Controls outside the measured band}

A control below the contact-active range was evaluated as well. Its forward history completes
on the common partition (released gauge compression $0.004186$, 26 intervals) and its path
derivative is available, $0.064968$. Before the correction described above, the exact
linearisation of the contact constraint was evaluated outside the positive-gap domain in
which it is defined and raised
\texttt{ValueError: contact tangent outside positive-gap domain: -5.951188168e-04}; that
record is kept with its own clock and message. A second control below the band,
$\theta = 0.55$, does not complete its forward history on the common partition at all: it
stops in the loading phase at interval 12 of 26 and records a
\texttt{predictor\_path\_}\allowbreak\texttt{requires\_substep:}\allowbreak\texttt{relative\_ccd}
request for a subdivision that the fixed partition does not perform, so it has neither a
released value nor a derivative. These two records bound the band in which one fixed
partition is usable rather than the existence of the derivative in general: the control task
treats a control with no released value or no derivative as infeasible instead of aborting,
and a sub-yield control is not a valid starting point for a task whose targets are plastic
released values.

\subsection*{Frozen settings of the control task}

The task is solved with a bounded quasi-Newton method on $\theta \in [0.8415, 0.8585]$, from
three starting controls ($0.84745$, $0.85$, $0.85255$) towards two targets frozen before the
first optimisation run: $Q^\star = 0.126076$ (the value at $\theta = 0.84745$) and
$Q^\star = 0.128256$ (the value at $\theta = 0.85255$). Each curve is limited to 12 outer
iterations and 14 evaluations, with a function tolerance of $10^{-14}$, a gradient tolerance
of $10^{-8}$ and at most 20 line-search steps; controls whose derivative is unavailable are
returned to the optimiser with an infeasible objective value. The limits and the stopping
reason of every curve are recorded. No other constitutive update of the comparison exposes a
path derivative through the same interface, so this task measures whether the released-state
derivative of the SECA chain is usable and how it behaves; it is not a comparison of
optimisation cost or of convergence speed against another method, and no such comparison is
claimed.

\subsection*{Recorded curves}

Every curve of the task is listed with the evaluations it made, in the order it made them
(table~\ref{tab:n2-curves} and table~\ref{tab:n2-curve-path}). The curves are the record of
one bounded search each on the common fixed partition, and the tables state where each search
started, how many complete histories it spent and what the last evaluation returned.

\input{common/n2_curve_table}

%% file: common/n2_curve_table.tex
% Generated by SECA_CL2.0/audit/cgf_n2_report.py from the recorded control curves.
\begin{table}[!t]
  \centering
  \footnotesize
  \setlength{\tabcolsep}{4pt}
  \begin{tabular}{@{}llrrrrr@{}}
    \toprule
    Target & Start $\theta$ & Iter. & Eval. & Loss at start & $|\Delta Q|/L_0$ at start & Wall (s) \\
    \midrule
    high & 0.84745 & 3 & 4 & $7.35\times10^{-7}$ & $1.21\times10^{-3}$ & 536.5 \\
     & 0.85000 & 3 & 4 & $1.84\times10^{-7}$ & $6.07\times10^{-4}$ & 268.5 \\
     & 0.85255 & 0 & 1 & 0 & 0 & 0.0 \\
    low & 0.84745 & 0 & 1 & 0 & 0 & 0.0 \\
     & 0.85000 & 3 & 4 & $1.83\times10^{-7}$ & $6.06\times10^{-4}$ & 265.0 \\
     & 0.85255 & 3 & 4 & $7.35\times10^{-7}$ & $1.21\times10^{-3}$ & 533.4 \\
    \bottomrule
  \end{tabular}
  \caption{Recorded control curves of the minimal control task. Each curve is a bounded
    search on the common fixed partition, from the listed start towards the target
    $Q^{\star}$ (\emph{low}: the value measured at $\theta = 0.84745$; \emph{high}: the
    value measured at $\theta = 0.85255$). The two middle columns report the objective
    value and the deviation $|\Delta Q|/L_0$ at the first evaluation of that curve, which
    is where the search starts; the targets are values of this scene's own measured
    response. Evaluated controls are cached to $10^{-6}$ in $\theta$, so the last
    evaluation of a curve returns the measured value of a control already evaluated, and
    the loss recorded at the last evaluation of every curve is zero in double precision.
    Two of the six curves start at their own target: they are consistency checks and they
    return loss zero after a single evaluation; the other four are non-trivial
    calibrations, each needing three iterations and four complete histories.
    Every curve stopped with \texttt{CONVERGENCE: NORM OF PROJECTED GRADIENT <= PGTOL}
    and none of them reached its iteration or evaluation limit.}
  \label{tab:n2-curves}
\end{table}

\begin{table}[!t]
  \centering
  \footnotesize
  \setlength{\tabcolsep}{4pt}
  \begin{tabular}{@{}llrrrr@{}}
    \toprule
    Target & Start $\theta$ & Eval. & $\theta$ & $Q$ & Loss \\
    \midrule
    high & 0.84745 & 1 & 0.847450 & 0.126076 & $7.35\times10^{-7}$ \\
     &  & 2 & 0.847738 & 0.126199 & $6.55\times10^{-7}$ \\
     &  & 3 & 0.852584 & 0.128270 & $3.33\times10^{-11}$ \\
     &  & 4 & 0.852550 & 0.128256 & 0 \\
    \midrule
    high & 0.85000 & 1 & 0.850000 & 0.127165 & $1.84\times10^{-7}$ \\
     &  & 2 & 0.850144 & 0.127227 & $1.64\times10^{-7}$ \\
     &  & 3 & 0.852538 & 0.128250 & $4.36\times10^{-12}$ \\
     &  & 4 & 0.852550 & 0.128256 & 0 \\
    \midrule
    high & 0.85255 & 1 & 0.852550 & 0.128256 & 0 \\
    \midrule
    low & 0.84745 & 1 & 0.847450 & 0.126076 & 0 \\
    \midrule
    low & 0.85000 & 1 & 0.850000 & 0.127165 & $1.83\times10^{-7}$ \\
     &  & 2 & 0.849856 & 0.127103 & $1.63\times10^{-7}$ \\
     &  & 3 & 0.847458 & 0.126080 & $1.86\times10^{-12}$ \\
     &  & 4 & 0.847450 & 0.126076 & 0 \\
    \midrule
    low & 0.85255 & 1 & 0.852550 & 0.128256 & $7.35\times10^{-7}$ \\
     &  & 2 & 0.852261 & 0.128132 & $6.54\times10^{-7}$ \\
     &  & 3 & 0.847480 & 0.126089 & $2.53\times10^{-11}$ \\
     &  & 4 & 0.847450 & 0.126076 & 0 \\
    \bottomrule
  \end{tabular}
  \caption{Every evaluation of every recorded control curve, in the order the
    optimiser made them. Each evaluation is one complete load--hold--unload--release
    history on the common fixed partition followed by one path derivative; every
    evaluation listed here completed and returned a derivative.}
  \label{tab:n2-curve-path}
\end{table}

%% file: common/appendix_selection.tex
\section{Bounded A1/A2 Error-Selection Analysis}
\label{app:selection}

\subsection{Scope and Prior-Art Boundary}
This appendix preserves the historical selector study. It is excluded from
the main contribution because
goal-oriented constitutive/discretisation error control and
tolerance-driven hierarchical selection already have direct prior art
\citep{LarssonRunesson2004}. The available publisher preview describes a
hierarchical finite-strain hyperelasto-Perzyna-viscoplastic setting,
localisation by time interval, and adaptive selection under a prescribed
goal tolerance. We do not infer unavailable equation-level details,
candidate ordering, stopping rules, or monotonicity assumptions.

The frozen studies below establish only an auditable implementation result.
They do not establish a deterministic or statistical error bound, a new
selector class, lower practical cost, or SECA superiority. The outer choice
$\boldsymbol\Lambda=(\mathrm{law},\delta,\eta,N)$ is immutable along the accepted
equilibrium-state sequence of each differentiated run; no derivative crosses a
candidate-index, integration-resolution,
fallback, or adaptive-partition change. The fixed-parameter local multiplier
product is Equation~\ref{eq:local-jvp}; it is not repeated here.

\subsection{Exhaustive Pilot}
The calibration set contains monotone tension, monotone compression, shear,
load--unload, cyclic hold/reversal, and one non-proportional path. A separate
twelve-path held-out manifest was frozen with seed 20260917. The pilot
enumerates
\begin{equation}
\delta/\sigma_y\in\{0.00625,0.025,0.10\},\qquad
\eta/(\sigma_y T)\in\{0.001,0.005,0.020\},\qquad
N\in\{64,128\}.
\end{equation}
For candidate state $\boldsymbol z_k^D$, one same-state exact-$J_2$ update
defines the local constitutive defect
\begin{equation}
\boldsymbol d_k=
\Phi_J(\boldsymbol\epsilon_{k+1},\boldsymbol z_k^D)
-\Phi_D(\boldsymbol\epsilon_{k+1},\boldsymbol z_k^D),
\end{equation}
which is transported through five symmetric-deviatoric coordinates and the
scalar history. A separate $N$ versus $2N$ integration supplies a
step-doubling indicator. Calibration-only maximum-ratio factors 1.33823 for
stress and 1.33640 for residual state multiply the summed indicators. They
are finite-sample calibrations, not bounds.

The joint budgets require normalised stress-trajectory and terminal plastic
state errors to remain below 0.5\%, 1\%, or 2\%. At 0.5\%, eleven of twelve
paths are predicted feasible and one invokes the declared least-violation
fallback; all twelve returned state sequences pass independent replay. At 1\%
and 2\%, all twelve are predicted feasible and pass. The maximum
selected-result actual-to-predicted ratio is 0.816. An inherited fixed
setting passes 0/12, 0/12, and 9/12 at the three budgets.

The cost taxonomy counts selection separately from the accepted
equilibrium-state sequence it selects and the independent audit. For one
held-out path, the finite-difference
exhaustive selector performs 1,728 exact-$J_2$ updates, 5,184 candidate
updates, and 29,376 total selector-local update calls after state-Jacobian
columns and doubled integrations are included. Reporting only the final
selected accepted equilibrium-state sequence would therefore be misleading.

\subsection{Analytic Products and Reduced Search}
Seed 20260918 freezes a second twelve-path set. The six-by-six symmetric
algorithmic stress tangent passes the declared 601-point continuity and
finite-difference-resolution checks at every width, but the narrowest
transition has about sixteen times the peak variation rate of the widest.
The best centred-difference relative discrepancy of the analytic local
product is $1.2391\times10^{-9}$, and JVP--VJP duality closes to reported
precision. These are fixed-$\boldsymbol\Lambda$ checks, not global smoothness of the
selector--simulator composition.

\begin{table*}[t]
\centering
\small
\caption{Frozen twelve-path results under joint 1\% stress and residual-state
budgets. ``Selector'' counts estimation and search calls; ``complete'' also
includes the selected accepted equilibrium-state sequence and explicit offline
exact-$J_2$ audit.
Shared calibration is not charged per path.}
\label{tab:selection-history}
\setlength{\tabcolsep}{4pt}
\begin{tabular}{lrrrrrr}
\toprule
Route & Pass & Mean $e_\sigma$ & Mean $e_p$ & Candidates &
Selector calls & Complete calls\\
\midrule
Inherited fixed & 0/12 & 1.6385\% & 0.6521\% & 1 & 512 & 1024\\
Fixed accuracy-fine & 12/12 & 0.2221\% & 0.0605\% & 1 & 512 & 1024\\
Fixed smooth & 12/12 & 0.7657\% & 0.2932\% & 1 & 512 & 1024\\
Exhaustive analytic & 12/12 & 0.4876\% & 0.1932\% & 18 & 6912 & 7360\\
Reduced monotonic & 12/12 & 0.4056\% & 0.1485\% & 5 & 2048 & 2560\\
Linear Perzyna selected & 12/12 & 0.3240\% & 0.1114\% & 5 & 1536 & 1984\\
\bottomrule
\end{tabular}
\end{table*}

The reduced search lowers median selector work from 6912 to 2048 local
updates relative to the analytic exhaustive policy, but both passing fixed
controls use only 512. The selected linear-Perzyna control also passes 12/12
with lower mean errors and lower selector work than the reduced SECA route.
Bounded fallback is invoked on 3/12 new paths and 3/12 regression paths when
candidate ordering fails. Across all candidates, maximum
actual-to-predicted ratios are 1.0401 for stress and 1.0065 for residual
state; approximately 0.34\% of each ratio population is underestimated.

\subsection{Second Released Structural Quantity}
The transfer protocol is a 48-tetrahedron small-strain block with a
bilateral finite-area top grip, not unilateral contact or a character scene.
It uses 11 load stages, four peak holds, ten grip-removal stages, and five
released holds. The setting
$(\delta/\sigma_y,\eta/(\sigma_yT),N)=(0.025,0.001,128)$ is not retuned.

The calibrated prediction is $4.9353\times10^{-5}L_0$ for released
full-surface RMS error, versus an independent audit of
$2.5835\times10^{-5}L_0$. The signed top-gauge magnitude is predicted as
$6.4747\times10^{-5}L_0$ and audited as
$2.9729\times10^{-5}L_0$, with matching sign. Prediction work comprises
1440 local $J_2$ queries, 2880 analytic JVPs, and 1650
equilibrium-residual calls. The factors come from one earlier clamp and are
tested on one frozen second clamp. This is an empirical goal-oriented
demonstration, not a structural bound.

%% file: common/appendix_contact_diagnostics.tex
\section{Historical Contact and Derivative Diagnostics}
\label{app:contact-diagnostics}
This appendix preserves historical diagnostics that are not evidence for the
default barrier/CCD character method. In particular, vertex-penalty states
and derivatives retain their original identity and are never relabeled as
feasible-contact results.

\subsection{Plane-Block and Component Checks}
The preliminary contact scene is a 27-node, 48-tetrahedron small-strain block
under an analytic rigid plane. With plane height $h$ and current vertex
coordinate $z_i$, its gap is $g_i=h-z_i$. For $0<g_i<\hat d$ the barrier is
\begin{equation}
b(g_i)=-(g_i-\hat d)^2\log(g_i/\hat d),\qquad
b(g_i)=0\quad\text{for }g_i\geq\hat d,
\end{equation}
and evaluation is rejected at $g_i\leq0$. Elastic, exact-$J_2$, and SECA
routes complete all 49 load--remove stages. Their minimum triangle gaps are
0.04874, 0.05390, and 0.05388; maximum relative free residuals are
$1.94\times10^{-7}$, $1.92\times10^{-7}$, and
$1.15\times10^{-7}$; and minimum determinants are 0.8474, 0.8736, and
0.8736. The largest contact/support reaction imbalance is
$1.22\times10^{-6}$ in recorded model-force units. This is a special
plane-block validation, not character-scale evidence.

A separate IPC Toolkit 1.6.0 smoke test constructs two triangle surfaces,
uses the official normal-collision barrier, and asks CCD for a proposal that
crosses the opposing triangle. The full step is not collision free; the
maximum returned step is 0.49945, and a 0.99-guarded step is collision free.
This checks the binding and API only.

\subsection{Fixed-Support Failure Diagnosis}
An archived Bunny run placed six upper-tool active faces on fully fixed
boundary triangles at peak. Their $10^{-5}L_0$ gap was positive but inside
the barrier support $\hat d=0.04L_0$, producing a support-dominated upper
reaction. The corrected primary path increases only the upper tool's outward
rest offset by $0.045566436L_0$, the minimum correction satisfying
\begin{equation}
g_{\mathrm{fixed,peak}}\geq\hat d+10^{-5}L_0=0.04001L_0
\end{equation}
for every fully fixed face. Material parameters, support nodes, sphere
radius, control amplitude, CCD tests, timing, and trial/commit semantics are
unchanged.

\begin{table*}[t]
\centering
\small
\caption{Archived fixed-support failure and corrected support-clear path.
Compression uses the fixed rest-space torso gauge. The failure is retained
only as diagnosis and is not character evidence.}
\label{tab:support-diagnosis}
\setlength{\tabcolsep}{4pt}
\begin{tabular}{lrr}
\toprule
Quantity & Archived failure & Support-clear path\\
\midrule
Peak reactions ($10^6$ model-force units) & $1.111/673.558$ & $1.00456/1.00450$\\
Reaction ratio & 606.009 & 1.00006\\
Fixed active faces & 6 & 0\\
Peak/release/hold (\%) & 26.963/18.361/18.361 & 21.294/12.716/12.716\\
Recovery (percentage points) & 8.602 & 8.577\\
Minimum gap ($L_0$) & $1.000\times10^{-5}$ & 0.021513\\
Runtime (s) & 643.1 & 208.5\\
\bottomrule
\end{tabular}
\end{table*}

\begin{figure*}[t]
\centering
\includegraphics[width=\textwidth]{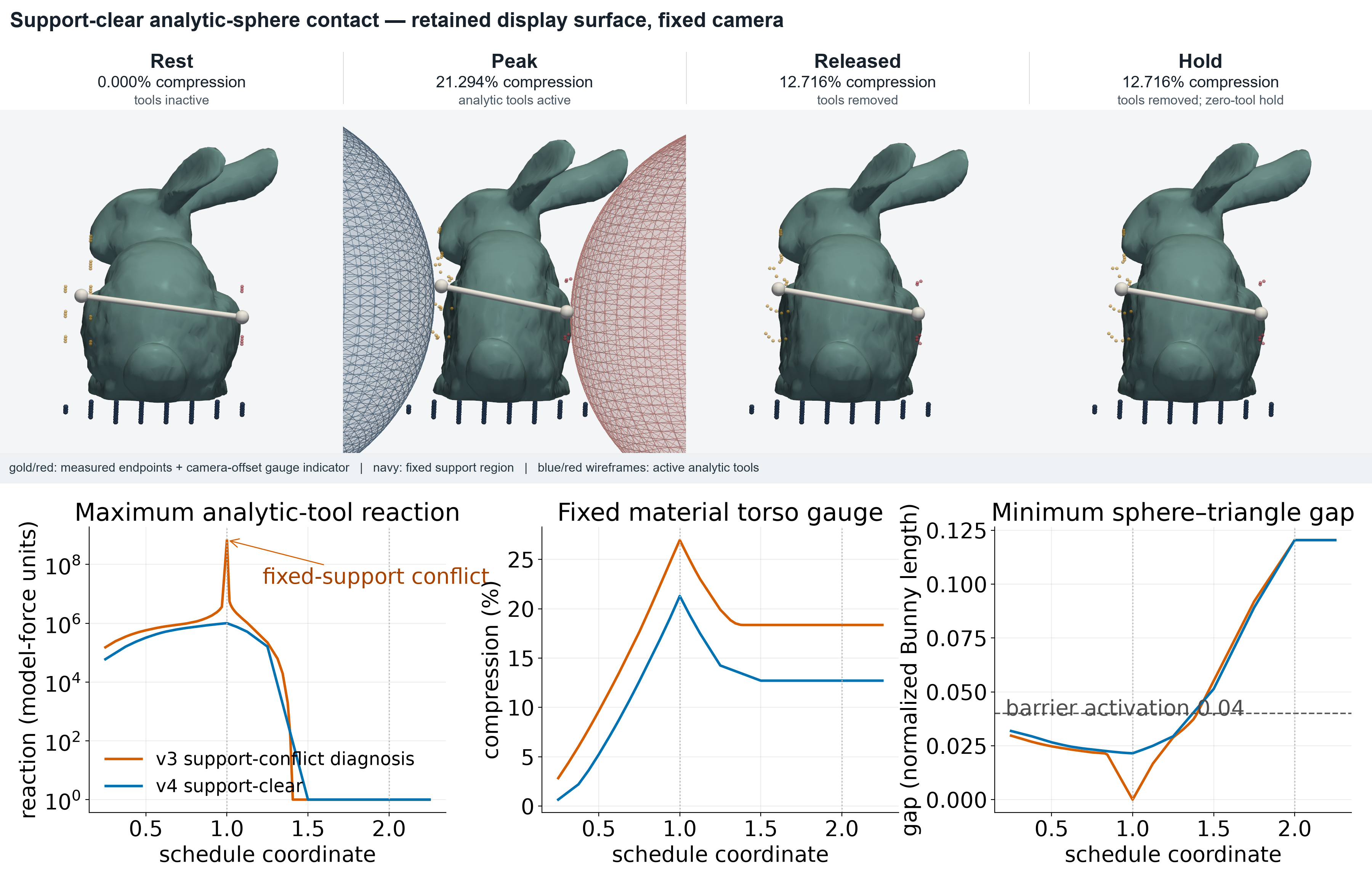}
\caption{Archived support-contact diagnosis and corrected path. The top row
shows the actual corrected rest, peak, release, and hold geometry; the lower
plots retain both histories so that the fixed-face failure is not hidden.}
\Description{Solved Bunny states and plots of reaction, torso compression,
and minimum gap compare the failed and support-clear paths.}
\label{fig:support-diagnosis}
\end{figure*}

The corrected run's forces equal $0.401825F_0$ and $0.401800F_0$, where
$F_0=\sigma_0L_0^2$; no metre assignment for $L_0$ is assumed. Moving the
tool out of impossible support contact lowers peak compression by a
disclosed 5.669 percentage points. The corrected geometry was not tuned to
reproduce the failed deformation.

\subsection{Vertex-Penalty Contact and Time-Step Diagnosis}
The archived penalty protocol uses the same Bunny volume mesh but smooth
vertex--sphere forces rather than the physical-triangle barrier. Its forward
sweep is retained only as response context:
\begin{table}[t]
\caption{Archived vertex-penalty control sweep. ``Released'' denotes the
immediate zero-tool-force stage, not a long-time stationary state.}
\label{tab:penalty-authoring}
\centering
\begin{tabular}{rrrr}
\toprule
Tool amplitude & Peak & Released & Recovery\\
\midrule
0.52 & 7.31\% & 5.01\% & 2.30\\
0.70 & 12.73\% & 10.03\% & 2.70\\
0.85 & 18.03\% & 14.96\% & 3.07\\
1.20 & 29.91\% & 26.26\% & 3.65\\
\bottomrule
\end{tabular}
\end{table}

At peak, physical sphere--triangle gaps for amplitudes 0.85 and 1.20 reach
$-0.0683/-0.0460L_0$ and $-0.0825/-0.0803L_0$, respectively. Thus a
converged penalty equilibrium can remain appreciably penetrated. Registered
$4\times$ and $16\times$ stiffness runs fail to complete at either
amplitude, so no stiffness-independent contact result follows.

For amplitude 0.85 at fixed duration, the 9-stage solve stops at stage 6.
Complete 17-, 33-, and 65-stage solves give immediate released compression
of 14.9616\%, 14.9541\%, and 14.9497\%. The failed point is not used as a
asymptotic-rate estimate. After removal, the 0.85 state is unchanged through a
four-unit hold, while the 1.20 case retains a small nonzero internal plastic
rate.
\begin{figure*}[t]
\centering
\includegraphics[width=\textwidth]{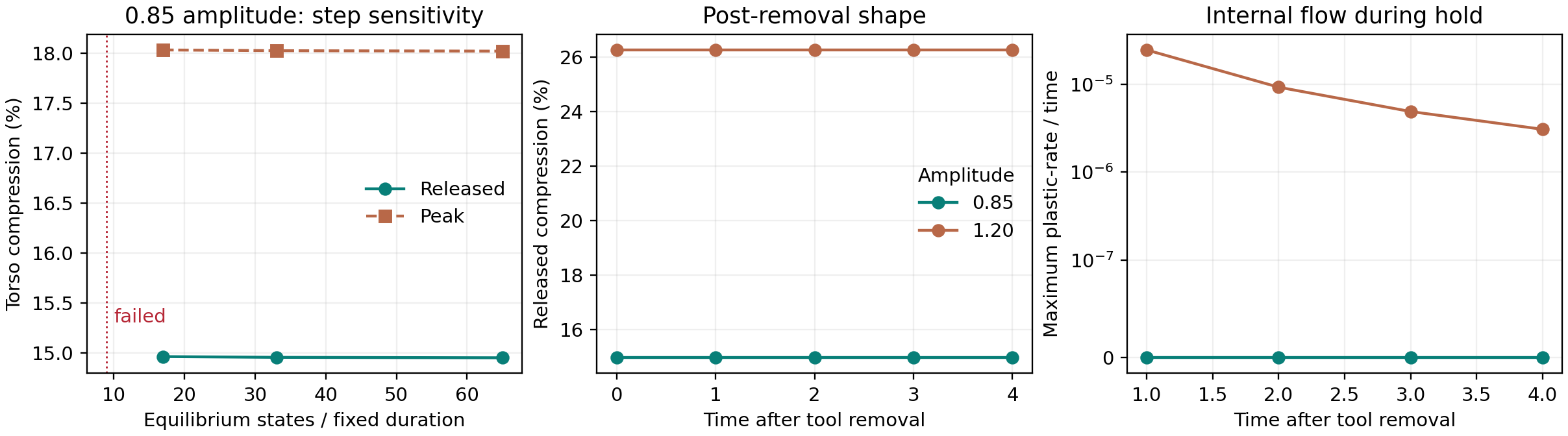}
\caption{Archived penalty-contact step refinement and post-removal holds.
Only completed runs contribute peak/release measurements.}
\Description{A failed coarse solve and completed refined solves are compared
with gauge and plastic-rate hold histories.}
\label{fig:penalty-stage-release}
\end{figure*}

\subsection{Penalty-Path Derivative and Negative Inverse Result}
On a cleared 17-stage penalty-contact path, a hybrid implicit derivative for
one symmetric amplitude was compared with independent complete forward
replays. At $h=10^{-3}$, the released torso-gauge derivative differs by
0.0277\%, and the largest of three fixed released-surface projection gaps is
0.283\%. The maximum stage linearisation residual relative to its
right-hand side is $5.55\times10^{-4}$ over 203 GMRES products. This check
does not validate the barrier/CCD path, a six-control gradient, or a native
all-AD finite-element implementation.
\begin{figure*}[t]
\centering
\includegraphics[width=\textwidth]{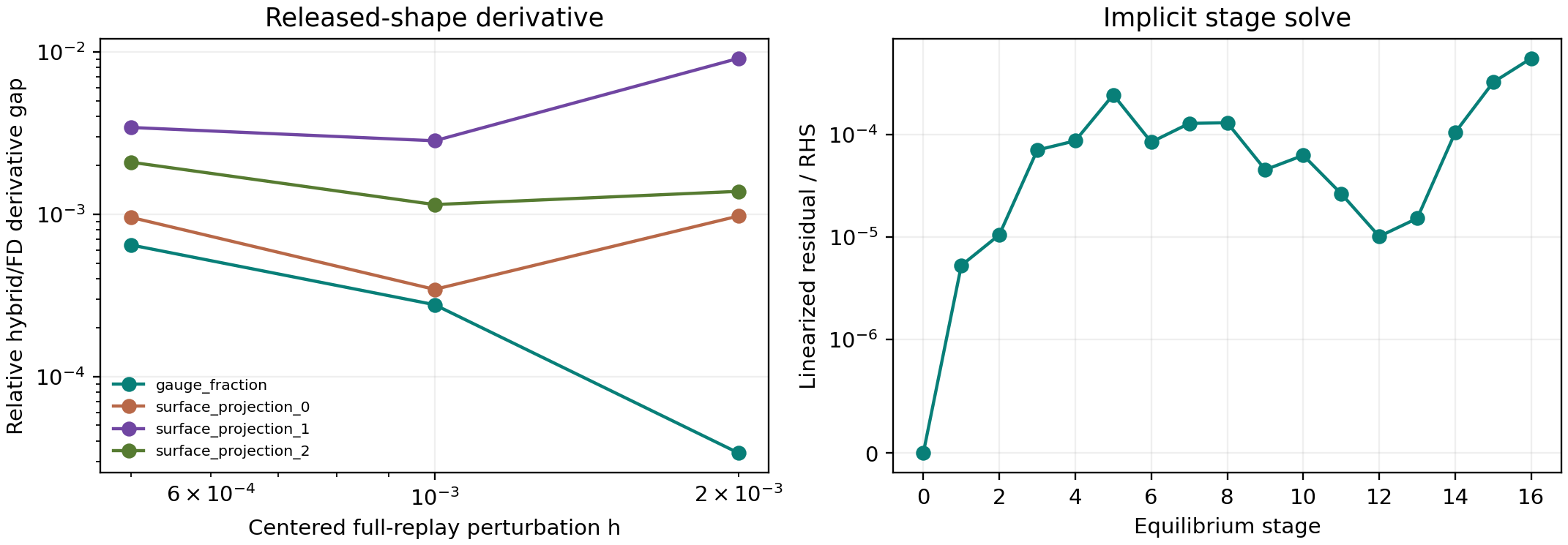}
\caption{Archived finite-$\boldsymbol F_p$ directional derivative under
vertex-penalty contact. It is intentionally not presented as
barrier/CCD derivative evidence.}
\Description{Finite-difference derivative gaps and stage linearisation
residuals are plotted for the penalty-contact path.}
\label{fig:penalty-gradient}
\end{figure*}

A matched one-control inverse check uses the same mesh, exact-$J_2$ target,
start, schedule, and three-law scalar Gauss--Newton budget for each law.
The exact target-model advantage is explicit.
\begin{table}[t]
\caption{Archived released-surface inverse errors. Percentages are normalised
by target-to-rest RMS motion; the clean-rest target moves only
$0.000495L_0$ RMS.}
\label{tab:penalty-inverse}
\centering
\begin{tabular}{lrr}
\toprule
Law & Previously yielded & Clean-rest transition\\
\midrule
Exact $J_2$ & 0.00223\% & 0.01480\%\\
Perzyna & 0.02098\% & 0.68167\%\\
SECA & 0.17605\% & 0.89626\%\\
\bottomrule
\end{tabular}
\end{table}
SECA finishes behind the implemented Perzyna baseline in both cases. The
previously yielded case also has archived rest-space penetration; the
clean-rest target removes that confound but is visually small. These are
negative results. They do not support superior inverse authoring and are not
upgraded by the feasible forward Bunny state sequence.